\documentclass[]{tADP2e}

\begin{document}
\doi{10.1080/00018732.2013.808047}
 \issn{1460-6976}
\issnp{0001-8732}  \jvol{62} \jnum{02 (pages 113-224)} \jyear{2013} \jmonth{18 June}

\markboth{K. Ostrikov \& et al.}{Advances in Physics}

\articletype{REVIEW ARTICLE [Advances in Physics, v. 62, Issue 2 (18 June 2013), pp. 113-224 (2013);
DOI: 10.1080/00018732.2013.808047]}

\title{Plasma Nanoscience: from Nano-Solids in Plasmas to Nano-Plasmas in Solids
}

\author{K. Ostrikov$^{\rm a}$ $^{\ast}$
\thanks{$^\ast$Corresponding author. Email: kostya.ostrikov@csiro.au \vspace{6pt}},
E. C. Neyts$^{\rm b}$ \& M. Meyyappan$^{\rm c}$ \\\vspace{6pt}  
$^{\rm a}${\em{CSIRO Materials Science and Engineering, 
P.O. Box 218, Lindfield NSW 2070, Australia}} \\
$^{\rm b}${\em{University of Antwerp, Universiteitsplein 1,
B-2610 Wilrijk-Antwerp, Belgium}} \\
$^{\rm c}${\em{NASA Ames Research Center Moffett Field, CA 94035, USA}} \\\vspace{6pt}
\received{21 May 2013}
}

\maketitle

\begin{abstract}
The unique plasma-specific features and physical phenomena in the organization of 
nanoscale solid-state systems in a broad range of elemental composition, structure, 
and dimensionality are critically reviewed. These effects lead to the possibility to 
localize and control energy and matter at nanoscales and to produce self-organized 
nano-solids with highly unusual and superior properties. A unifying conceptual 
framework based on the control of production, transport, and self-organization of 
precursor species is introduced and a variety of plasma-specific non-equilibrium 
and kinetics-driven phenomena across the many temporal and spatial scales is 
explained. When the plasma is localized to micrometer and nanometer 
dimensions, new emergent phenomena arise. 
The examples range from semiconducting quantum dots and nanowires, 
chirality control of single-walled carbon nanotubes, ultra-fine manipulation of graphenes, 
nano-diamond, and organic matter, to nano-plasma effects and  
nano-plasmas of different states of matter.\bigskip

\centerline{\bfseries Contents}\medskip

\hspace*{-12pt} {1.}    Introduction \\
\hspace*{7pt} {1.1.}  Scope of plasma nanoscience research\\
\hspace*{7pt} {1.2.}  Aims, focus, and unifying physics\\
\hspace*{7pt} {1.3.}  Organization of the review\\

\hspace*{-12pt} {2.}  Self-organized plasma-solid systems  \\
\hspace*{7pt} {2.1.}  Deterministic requirements at nanoscales \\
\hspace*{7pt} {2.2.}  Spatial scales and controls \\
\hspace*{7pt} {2.3.}  Self-organization pathways and plasma effects \\
\hspace*{24pt} {2.3.1.} General formalism \\
\hspace*{24pt} {2.3.2.} Plasma effects \\
\hspace*{24pt} {2.3.3.} Ion-related effects \\
\hspace*{24pt} {2.3.4.} Temporal dynamics \\

\hspace*{-12pt}{3.}  Plasma controls in nano-solid formation: fundamentals  \\
\hspace*{7pt}{3.1.} Generation of building units \\
\hspace*{7pt}{3.2.} Plasma sheath and surface conditions \\
\hspace*{7pt}{3.3.} Nanoscale plasma-surface interactions   \\
\hspace*{7pt}{3.4.} Nanostructure nucleation \\
\hspace*{7pt}{3.5.} Nanostructure growth \\
\hspace*{7pt}{3.6.} Self-organized pattern formation \\

\hspace*{-12pt} {4.}  Nano-solids from plasmas: plasma-specific effects and physical, chemical,
and functional properties  \\
\hspace*{7pt}{4.1.}   Quantum dots, nanocrystals and nanoparticles\\
\hspace*{24pt} {4.1.1.} Quantum dots \\
\hspace*{48pt} {4.1.1.1.} Semiconducting quantum dots \\
\hspace*{48pt} {4.1.1.2.} QD formation and properties: numerical simulations \\
\hspace*{48pt} {4.1.1.3.} C-dots \\
\hspace*{24pt} {4.1.2.} Nanocrystals and nanoparticles \\
\hspace*{48pt} {4.1.2.1.}  Nanocrystals from low-pressure plasmas \\
\hspace*{48pt} {4.1.2.2.}  Nanocrystals from microplasmas \\ 
\hspace*{48pt} {4.1.2.3.}  Nanocrystals from plasmas in liquids\\ 
\hspace*{24pt} {4.1.3.} Nanoarrays \\
\hspace*{7pt}{4.2.} One-dimensional nanostructures  \\
\hspace*{24pt} {4.2.1.} 1D carbon nanostructures \\
\hspace*{24pt} {4.2.2.} Silicon nanowires \\
\hspace*{24pt} {4.2.3.} Other inorganic nanowires \\
\hspace*{7pt}{4.3.} Two-dimensional nanomaterials \\
\hspace*{24pt} {4.3.1.} Graphene, graphene nanoribbons, and graphene oxide \\
\hspace*{48pt} {4.3.1.1.} Graphene \\
\hspace*{48pt} {4.3.1.2.} Graphene nanoribbons\\
\hspace*{48pt} {4.3.1.3.} Graphene oxide\\
\hspace*{24pt} {4.3.2.} Graphene nanosheets \\
\hspace*{48pt} {4.3.2.1.} Surface-supported vertical graphenes \\
\hspace*{48pt} {4.3.2.2.} Unsupported graphenes from gas phase \\
\hspace*{24pt} {4.3.3.} Other 2D nanomaterials \\
\hspace*{7pt}{4.4.} Hybrid nanomaterials \\
\hspace*{7pt}{4.5.} Three-dimensional nanomaterials  \\
\hspace*{24pt} {4.5.1.} Nanoscale plasma etching \\
\hspace*{24pt} {4.5.2.} Self-organized arrays of inorganic nanotips \\
\hspace*{24pt} {4.5.3.} Carbon nanotips \\
\hspace*{24pt} {4.5.4.} Nanodiamond \\
\hspace*{24pt} {4.5.5.} Soft organic matter \\
\hspace*{24pt} {4.5.6.} Biological objects \\

\hspace*{-12pt} {5.}  Nano-plasmas: interplay of the size and the fourth state of matter  \\
\hspace*{7pt}{5.1.} Non-equilibrium nano-plasmas of solids \\
\hspace*{7pt}{5.2.} Nano-plasmas meet plasmons \\

\hspace*{-12pt} {6.}  Conclusion and outlook  \\

Acknowledgments  \\

References \\

Appendix A. Alphabetic list of acronyms used in the review \\

\end{abstract}

\bigskip

\section{\label{intro} Introduction}

Nanoscience is a commonly known research field which deals with objects of nanometer and sub-nanometer 
dimensions, the basic organizing principles of these objects, and the unique properties these objects possess compared to bulk objects made of the same material(s) \cite{BaderRMP2006}. But what is {\em plasma nanoscience}? From the combination of 
the two terms {\em plasma} and {\em nano}, it is clear that 
this is a field at the intersection of plasma science and nanoscience. 
A more in-depth understanding comes from focusing on {\em dimensions}, {\em organization}, and {\em properties}. 
One should also note that matter exists in 4 basic states, and the plasma state is one of them. 
From basic symmetry considerations, nanometer-sized objects should exist 
in these 4 basic states, namely solid, liquid, gas, or plasma.  
 
From merely the dimensional perspective, plasma nanoscience should deal with the plasma state of matter confined 
to nanometer dimensions. However, from the organizational perspective, there is a variety of 
options of assembly of nanoscale objects using precursors in different states. 
Indeed, one can assemble solid nanoparticles using (e.g., precipitating) 
precursors in the liquid, gaseous, or plasma states. 
Liquid nano-droplets can be produced by manipulating (e.g., heating or condensing) 
precursors in the solid, gaseous, or plasma states, etc.  
Depending on the specific way of organization, the arrangement of atoms in nanoscale objects may also be 
different. This, in turn, leads to differences in the physical and chemical {\em properties} of the objects produced.       

Hence, there arises another possibility - to use precursors in the ionized (plasma) state and follow how they organize 
into solid nanoscale objects. 
A plasma typically produces a plethora of different precursor 
species (molecules, atoms, and radicals), in a large number of different 
energetic states (ionized, excited, metastables, and ground states) compared to 
solid, liquid or gaseous precursors. These plasma precursors are thus very likely to
assemble into solid objects with quite different 
atomic arrangements. This in turn will lead to the different properties of these nano-solids.  

This possibility also exists for bulk objects. However, when dealing with small objects, greater 
precision is required, simply because of the basic requirement to confine matter to nanometer or even subnanometer 
dimensions. This is why the way of organizing matter at the nanoscale should be more precise,
more ``careful". 
These basic considerations also apply to modifying existing 
nanoscale solid objects. In this case, there are numerous options to modify such objects using species in the same or different 
states of matter. Therefore, using plasma-produced species should lead to quite different outcomes compared to other 
environments.

\subsection{\label{introA} Scope of plasma nanoscience research}

In basic terms, what does plasma nanoscience study? On one hand, it studies the unique way 
of using ionized gases to produce or modify nanoscale solids, thereby giving them different 
properties compared to other ways of organization that use precursors in other states of matter. 
This is why the title of this review contains ``nano-solids in plasmas". 
On the other hand, if it is possible to confine the plasma to nanometer dimensions, it will almost certainly have 
very different properties compared to bulk plasmas. This focus is denoted by the ``nano-plasmas" in the title. 

As will be discussed in this review, the ability to control nanoscale 
localization of energy and matter delivered from the bulk plasmas to 
developing nano-solids, is the key to achieve the desired 
morphological, structural, and functional properties. On the other hand,
the energy and matter can also be confined to nanoscales if the plasma 
itself is confined to the nanoscale, in other words, when nano-plasmas are 
generated. These nano-plasmas can be sustained near the surface 
of a bulk solid, for example under extremely non-equilibrium conditions 
of very dense plasmas of physical vapors of solids. 
The bulk plasmas  can be used to produce nanoscale solids
which in turn can confine nano-plasmas, or have nano-plasmas in their 
vicinity. In all these cases, the plasma state is in contact with a surface 
with nanoscale features or dimensions.
The physics of these nanoscale interactions is what makes this field particularly
interesting.     
This is the reason why the title of this review bridges 
nano-solids (produced) from plasmas 
with nano-plasmas (sustained) in or near solids. 
  
In general,       
plasma nanoscience focuses on the specific roles, purposes, and benefits of 
the ionized gas (plasma) environments in assembling, 
processing, and controlling microscopic and 
nanoscale (including biological) objects in natural, 
laboratory, and technological environments and to 
find the most effective ways
to ultimately bring these unique plasma-based processes to 
the deterministic level \cite{my-plasma-nanoscience-book}. 
One of the primary aims of this research is to find optimum plasma 
process parameters to minimize the number 
of experimental trials one needs to undertake
to achieve nanostructures and nanomaterials with the desired properties, which
in turn determine their performance in practical applications. Numerous 
effects of microscopic and nanoscale localization of energy and matter 
in plasmas produced from gases, liquids, and solids and the associated 
interactions of such plasmas with matter in different phases are also of
primary interest.

Plasma nanoscience is a relatively new research field which emerged in the last two decades because of the rapid advent 
of nanoscience and nanotechnology and continuously increasing demand for applications of low-temperature plasmas 
in nanoscale synthesis and processing. Such plasmas have been used in materials science and microelectronics for decades as a 
viable fabrication tool. 

However, as the number of nanomaterials of different structures, composition, morphology, and 
dimensionality experienced a rapid surge, it became increasingly clear that nanoscale synthesis and processing is 
very different compared to bulk solids. 
From the fundamental perspective, this refers to the much higher required precision in the ability to control energy and matter 
during the production or modification of nanomaterials. 

Following the pioneering discovery of fullerenes 
in laser-generated vapor plasmas of graphite \cite{KrotoNature1985} and 
carbon nanotubes produced from arc discharge plasmas \cite{Iijima}, the 
number of successful applications of low-temperature plasmas in nanofabrication has skyrocketed.
From merely carbon nanotube synthesis and microstructuring of semiconductors in early 1990s, the plasma 
success stories have expanded into a plethora of pure and hybrid nanoscale objects of virtually any 
dimensionality spanning across a very large number of materials systems. 
More importantly, in some cases it became possible to achieve certain objectives 
only using tailored plasmas while other techniques failed.  
This is why a clear understanding on how exactly plasma-specific effects work 
in such nanoscale processes, was urgently required.

After years of experimental trials, theoretical studies, and numerical simulations,
it became possible not only to use low-temperature plasmas for nanoscale processing with the 
level of confidence comparable with processing of bulk materials but also, in many cases, 
achieve outcomes superior to those achieved by other common techniques, e.g., based on thermal 
chemical vapor deposition (CVD), wet chemistry-based synthesis and processing, laser-assisted
microfabrication, etc.   
This review will focus on such examples and on the underlying physics of the 
elementary processes involved.     

Plasma nanoscience is intrinsically multidisciplinary. It involves elements of basic sciences such as 
physics and chemistry of solids, liquids, gases, and plasmas, materials science, nanoscience, optics, astrophysics, 
molecular, radical, and cell biology, etc. From the applications perspective, it also 
involves knowledge from related engineering and technological sciences such as nanotechnology, 
materials and device engineering and design, as well as chemical, mechanical, electrical, electronic, and optoelectronic 
engineering, etc. 

There is a large number of relevant monographs and reviews published to date
(see e.g., \cite{MeyyappanPSST03, RMP2005, Gordillo-Vazquez2007, HoriGoto2007, Kortshagen-review, Meyyappan2009, Mariotti-Sankaran, 
XingguoLi, HatakeyamaJPD2011, ZFRen-AdvPhys2011, Editorial_JPD2011, 
Mohan-Book-2012, NeytsJVSTB2012, reserved1, reserved2} and references therein). These publications mostly focus on the 
types of plasmas used, materials systems, envisaged applications, etc. 
The {\em scope of this review} is unique as it aims to explain, from the unifying 
physical principles (e.g., involving concepts of 
nanoscale localization, determinism, self-organization, 
non-equilibrium, complexity, etc.), how plasmas can be used to produce or process 
matter (including hard, soft, and living matter)  with 
spatial localization in the sub-nanometer to micrometer range, 
as well as to discuss the unique properties of nano-plasmas. 

\begin{figure}[t]
\begin{center}
\includegraphics[width=13cm,clip]{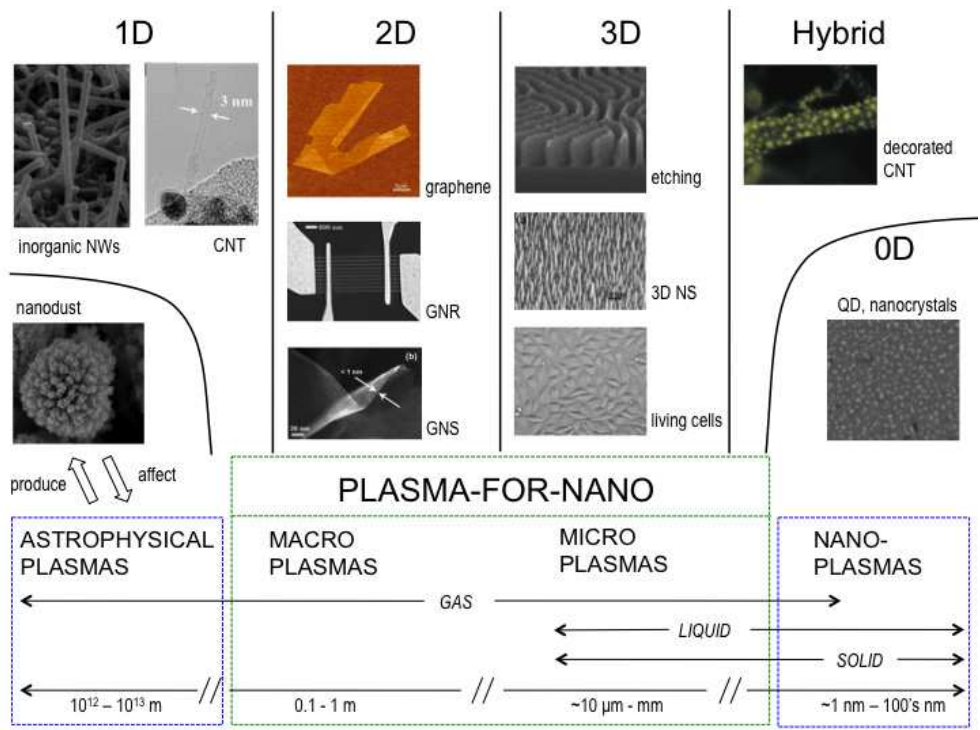} 
\caption{ \label{f1} Scope of plasma nanoscience research. Plasmas in gases and
liquids are used to produce nanoscale solid objects of various dimensionality, 
structure, and elemental composition (denoted ``plasma-for-nano" in the figure).
Plasmas with micrometer and nanometer spatial localization 
(denoted ``micro-" and ``nano-plasmas" in the figure, respectively)  
are possible in gases, liquids, and solids. Generation and 
collective behavior of dust nanoparticles 
in astrophysical and space environments (denoted ``astrophysical plasmas" in the figure) 
is also of interest but outside the scope of this review. 
Fragments of images are reproduced with permissions from  
\cite{KatoCVD2006} [Copyright \copyright 2006 Wiley-VCH Verlag GmbH \& Co. KGaA, Weinheim],
\cite{NovoselovPNAS} [Copyright \copyright (2005) National Academy of Sciences, USA],
\cite{MuraliAPL09} [Copyright \copyright (2009), American Institute of Physics], 
\cite{GNS-Carbon2007} [Copyright \copyright (2007), with permission from Elsevier],
\cite{XuJPD2011-SpecIssue,Morris} [Copyright \copyright IOP Publishing. Reproduced by permission of IOP Publishing. All rights reserved].  
All other fragments are from original unpublished research in K. Ostrikov's laboratories. 
All relevant team members are gratefully acknowledged. Acronyms: 0D -- zero-dimensional, 1D -- one-dimensional, 
2D -- two-dimensional, 3D -- three-dimensional,
NS -- nanostructure, NW –- nanowire, QD -- quantum dot, CNT –- carbon nanotube, 
GNR –- graphene nanoribbon, GNS –- graphene nanosheet. }   
\end{center}
\end{figure}

Figure~\ref{f1} shows the scope of plasma nanoscience research. 
In addition to the already mentioned nano-solids from plasmas (denoted and used hereafter as {\em plasma-for-nano} 
for brevity and simplicity) and {\em nano-plasma} domains, it also involves {\em astrophysical plasmas}. These domains, although vastly different 
by their spatial localizations (from 10$^{12}$--10$^{13}$m for plasmas at astrophysical scales down to nanometer-sized plasmas), 
can have one crucial common attribute - nanoscale objects. 

Indeed, nano- and micrometer-sized solid dust nucleates in the 
relatively cold environments of stellar envelopes and plays a critical role in the energy and mass transfer on the scale of the 
Universe, where plasmas constitute more than 99\% of visible matter \cite{Mann}. This relation between the microscopic 
solid objects and plasmas is reflected by the arrows on the left hand side of Fig.~\ref{f1}: plasmas produce (dust) nanoparticles 
which in turn affect the plasma. This interdependence persists both locally, over microscopic scales, and globally, over
macroscopic scales, which can be as large as the dimensions of astrophysical objects such as galaxies.
 
This leads to the two essential attributes of the {\em plasma-solid systems} of our interest. 
The first attribute is 
{\em self-organization} which results from interplay between the plasma and solid components involved. The other 
attribute is the {\em multi-scale nature} of this system, where very diverse 
physical phenomena may take place at different spatial and temporal scales. 

These basic attributes are also common to other phase combinations 
and  to the {\em plasma-for-nano} and {\em nano-plasma} domains in Fig.~\ref{f1}. The plasma-for-nano domain 
involves plasmas in gases and liquids  localized across the dimensions of plasma reactors commonly used 
for nanomaterials synthesis and processing (0.1--1 m) and more recent microplasma discharges ($\sim$10$ \mu$m--1 mm).  
Plasmas in both gases and liquids have been used for the synthesis and processing of 
nanoscale objects of different elemental composition, structure, and dimensionality, as shown in Fig.~\ref{f1}. 
These objects include zero-dimensional (0D) quantum dots (QDs) and nanocrystals, one-dimensional (1D) nanowires (NWs) and
carbon nanotubes (CNTs), and two-dimensional (2D) flat (e.g., graphene, organic monolayers) 
and bent (e.g., graphene nanoribbons (GNRs), graphene nanosheets (GNSs), nanowalls, etc.) structures. 

Further examples include 
more complex three-dimensional (3D) objects such as vertical pyramid-like pillars produced by plasma etching, other 
3D nanostructures (NSs), and also living cells. Many of these objects have intricate internal and surface 
structures. For example, living cells have organelles of sub-micrometer 
dimensions both in the intracellular space and on the surface, CNTs have narrow channels along their length,
whereas many other nanomaterials have similarly small features both in their interior (e.g., embedded quantum dots or pores, 
regular arrays of atomic vacancies, etc.) or on the surface (e.g., nanometer surface texture).  
 
These objects may also be arranged in complex patterns and arrays. For instance, CNTs or inorganic NWs
can be arrayed through self-organized growth while  
nano-pillars can be carved out of a monolithic piece of crystalline Si wafer by reactive etching. These are the two 
{\em fundamental approaches to nanoscale synthesis}. The first case refers to the bottom-up, self-organized 
growth or nano-assembly while the second one is an example of top-down nanofabrication. 
Importantly, not only can a plasma be used in both cases, but in many cases it also shows superior performance compared to 
other (e.g., thermal CVD and wet-chemistry-based) nanoscale synthesis and processing techniques. In some cases, plasmas 
can even enable nanoscale processes which are not possible otherwise under the same conditions. 
This review will show examples of specific benefits of using plasmas in nanoscale 
synthesis and processing and reveal the underlying physical processes that make these benefits possible. 

Nanoscale structures can also be organized on the surface or inside other objects of different dimensionality.
Figure~\ref{f1} shows an example of a hybrid structure made of a 1D carbon nanotube decorated by small 0D 
metal nanoparticles. Such hybrid and multidimensional arrangements are used to modify properties 
of the nanomaterials involved and produce physical effects (e.g., electron transport, 
energy band structure, interaction with light, etc.) which are difficult 
or impossible to achieve using the constituent materials separately.   
Plasmas have also been used to produce and modify such (and many other) 
hybrid multi-dimensional objects, also with several advantages 
compared to other approaches.    

The last (but not the least) domain in Fig.~\ref{f1} is {\em nano-plasmas}. 
By nano-plasmas, we refer to plasmas confined to sub-micrometer dimensions. 

Here we would like to stress two important features shown in Fig.~\ref{f1}.
First, plasmas can be produced from gases, liquids, and solids. 
The fourth, plasma state is achieved through state transformations from 
gas, liquid, or solid phases. When such transformation is complete, 
the same bit of original matter cannot simultaneously be in the plasma 
and some other (e.g., solid, gas, or liquid) state. However, incomplete 
phase transitions may result in mixed phases, e.g., partially ionized gases where the 
ionized (plasma) and charge neutral gaseous phases coexist.

Second, the plasmas feature very different sizes. 
The denser and cooler the matter, the smaller is the minimum size of the plasma which 
is determined by a few multiples of the electron Debye length 
\begin{equation}    
\label{Debye-length}
\lambda_D = V_{Te}/\omega_{pe},
\end{equation}
which decreases with the plasma density and increases with the electron temperature. 
Here, $V_{Te}=(k_B T_e/m_e)^{1/2}$ is the electron thermal velocity, $\omega_{pe} = (4\pi n_{e0} e^2/m_e)^{1/2}$ is the Langmuir (plasma) frequency, 
$e$, $T_e$, $m_e$, and $n_{e0}$ are the charge, temperature, effective mass, and equilibrium number density of electrons,
respectively, and $k_B$ is Boltzmann's constant.   
According to the classical plasma definition, charge neutrality has to be preserved well beyond the Debye sphere 
with $\lambda_D$ as a radius. 
However, the principle of the Debye length does not apply to all plasmas. This 
is the case for electronegative or electropositive plasmas with unbalanced electric charges or microplasma 
discharges with delocalized charge distributions 
where charge balance is sustained dynamically.   

The minimum plasma size can be in the nanometer range, e.g., under conditions of  
ultra-high-energy-density laser ablation of solid targets. 
This is why ultra-small laser plasmas may be generated in localized areas on solid surfaces thereby producing 
tiny features on the surface. The  possibilities and specific mechanisms to enable these 
effects will also be considered in this review.

\subsection{\label{introC} Aims, focus, and unifying physics} 

This review aims to critically review the physical phenomena related to:

\begin{itemize}

\item[(1)]	{\em nanoscale organization} of solid-state matter using plasmas in gases and liquids, and 

\item[(2)]	{\em nano-plasmas} confined in sub-micrometer volumes and possibly surrounded by gas, liquid, or solid matter or a combination thereof, 

\end{itemize}
by focusing on the plasma-specific organizing principles, nanoscale 
energy- and matter localization effects, as well as the arising structural, 
morphological, and functional properties of the nano-solids involved.

In other words, here we focus on highly-unusual features of 
non-equilibrium plasmas and physical mechanisms 
of using these features to achieve the required  
nanoscale solid objects and their arrays and networks. 
The particular highlight here is on {\em nanoscale localization of 
energy and matter} and {\em non-equilibrium and kinetic features} 
that determine the physics of nanoscale interactions in the 
plasma-solid systems most common to nanoscience.   
We also critically examine the physics of nano-plasmas generated 
near solid surfaces.

In particular, this review aims to clarify the following questions: 
\begin{itemize} 
\item{what can be achieved at the nanoscale using plasmas?} 
\item{what unique plasma-specific physical processes (e.g., non-equilibrium, complexity, 
kinetics, self-organization, etc.) are involved?}
\item{how can plasma-specific effects be used to tailor the  
organization and properties of the nanoscale solid objects to meet the requirements in applications?} 
\item{which plasmas to use to achieve superior outcomes?}
\item{can the plasma be confined to nanoscales and how can it be used?}
\end{itemize}

In a simple catchphrase, the aim is to show {\em how and why plasma-specific effects make a difference at nanoscales}.

The unifying physics-based approach to describe these phenomena 
is based on the concepts of {\em nanoscale localization},  {\em determinism}, {\em organization}, {\em non-equilibrium}, and {\em complexity}. 
As mentioned above, the nanoscale localization of energy and matters 
arises when the dimensions of either the solid objects or the plasmas 
are in the nanometer range, or when both are within nanoscales. 
By nanoscales here we broadly imply spatial scales ranging from 
fractions to hundreds of nanometers; outside of this range we 
use terms ``atomic" and ``micrometer" to determine the relevant scales.  

Determinism stems from Feynmann's vision to ``arrange atoms, 
the way we want them", which is one of the guiding principles of 
nanoscience. This usually refers to achieving 
the required effects by customizing the nanoscale organization of matter. 
The latter can be achieved using either a ``top-down" fabrication or 
``bottom-up" self-organization or a combination thereof. The importance of 
controlled self-organization is rapidly gaining momentum 
as the feature sizes of nanoscale objects continuously 
decrease and the ``top-down" techniques are rapidly reaching their intrinsic 
physical limits. Non-equilibrium conditions are critical  
for nanoscale synthesis while complexity determines the available controls of organization.

The required effects can be achieved through nanoscale organization of matter 
using plasmas. Plasmas feature a higher degree of complexity and allow nanoscale synthesis 
under unique non-equilibrium conditions compared to neutral gases.
Under such conditions the structure and properties of the produced nanoscale solid objects are determined by the 
complex interplay of the process kinetics and equilibrium thermodynamics.  
These objects find themselves in exotic metastable states, which offer truly unique 
structural, electronic, optical, and other properties hardly achievable in the 
equivalent bulk materials. This is why one can expect many distinctive possibilities 
for the highly-controlled synthesis of more complex and ``non-equilibrium" 
nanoscale objects from non-equilibrium plasmas.

Below, we will discuss how plasma-specific, long-range 
electromagnetic field-related effects can be used to control 
the structure and hierarchical arrangements of many nanoscale solid objects and their patterns.
Plasma-surface interactions also enable unique control over the surface energy, 
shaping, faceting, etc. of these objects by significantly affecting the 
multi-stage nanopattern development from nucleation of small clusters to nanostructure
self-organization in 3D mesoscopic patterns.  

Therefore, one can expect that exotic non-equilibrium plasma conditions
may lead to the fast, dynamic, non-equilibrium way of organization of solid 
matter. Moreover, the specific complexity of the plasma state should lead to many diverse options for  
effective control of self-organization, which may not be available for
other states of matter. Ultimately, the self-organization which is controlled/guided 
by the plasma-specific effects, should reach the deterministic level, when it is possible to produce  
nanomaterials with the desired properties. Plasma nanoscience research helps finding and utilizing this control.

\subsection{\label{introE} Organization of the review} 

The aims, focus, and the unifying physics defined above determine the structure of the review.
The main part is composed of the 4 main sections.  
The {\em overarching logic} is to first identify the most relevant plasma-specific control mechanisms in 
the assembly of nanoscale solids, then show how they work at every stage of this assembly, 
and then show specific examples from a broad range of materials systems, followed by 
discussing the physics of localizing plasmas to nanometer dimensions. 

Section~\ref{Sec2} will introduce a conceptual approach to deterministic 
nanoscale synthesis and processing 
through the effective control of self-organization pathways in 
plasma-solid systems (PSSs).  
This section will first introduce the PSSs of interest, the 
spatial scales and basic control 
mechanisms involved. The key focus here is on the most relevant 
self-organization pathways and the plasma-specific effects to control them. 

In Sec.~\ref{Sec3} 
we will follow the nanoscale assembly, step-by-step, 
from precursor species generation to nucleation and growth 
of individual solid nanostructures and their patterns, and 
show uniquely plasma-specific effects during each   
step. This section will use a few focused examples of the 
most common nano-solids. 

Section~\ref{Sec4} provides a broad coverage of nanoscale 
objects and features assembled or modified using plasmas. 
Each subsection discusses the intended applications of related 
solid materials, existing problems in their synthesis and processing using 
other approaches, plasma-specific effects that enable the desired 
effects (e.g., advantages in the synthesis process, 
superior properties, etc.), and the remaining challenges.
Wherever appropriate, these effects are linked to 
the physical effects from Secs.~\ref{Sec2} and \ref{Sec3}.
Because of the very large number of examples that 
cover dielectric, semiconducting, and metallic materials in 
all dimensionalities (from 0D quantum dots from 
3D nanoarrays, as well as structures of a hybrid 
dimensionality), it is impossible to provide the 
exhaustive and in-depth coverage of all structural, 
morphological, and functional (e.g., electronic, optical, magnetic, etc.) 
properties of all the materials involved. 
While we discuss the most essential of such properties
(most prominently those that are affected by the plasma), 
wherever possible, we refer the reader to relevant 
publications which focus on some of these properties. 
  
The scope of Sec.~\ref{Sec5} is the plasmas localized to 
nanometer dimensions. Examples include very dense 
plasmas of physical vapors of solids and 
plasmon-assisted nano-plasma formation near the 
surface of the nanoparticle.

The review concludes with the summary of the 
most interesting and generic physical effects  
and a brief outlook for the future research in this 
field. For the convenience of readers, the acronyms used in this 
review are summarized in Appendix~\ref{acronyms}.               

Given the breadth of the topic, it will be a futile attempt to provide 
the fully exhaustive coverage of 
the relevant existing knowledge. This is why we apologize to the 
authors of any relevant publications which are 
not mentioned due to the obvious space and time limitations. 
This is why we cover the {\em plasma-for-nano} and {\em nano-plasma} 
domains in Fig.~\ref{f1}, primarily focusing on the 
plasma-specific physical effects that make a difference 
at nanoscales while keeping in mind the 
unifying physical principles discussed in Sec.~\ref{introC}.  

The analysis and comments in this review may not always be 
the same with the opinions of other researchers.
This is why we welcome any comments or discussions on any relevant topic and hope that our effort 
will stimulate further development of this  very interesting field.
Just like a {\em difference} in chemical potentials between 
two phases drives self-organization (as will be discussed in the following section),  
a difference in scientists' opinions drives progress in the field.

\section{\label{Sec2} Self-organized plasma-solid systems}

Having mentioned the issue of controlled self-organization, we can now clarify how to achieve this in
self-organized plasma-solid systems (PSSs). 
A PSS is a system where a plasma faces a solid and affects 
processes on its surface or within its bulk. Such systems are non-equilibrium in the presence of significant 
transfer of energy and matter. This happens, e.g., when the system (e.g., a solid nanostructure on the surface) 
cannot reach its thermodynamically defined equilibrium (minimum free energy) state while new material is delivered. 
This happens during the growth or processing of nanoscale objects and the rates of dynamic material re-organization 
determine the nature and degree of non-equilibrium, which in the plasma can be very different than in 
the equivalent neutral gas. 
The system is self-organized when it has the ability to rearrange its constituents 
when it is left on its own, i.e., without any external action \cite{BarthNature2005}. 

The most critical 
factor that enables this ability is the {\em driving force}, which in many cases is related to significant 
variations of a physical quantity (e.g., density, temperature, etc.) in space and a mechanism capable 
to sustain any rearrangement within the system which can reduce these variations. 
For example, for multicomponent mixtures of different species on the surface, a generic 
expression for the driving force for nucleation \cite{drforceJCP2004}
\begin{equation}
\label{drivforce1}
\Delta g = \Sigma_i n_i \Delta \mu_i
\end{equation}
relates the supersaturation $\Delta g$, the number of species $n_i$, and the driving forces
\begin{equation}
\label{drivforce2}
\Delta \mu_i = \mu_{\alpha i} - \mu_{\beta i}
\end{equation}
for each individual component (species), where $\mu_{\alpha i}$ and $\mu_{\beta i}$ are 
the chemical potentials of the old (e.g., gas) and the new (e.g., solid) phases, respectively.
These chemical potentials depend on the prevailing process conditions, such as gas pressure, surface 
temperature, interface energy, plasma-specific parameters (e.g., ionization degree), etc. 
Importantly, the overall driving force (\ref{drivforce1}) must be positive for a new phase to form.
The composition of the new phase is affected by the driving forces of individual components 
(\ref{drivforce2}); not all of these species may necessarily form a new phase.  

The driving force formalism may be generically used during solid nanostructure 
growth on a surface from a plasma phase. In this case, building material can be delivered to the surface 
non-uniformly, and the difference between the local densities of adsorbed atoms (adatoms) sustains 
surface diffusion, which aims to minimize the initial non-uniformity of the adatom density. 
However, the rates of this reorganization will depend on the amount of energy the 
adatoms have, which in most cases is  
determined by the surface temperature, which also depends on the plasma conditions. 
Hence, the rates of thermal hopping of these adatoms will determine 
how fast the system can equilibrate. However, it already becomes clear that plasma-specific 
effects (e.g., surface heating and ion impact) will definitely affect adatom migration and 
consequently, the equilibrium states of the plasma-solid system. 

In this section, we will consider the 
basics of diffusion-driven self-organization and how exposure of a solid surface to a 
plasma can lead to faster and more exotic 
self-organization. While the focus here is on the cases involving mass relocation 
about the surface, ion-assisted surface nanostructuring (e.g., rippling) without 
significant mass relocation, is considered as well. 

One should always keep in mind that any rearrangement of atomic matter at nanoscales should be guided  
to achieve some specific purpose. 
In other words, such organization should ultimately be 
deterministic no matter which specific nanoassembly approach (e.g., bottom-up or top-down 
or a combination of the both) is used.    
This is why we will first show some examples of deterministic arrangement 
of nanoscale objects and will then discuss how to control 
self-organization in the plasma-solid systems of interest. 

\subsection{\label{basicA} Deterministic requirements at nanoscales}   

Without trying to give an exhaustive definition of determinism, it would suffice to say that this term refers to the ability to
achieve pre-determined outcomes - for example, produce specific nanostructures by customizing the plasma-based 
process and reduce the number of trials to the absolute minimum. These outcomes are usually determined by the 
specific requirements of the envisaged applications.  

For example, by controlling vapor transport, one can achieve size- and shape-selective growth of inorganic 
nanowires, nanoribbons and other related structures for electronic and optoelectronic applications  
\cite{Determinism2}, produce self-assembled, deterministic networks of wired nanotubes to achieve 
nanoscale electrical connectivity \cite{Determinism3}, biomaterials by design with the desired  
biological response \cite{Mater-By-Design2}, or customize crystalline structure of materials 
to achieve the desired properties by using first-principles calculations \cite{Mater-By-Design1}. 

In many cases, structural hierarchy at different spatial scales determines physical properties 
of nanoscale objects. For example, precisely designed molecular building blocks can be used to 
create hierarchical arrays and networks through reticular synthesis where structural integrity of these blocks remains 
unchanged during the assembly \cite{Hierarch1}. Hierarchically structured carbon nanotube arrays 
can be tailored to allow a strong shear binding-on and easy normal lifting-off thereby 
mimicking the ``gekko feet" effect \cite{Hierarch2}. Hybrid materials usually feature a significant hierarchy in the sizes of different components 
which is very common to nature-designed biological objects \cite{Hierarch3}. 

A certain degree of ordering of nanoscale objects is required in many applications. For example, two- or
three-dimensional order is required to improve the density of information storage \cite{ordered_dense_arrays_data_storage},
improve photoconversion efficiency of third-generation photovoltaic solar cells \cite{ordered_dense_arrays_PV},
or produce regular arrays of size-uniform metal nanoparticles for deterministic 
nanoscale synthesis  (e.g., CNT or QD arrays) \cite{regular_catalyst_arrays_CNT}.  
This order can be periodic or custom-designed aperiodic, or even irregular, and can apply to nanoparticle 
sizes, shapes, and other characteristics  \cite{ordered_arrays_plasmonics}.

\begin{figure}[t]
\begin{center}
\includegraphics[width=12cm,clip]{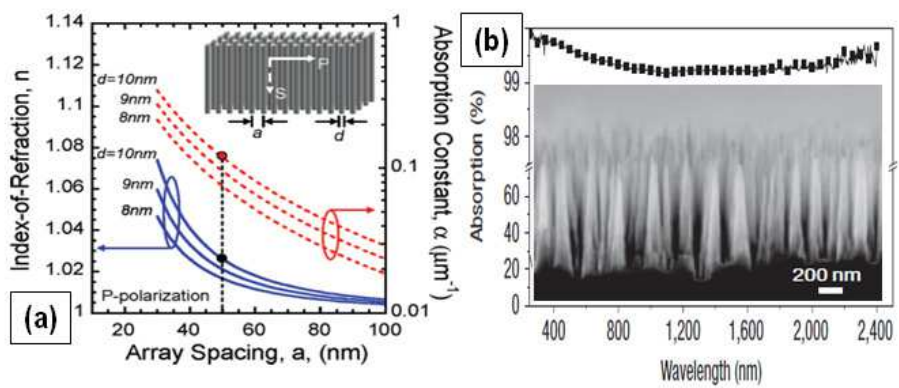} 
\caption{ \label{f2}  Two examples of deterministic requirements 
for anti-reflecting nanoarrays. 
By tailoring the spacing between a few-wall-thick carbon nanotubes,
it is possible to maximize anti-reflection properties of CNT arrays (a)
[Reprinted with permission from \cite{Darkest-Matter}. Copyright \copyright (2008) American Chemical Society]. 
A quite similar effect can be achieved 
by mask-less self-organized plasma etching of Si (b) [Reprinted/adapted by permission 
from Macmillan Publishers Ltd \cite{LCChenNatureNT2007}, Copyright \copyright (2007)].}  
\end{center} 
\end{figure} 

Determinism also implies selecting the best process that suits the requirements 
for specific applications. Figure~\ref{f2} shows two examples of 
deterministic arrangements of solid nanostructures to 
minimize light reflection from the surface; one process (a) relied on catalytic thermal CVD, 
while the other one (b) on plasma etching. The aim in both cases is to achieve 
a pre-determined level of light absorption. 
In the first case which involved 
multi-stage, highly-precise, and sophisticated fabrication of ordered carbon nanotube arrays,
a record low light reflection of a small fraction of a percent was achieved \cite{Darkest-Matter}. 
The light reflection in the second case was slightly higher, yet remained below 
1\% in a very broad spectral range \cite{LCChenNatureNT2007}. 

However, the arrays of silicon nanocones 
that enabled this functional performance, were produced in a very simple, cheap, and 
environment-friendly single-stage plasma process. 
Therefore, for anti-reflection devices with $\sim$1\% tolerance within the broad spectral 
range, the second, plasma-based process may be considered as a simpler and viable alternative. 
However, precise fabrication of CNT arrays will certainly be advantageous for 
anti-reflection applications with tolerance levels of just a fraction of a percent.

In the first case (Fig.~\ref{f2}(a)), thin 
(few-wall) carbon nanotubes of thickness $d$ were arranged in a regular array using 
pre-patterned metal catalyst \cite{Darkest-Matter}. The inter-nanotube spacing was adjusted to achieve 
the extremely low coefficient of optical reflection $R^p = 0.02$\% for $p$-polarized light. 
Numerical results in Fig.~\ref{f2}(a) suggest that this can be achieved when the array spacing $a$ 
between $\sim$10 nm thick CNTs is $\sim$50 nm. In this case the effective refraction index 
$n^p_{eff}=$1.026 and the effective absorption constant $\alpha^p_{eff}=$0.12. 
$R^p$ can be further decreased by adjusting the array spacing. 
However, any such adjustment requires changing the entire catalyst pattern, which in turn involves 
several nanofabrication steps, such as using masks, pattern delineation, catalyst material evaporation through 
the mask, etc. Carbon nanotube growth is usually carried out after the catalyst pattern is prepared. Different 
nanotube patterns usually require new masks or templates with different pore sizes and spacings between the pores. 

In the second case, excellent anti-reflection properties were
achieved through a much simpler plasma-based process, which does not require any masks,
pattern delineation, catalyst or material deposition \cite{LCChenNatureNT2007}. Figure~\ref{f2}(b) 
shows antireflection properties of a fairly regular array of high-aspect-ratio Si  
nanotips produced through the self-masked plasma etching of a 6-inch Si wafer. A typical length and the base width of the Si nanotips are
approx. 1.4--1.6 $\mu$m and 100--200 nm, respectively. These large-area nanoarrays produced in a simple, single-step, 
environment-friendly process feature excellent anti-reflecting properties over a broad spectrum.
Indeed, the effective light absorption appears to be higher than 99\% over the $\sim$200--2400 nm spectrum.
Surface nanostructuring allows one to dramatically improve the antireflection properties compared to a flat Si surface. 
These Si nanotip arrays can also be deterministically tailored to mimic optical response of biological 
objects (e.g., moth eye, butterfly wing, etc.) to develop next-generation biomimetic photonic 
nanostructures \cite{LCChenMatSciEng2010}.

The examples in Fig.~\ref{f2} were also chosen to introduce two very different 
approaches to nanoscale fabrication to achieve the desired functional (in this case anti-reflection) properties.
The first approach involves preparation of regular patterns using customized masks followed by 
(bottom-up) CNT growth at pre-determined positions. It involves a larger number of steps
and offers a higher precision of nanotube positioning. 
However, care should be taken to grow the carbon nanotubes to the same length and ensure that  
nucleation takes place on every catalyst nanoparticle to avoid major defects in the array. 
Plasma-assisted CNT growth can be used to help achieving higher growth rates and better 
uniformity of nucleation across the pattern as predicted 
by numerical simulations \cite{IgorAPL08-Catalyst}; however, this possibility 
still awaits its experimental realization.
The second approach uses low-temperature reactive plasmas of CH$_4+$SiH$_4$ gas mixtures to generate small 
SiC clusters which serve as self-organized masks for the subsequent Si wafer etching in H$_2+$Ar plasmas. 
This essentially self-organized process is considered in Sec.~\ref{Si-nanotips} 
\cite{LCChenNanoLett2004}.

Both processes aimed to achieve the desired optical properties and used a combination 
of different approaches. Pre-patterning and guided bottom-up growth was used in the first 
case while self-organized, bottom-up self-masking 
and top-down etching were used in the second case. These synergistic combinations reflect the modern trend in 
nanofabrication which relies on various combinations of the bottom-up and top-down
fundamental approaches rather than using them separately.  
Indeed, the top-down approaches based on 
lithography, masking, etc. are rapidly nearing their physical limits while the 
bottom-up nanoassembly is difficult to control.              

This is why, without trying to be exhaustive, in this review we will focus on the mechanisms 
that are particularly effective in a plasma and 
highlight the cases when the synergism between the different approaches helps achieving the desired outcomes 
(e.g., structure, properties, etc.).   
These approaches involve a large number of elementary processes that 
unfold at multiple spatio-temporal scales. 
Therefore, effective control of these elementary processes is central to the ability to achieve deterministic outcomes,
no matter which approach or combination of approaches is used.  
For this reason, in the following section we will discuss the different spatial scales involved and indicate some examples of 
the elementary processes that can be controlled in each case.

\subsection{\label{basicB} Spatial scales and controls}

\begin{figure}[t]
\begin{center}
\includegraphics[width=11.5cm,clip]{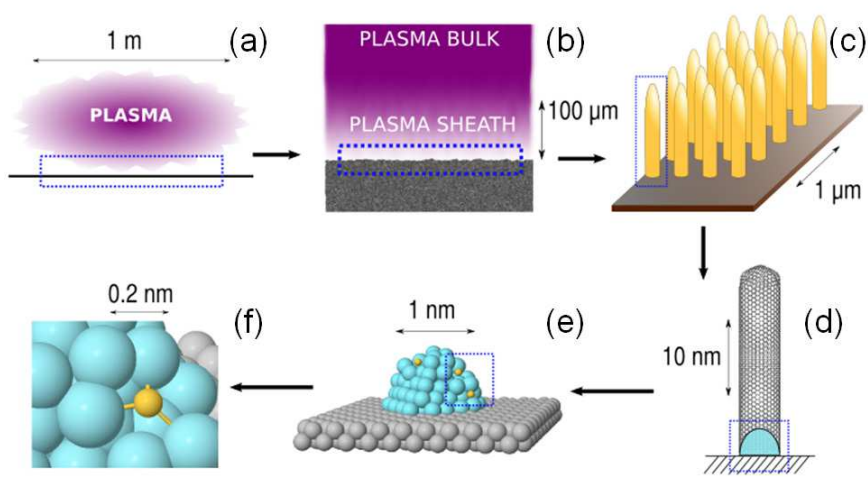} 
\caption{ \label{f3} Spatial scales used to study 
self-organized nanoscale plasma-solid systems: 
(a) plasma in a reactor chamber; (b) plasma sheath
near a solid surface; (c) pattern (array) of solid nanostructures; (d) individual 
nanostructure; (e) catalyst nanoparticle; (f) atoms and interatomic bonds.} 
\end{center}  
\end{figure}

A typical plasma-solid system involves a gaseous plasma in contact with a nanostructured solid surface. 
These systems are highly complex because of the very strong hierarchy of the spatial 
scales involved and a large number of physical and chemical processes that take place at each of these scales. 
The spatial scales of particular interest to the formation of nanoscale objects in plasmas are shown in 
Fig.~\ref{f3}, where each panel represents a higher-magnification zoom of the previous one. 
To achieve effective control at the atomic level in nanostructures, 
every relevant elementary process should proceed in the ``correct" way.  
Here we summarize the most typical effects at each spatial scale, with more specific details 
following in Sec.~\ref{Sec3}. 

At the scales of typical plasma reactors (up to $\sim$1 m, panel (a)), one has to produce 
species (building units, BUs) to use
in the nanostructure (NS) production and also to sustain appropriate 
channels of heat exchange between the solid surface and the plasma bulk. 
Ignition and stability of the plasma discharge is determined 
by the power deposition  and the balance of species and energy. 

The surface of a solid substrate that is used for the NS production (CNTs in this example) is in direct contact 
with the plasma as shown in Fig.~\ref{f3}(b). Upon this contact, a self-organized plasma sheath is formed. Typical dimensions 
of the sheath range from tens of microns in dense (typically $n_e \sim$10$^{11}$--10$^{12}$ cm$^{-3}$) 
plasmas to $\sim$1--2 cm (typically $n_e \sim$10$^{8}$--10$^{9}$ cm$^{-3}$) in low-density plasmas 
used in materials synthesis and processing. The sheath width 
\begin{equation}
\label{sheatheq1}
\lambda_S \sim \lambda_D \left( \frac{2eU_0}{k_BT_e} \right)^{3/4}
\end{equation}
can also be controlled by applying a voltage $U_0$ to the surface and 
by modifying the discharge conditions (e.g., input power, gas pressure and composition) which affect the electron temperature $T_e$.
The higher the plasma density, the smaller is $\lambda_D$ (see Eq.~(\ref{Debye-length})), 
and the thinner is the plasma sheath. Conversely, a larger surface bias and higher electron temperatures result 
in thicker plasma sheaths. Equation (\ref{sheatheq1}) is valid for a collision-less sheath 
(e.g., of low-pressure plasma discharges) when the pulse duration of the applied bias 
$\tau_0$ is much longer than the ion traverse time $\tau_i$ through the sheath \cite{RMP2005}. 

The sheath formation is a self-organized process which shields the 
plasma from the electric field on the solid surface. This field arises when 
more mobile electrons deposit on the surface much faster than the ions. 
Therefore, to maintain charge neutrality in the plasma bulk and shield it from the electric field on the surface, an ion ``coat" is formed 
and the ion density within the sheath is higher than the electron density. 
The electric field peaks at the surface and vanishes in the plasma bulk, where the plasma is charge neutral. 
The ions are thus driven towards the surface and transfer energy upon impact.    

Low-temperature plasmas are usually weakly ionized, $n_n \gg n_e$, 
where $n_n$ is the density of neutral species such as molecules, radicals, and atoms. 
In thermally non-equilibrium plasmas 
\begin{equation}
\label{thermnoneqpl}
T_e \gg T_i \sim T_n, 
\end{equation}
where $T_e$, $T_i$, and $T_n$ are 
the temperatures of electrons, ions, and neutrals. This is why electron-impact reactions play a major role in the
species (e.g., radical) production, which takes place in the plasma bulk, within the sheath, and also on 
the nanostructured surface. 
The plasma-produced species (ions, radicals, atoms, etc.) in turn interact with the surface and lead to numerous processes of exchange of 
energy and matter such as heating, deposition, etching, recombination, and several others.  
 
Of particular importance are the processes of the energy and matter exchange over the dimensions of the patterns and surfaces of 
nanostructures. These two spatial scales are shown in Fig.~\ref{f3}(c,d). 
At the nanopattern scale ($\sim$1 $\mu$m in this example), control should be exercised over the amount of building units 
and heat delivered to the pattern, either 
directly from the plasma or from the surface. Indeed, BUs can be created from the gas-phase precursors or via extraction from the surface
(e.g., by sputtering or etching), whereas the surface can be heated both externally (e.g., substrate heating from underneath or focused 
laser from above) and through the plasma-surface interactions (e.g., upon ion recombination on the surface). 
It is also critical to determine the nanostructure positions within the pattern as well as the 
amounts of BUs and energy per nanostructure. 
The delivery of energy and matter in turn strongly depends on the NS sizes and 
their positioning within the array.         

The next hierarchical level is represented by individual nanostructures (Fig.~\ref{f3}(d)), which also 
have own hierarchical structure. For example, single-walled carbon nanotubes (SWCNTs) of a millimeter length develop from 
tiny nanometer-sized metal catalyst nanoparticles (CNPs), which sustain the nanotube nucleation and growth processes.  
This is why processes at the scales of the SWCNT length 
are different compared to the processes on the surfaces and within the small CNPs. 
Although the BU (carbon atom) incorporation takes place through the catalyst, the 
CNT lateral surfaces play the key role in the building unit collection, transformation, and redistribution.

These surfaces significantly contribute to the production of carbon atoms (e.g., by ion impact reactions) and their transport to the
CNP for the subsequent incorporation into the nanotube wall. The shorter and the hotter the SWCNT, the easier it is for the 
adsorbed carbon atoms or radicals to migrate to the catalyst and then incorporate into the wall after overcoming a certain energy barrier. 
This is why it is very important where and how carbon precursor species are delivered on the nanotube surface. 
The plasma effects can be used to control the species delivery and heat transfer to the developing nanotubes, 
see Secs.~\ref{nano-plasma-surface}, \ref{nucleation}, and \ref{growth}.

Similar processes take place during the nucleation of the CNT cap on the CNP surface. These processes are 
confined to the catalyst nanoparticle dimensions, e.g., $\sim$1.0 nm (or even smaller) for SWCNTs. The cap nucleation 
kinetics is determined by the surface and bulk diffusion of carbon atoms and the formation of a graphene monolayer which 
bends to detach from the surface and form a stable SWCNT cap \cite{Ding2004,NeytsACSNano2010}. 
Numerous plasma-assisted processes (e.g., species production, hopping, detachment, etc.)
on the catalyst nanoparticle surface have a major effect on the nanotube nucleation and growth \cite{NeytsJVSTB2012}.   
For very thin nanotubes, the quasi-classical 
diffusion treatment based on the densities of carbon atoms may fail. In this case, movements of each individual atom 
become decisive and could be predicted using sophisticated atomistic 
simulations \cite{NeytsJACS2011, NeytsJACS2012}. In Fig.~\ref{f3}(e), gray spheres 
represent substrate material, blue spheres - atoms of metal catalyst (e.g., Ni), while C atoms are 
shown by  yellow spheres. 

The last (Fig.~\ref{f3}(f)) spatial scale corresponds 
to the sizes of atoms and atomic bonds, which are typically in the $\sim$0.1--0.2 nm range.       
These atomic scales are characterized by ultra-fast dynamics, e.g., atomic vibrations in the lattice,
bond formation, etc. The effectiveness of  atomic bond formation is determined by the ability of interacting 
atoms to share electrons and establish stable orbitals. This is why small relative motions of atoms with respect to each other 
as well as their bonding are perhaps the most important processes that determine self-organization of matter at these scales.

The plasma may affect the processes even at the atomic-level scales. 
Indeed, the energy states of the atoms involved may be modified through the interaction with the plasma 
ions and electrons; this will affect the minimum energy configurations within the catalyst nanoparticle or in the nanotube cap.
Moreover, microscopic electric fields in the plasma may induce polarization (charge redistribution) along the 
dangling bonds of the carbon and metal atoms, which in turn affect their ability to form stable bonds \cite{NeytsJACS2012}.

Self-organized processes and their driving forces in plasma-solid systems are thus numerous and vary from one spatial scale 
to another. 
Hence, the growth of single-walled carbon nanotubes (e.g., thickness, chirality, length, etc.) 
in the plasma will be different compared to other processes. 
This conclusion is supported by numerous results of theoretical, computational and experimental studies discussed 
in this review and elsewhere in literature. 

For example, a self-organized electric field near the surface is 
one of the major driving forces for the redistribution of ion fluxes over the micrometer nanopattern scales.
This driving force is stronger at higher ion energies and fluxes. 
The non-uniformity of density of adsorbed carbon atoms along the lateral surface of the 
nanotube is a driving force for adatom re-distribution over the surface. One can control this driving force by 
tailored, non-uniform deposition of carbon species from the plasma and adjusting 
the rates of C atom production in different areas on the carbon nanotube surface; both these factors are related to the 
non-uniformity of microscopic electric fields near the nanotube surface, 
which can also be controlled \cite{IgorJPD07-SpecIssue}.

Likewise, the driving force for the SWCNT cap nucleation on the catalyst surface is the difference in chemical 
potentials within and on the surface of the supersaturated nanoparticle. In this case, plasmas can also  
control the driving force, e.g., by localized exchange of energy and matter on the catalyst nanoparticle surface, see 
Sec.~\ref{nano-plasma-surface}. 
The next step is to review the most relevant self-organization pathways and  
how   plasma-specific effects can be used to control them.

\subsection{\label{basicC} Self-organization pathways and plasma effects} 

\subsubsection{\label{genform} General formalism}

Solid nanostructure formation usually proceeds through a sequence of events which starts from deposition of building units, their 
redistribution, clustering, formation of stable nuclei, followed by the growth, shape and structure formation and relaxation.  
Although these events also take place in the formation of nanostructures in the plasma bulk
(see, e.g., Sec.~\ref{Nanocrystals}), here we focus on surface-supported growth. 
Low-temperature growth is particularly important from the energy efficiency perspective and also allows nanostructure formation 
on temperature-sensitive substrates such as polymers and plastics. Under such conditions, the rates of redistribution of the 
deposited material due to random thermal motion are quite low and non-uniform distributions of atoms on the surface usually form. 

In the presence of these deposits, the surface energy is higher compared to a clean surface, which indicates that the 
system is {\em non-equilibrium}. Therefore, the system will attempt to 
invoke some mechanism to reduce its surface energy. This mechanism is due to the 
non-uniformity of adatom density on the surface
and creates a {\em kinetic} force that drives these atoms to self-organize into more 
compact clusters thereby minimizing the surface energy.
Hence, nucleation is an energetically-favorable process and is determined by an interplay of 
thermodynamic and kinetic factors. As the nanostructures  
inevitably shrink, they may form in a single nucleation event \cite{Tersoff-Nucleation-Science08} 
and the ability to control the nucleation becomes critical.  
  
Surface diffusion is an effective kinetic pathway of self-organization 
of solid nanostructures, especially at low temperatures \cite{diffusion-pathways-self-org}.  
Here we will consider two cases of diffusion, with and without significant mass 
relocation about the surface. The first case (the main focus of this 
section) originates when species of 
a solid material are deposited onto the surface (e.g., from a vapor phase) 
while the second one is typical for ion-assisted nanoscale surface structuring (e.g., 
surface sputtering and rippling). Both cases bring about a plethora of 
interesting physical effects at low temperatures.  
Since low-temperature operation is one of the key advantages of plasma-based nanoscale synthesis, one can thus expect 
a particular importance of kinetic, e.g., diffusion-related self-organization pathways.  
This in turn allows the formation of metastable NSs or nanometer surface textures, which feature  
highly-unusual morphological, structural, electronic and other properties.

\begin{figure}[t]
\begin{center}
\includegraphics[width=8cm,clip]{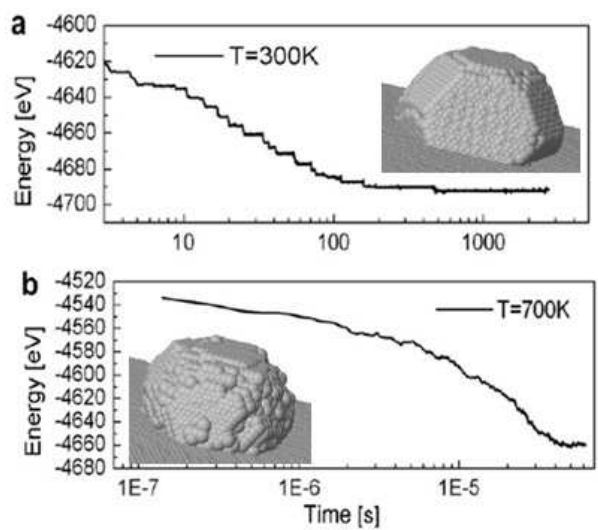} 
\caption{ \label{f4} Dynamics of nanocrystal faceting  
at different temperatures. Lower-temperature process (a), although produces better faceting, 
is several orders of magnitude longer than at the higher temperature (b) 
[Reprinted from \cite{island-shape-evolution-LaMagna}, 
Copyright \copyright (2007), with permission from Elsevier]. } 
\end{center}  
\end{figure} 

Temporal dynamics of deposition, redistribution, and relaxation therefore becomes critical for the 
NS formation. After the nucleation process is complete, the as-formed cluster increases in size and an island forms 
as the new atoms attach to it during the deposition process. Meanwhile, the atoms in the island tend to 
rearrange to reshape the cluster and form a more regular, minimum-energy structure. One example of such 
rearrangement is shown in Fig.~\ref{f4}. Bond-counting kinetic Monte Carlo simulations were used to study the 
equilibration processes of an initially elongated nanoisland of 8232 atoms under different  
surface conditions \cite{island-shape-evolution-LaMagna}. At room-temperature conditions, a 
clear faceting (nanocrystal attribute) of the island was observed within $\sim$100 s, 
as can be seen in Fig.~\ref{f4}(a). At higher temperatures ($T=$700 K), the reshaping takes place much faster, 
within only $\sim$1 $\mu$s (Fig.~\ref{f4}(b)). However, the island structure is amorphous and crystalline 
facets do not form. The observed reshaping of the island proceeded by diffusion of atoms 
over the surface, e.g., from a facet to an edge or another facet, etc. 

This process is characterized by the 
kinetic barrier $E_{\rm dif}(n_{na},E_{ad})$ which is determined by the number of neighboring 
atoms $n_{na}$ and the adhesion energy $E_{ad}$, which vary between different crystal facets.      
This kinetic barrier and the surface temperature are the decisive factors that determine the hopping frequency
\begin{equation}
\label{hopping-freq}
\omega(n_{na},E_{ad},T) = \nu_D \exp \left[-\frac{E_{\rm dif}(n_{na},E_{ad})}{k_B T} \right],  
\end{equation}
where $\nu_D$ is the attempt frequency, and $k_B$ is the Boltzmann constant. 
If the temperature is lower, the kinetic barrier should also be reduced to 
maintain the same hopping frequency as at higher $T$.
Importantly, lower kinetic barriers have resulted in faster nanostructure relaxation, especially 
at lower temperatures. This result is consistent with experimental observations that metastable 
self-organized NSs can form within unexpectedly short times, thus indicative of some unique 
and effective kinetic self-organization pathways and a possible unusually high degree of 
self-organization at low temperatures \cite{diffusion-pathways-self-org}, which is due to strong diffusion currents 
\begin{equation}
\label{diff-current}
J_D = - \aleph \nabla \mu (r) ,  
\end{equation}
where $\mu (r)$ is the chemical potential, and $\aleph$ is a phenomenological transport coefficient.
The gradient of the chemical potential $\nabla  \mu (r)$ is therefore a driving force for the 
diffusion-driven self-organization. The stronger the non-equilibrium (gradient), 
the stronger is the driving force, and the more effective is the self-organization. 

Diffusion currents (\ref{diff-current}) and the continuity equation are commonly used to introduce 
the phenomenological diffusion coefficient  
\begin{equation}
\label{diffcoeff}
D = D_0 \exp \left(- \frac{{\cal{E}}_d}{k_B T} \right),  
\end{equation}
where ${\cal{E}}_d$ is the diffusion barrier energy which is determined by  
the difference between the minimum and the maximum of the periodic 
surface potential energy. Equation (\ref{diffcoeff}) is more accurate at lower temperatures 
when the residence time of adatoms in the potential well is much longer than the 
characteristic times of surface phonon vibrations \cite{diffusion-pathways-self-org}. 
In this case the solid surface may be treated as rigid. 

\subsubsection{\label{plaseffdiff} Plasma effects}

Therefore, when trying to engage the plasma to custom-made self-organized solid nanostructures  
at low temperatures, one has to understand how it may affect the surface diffusion processes. 
First, larger gradients of the chemical potential $\nabla  \mu (r)$ can be produced through a more localized, non-uniform deposition 
of BUs from the plasma, for example, by using ion-focusing effects and a very fast material 
delivery \cite{IgorJPD07-SpecIssue}. 
Second, the temperature of the top layer on the surface can be increased by the plasma exposure, which 
will increase the hopping frequency (\ref{hopping-freq}).  
Third, the diffusion barriers ${\cal{E}}_d$ may also be reduced through the unique, electric-field mediated 
polarization effects \cite{NeytsJACS2012} which may modify (e.g., reduce) the adhesion/bonding energy of 
an atom (e.g., a yellow atom in Fig.~\ref{f3}(f)) on a 
plasma-exposed surface \cite{plasma-control-self-org}. This energy may be quite 
different in the crystal facet, edge, or corner locations \cite{Barnard-shapes1}.  

Using a macroscopic treatment, it was shown that the presence of a non-uniform
electric field $E (r)$ in the vicinity of a nanostructured 
plasma-exposed surface leads to the effective increase of the 
diffusion coefficient
\begin{equation}
\label{diffcoeffpolar}
D = D_0 \exp \left(- \frac{{\cal{E}}_d - {\cal{E}}_{pl}}{k_B T} \right),  
\end{equation}
through the reduction of the potential barrier by 
\begin{equation}
\label{diffbarpolar}
{\cal{E}}_{pl} = \lambda_{\rm lat} [\tilde{p}  + \tilde{\alpha} E(r)] \frac{\partial E(r)}{\partial r}, 
\end{equation}
where $\tilde{p}$ and $\tilde{\alpha}$ is the dipole moment and polarizability 
of an adsorbed species on the surface, and  $\lambda_{\rm lat}$ is the lattice constant 
\cite{plasma-control-self-org}.

Fourth, the plasma sheath electric field (which is normal to the surface away from it and has a horizontal 
component upon approaching the nanostructures/nanofeatures) can be used to tune the NS morphology and 
arraying. Indeed, the interplay between the surface energy, energy of the substrate-nanostructure 
interface, and the electrostatic energy determine a driving force for mass relocation through surface 
diffusion \cite{ParkAPL2009}. 

In this case, the variation of the free energy of the system $G$  with mass relocation 
$\delta c({\bf r},t)$ determines the chemical potential 
\begin{equation}
\label{chempot3}
\mu = \delta G / \delta c 
\end{equation}
in (\ref{diff-current}). 
The free energy of a charged nanoparticle developing on a dielectric surface 
\begin{equation}
\label{freeen4}
G = G_{\rm phs} + G_{\rm el} + G_{\rm int}^{\rm g} + G_{\rm int}^{\rm s}  
\end{equation}
in turn includes contributions from the chemical energy of phase separation $G_{\rm phs}$, 
electrostatic energy $G_{\rm el}$, energy of the nanoparticle-gas interface $G_{\rm int}^{\rm g}$,
and the energy of the nanoparticle-substrate interface $G_{\rm int}^{\rm s}$ \cite{ParkAPL2009}.
The latter may include surface stress-related contributions, which can also be 
modified by the plasma-specific effects (e.g., ion impact).  
It was also shown that electric field effects significantly contribute to 
morphological stabilization of stressed surfaces \cite{TomarPRL2008}.
Therefore, electric fields may control diffusion-based self-organization, 
which may in turn lead to effective control of shapes, sizes, 
and positioning of solid nanostructures on plasma-exposed surfaces (see Sec.\ref{Sec3}).

Surface diffusion involving mass relocation is one of the main contributors 
to the growth and reshaping of surface-supported solid
nanostructures. For example, the amount of BUs that can be collected 
from the surface and then used in nanowire growth, 
has a Gaussian distribution with the characteristic length, the migration length 
$\lambda_{\rm mig}$ \cite{Persson-Surf-Diff}. Each circular surface area $dA=2\pi r dr$ contributes material to the 
growth, and the NW radius increase is  
\begin{equation}
\label{NW-growth-rate1}
dR_{\rm growth} \sim (\sqrt{2 \pi})^{-1} \lambda_{\rm mig} \exp[-r^2/2\lambda_{\rm mig}^2]dA, 
\end{equation}  
whereas the total growth rate can be obtained by integrating (\ref{NW-growth-rate1}) over the entire 
material collection (e.g., adatom capture) area $A$. Plasma exposure strongly affects 
material collection from the surface. For example, long-range 
interactions in plasma-solid systems modify adatom densities around the NSs thereby 
enhancing their growth.

\subsubsection{\label{ioneffdiff} Ion-related effects}

Since plasma-exposed surfaces are intrinsically subjected to ion fluxes, the 
effects of ion bombardment on the surface processes are of particular interest.  
The key effects of the ion-surface interactions are the modified surface diffusion and 
morphology. These effects are primarily determined by the ion energy and fluxes. 

When an ion impinges on a solid surface, it transfers energy to the surface atom.
This leads to several interesting effects. If the ion hits an adatom, the 
hopping frequency (also frequently termed migration probability) of 
the adatom 
\begin{equation}
\label{hopping-freq-ion}
\omega(n_{na}^i,E_{ad}^i,T^i) = \nu_D^i \exp \left[-\frac{E_{\rm dif}(n_{na}^i,E_{ad}^i) - \delta E_i}{k_B T^i} \right],  
\end{equation}
is significantly modified compared to (\ref{hopping-freq}). First of all, 
as the adatom receives a fraction of the ion energy $\delta E_i$, the 
effective kinetic barrier $E_{\rm dif}(n_{na}^i,E_{ad}^i) - \delta E_i$ becomes lower than 
$E_{\rm dif}(n_{na}^i,E_{ad}^i)$, leading to the higher hopping frequency.   
The reduction of the effective kinetic barriers is perhaps one of the 
most common effects of the ion-adatom interactions.  

However, because the atoms firmly bonded in the top surface layer are 
also exposed to ion impacts, this also changes the background conditions 
for the adatom diffusion. This is why all the key parameters in  
(\ref{hopping-freq-ion}) have been labeled with superscript $i$. 
Indeed, sufficiently intense ion fluxes lead to higher surface
temperatures $T^i > T$, which enhances the adatom surface diffusion 
through stronger phonon vibrations associated with the atoms 
in the near-surface layer. Since the energies of the ions 
generated by the plasma always feature relatively broad energy distributions,
some ions from the high-energy tail of the distribution 
may sputter some surface atoms \cite{Sigmund-sputtering}. 
This sputtering may reduce the number of 
neighboring atoms $n_{na}^i < n_{na}$, which will also 
lead (after a relaxation) to a distortion 
of the lattice constant $\lambda_{\rm lat}$, which affects both $\nu_D^i$ and  $E_{ad}^i$.   

Therefore, sputtering, which occurs when the ion energy exceeds the 
surface atom displacement threshold (which depends on the bond strength
of the surface atoms and the relative masses of the impinging ions and the 
surface atoms), may significantly affect the surface diffusion. 
Importantly, the sputtering may lead to the two major effects \cite{Makeev,Marton}:
(i) effective surface diffusion without 
mass relocation; (ii) the possibility to both enhance and suppress 
the diffusion.   

The first effect stems from surface erosion in selected areas, 
which does not lead to mass transfer about the surface and is nearly 
temperature-independent. The second effect is determined by 
the ion energy and flux, angle of incidence, as well as the surface 
material and temperature. Under certain conditions \cite{Makeev}, 
the ion-induced effective diffusion may dominate over thermally activated
diffusion. This effect is quantified by comparing the 
coefficients of the thermally-activated $D_T$ and ion-induced $D_I$ diffusion, a 
combination of which determines the total diffusion constant
\begin{equation}
\label{totdifconst}
D_{\rm tot} = D_T + D_I. 
\end{equation}

Typically, ion sputtering leads to the pronounced morphological 
rearrangement of the surface layer due to 
preferential surface erosion in specific areas. 
This surface structuring is quantified by rather complex 
morphological phase diagrams which incorporate the effects of 
several factors such as effective surface diffusion (including both 
ion- and thermally activated) rates, surface tensions, 
effects of anisotropy, and surface temperature. 

For example, morphological instabilities may lead to the 
surface ripples \cite{ion-beams-self-org}, when 
valleys are eroded faster than crests. 
These ripples may be oriented in different directions, for example either 
along the $x$ or $y$ direction in the surface plane. 
The wavelength of these ripples \cite{Makeev} 
\begin{equation}
\label{ripples}
\lambda_{\rm ripple} = 2 \pi \sqrt{\mid D_{\rm tot} / \varpi \mid}  
\end{equation}
is determined by the total diffusion constant (\ref{totdifconst})
and the absolute value of the largest of the surface tension 
coefficients $\varpi$. Interestingly, the periodicity 
of the ripples (\ref{ripples}) often appears 
at the nanometer scales. As such, this effect is
useful for solid surface nanostructuring with periodic linear structures.

However, ion-induced diffusion without mass relocation 
dominates at high ion energies, typically in the kilovolt range. 
This energy range is commonly used for plasma 
immersion ion implantation (PIII) which is presently a common tool 
in plasma processing of a broad range of solid materials \cite{AndersJPD07}. 
Because of the high ion energies and the 
associated surface erosion, significant caution should be exercised 
to avoid damage to the developing delicate nanostructures, such as 
graphenes, single-walled nanotubes, or polymers.

Nevertheless, controlled ion damage may in some cases be used intentionally 
to enable a certain functionality.
For example, the top sections of multiwalled carbon nanotubes were intentionally 
converted into amorphous phase by impact of $\sim$10 kV energetic Ar$^+$ ions 
to enable interesting superhydrophobic properties in  
carbon nanotubes \cite{HanAPL09}. 

Unless specified otherwise, most of the examples in this 
review focus on cases when ion-sputtering effects are not 
dominant. In most cases, this is achieved by the appropriate 
selection of the most suitable plasmas and the associated 
ion energy distributions.

\subsubsection{\label{tempdynam} Temporal dynamics}
The temporal dynamics of the nano-solid nucleation and growth processes is 
characterized by several time scales. These time scales correlate the deposition of building material 
and the processes of redistribution of this material over the surface. 
For example, when solid nanostructures are created by cluster deposition, 
the clusters may diffuse, coalesce 
and even evaporate from the surface before they dissociate \cite{space-time-scales, Jensen2}. 
What is important is how fast  these clusters are 
delivered to the surface. For example, if the deposition time $t_{\rm depos}$ is larger than the 
time needed for surface diffusion $t_{\rm dif}$ but 
is still shorter than the cluster coalescence time $t_{\rm coalesc}$, i.e., 
\begin{equation}
\label{clustbeam5}
t_{\rm dif} < t_{\rm depos} < t_{\rm coalesc}, 
\end{equation}
the nanostructure formation will be dominated by the 
cluster diffusion rather than the cluster coalescence. In this example, a fairly slow evaporation will also likely lead 
to piling up of material on the surface, unless the diffusion-driven material redistribution 
processes are effective. 

In other words, the nanostructure formation strongly depends on the {\em demand, supply, and redistribution 
of building units} on the surface. The demand is determined by the NS arrays to be produced, the supply - by the 
rates of BU creation and delivery, and the redistribution - by the strength of the driving force 
for self-organization on the surface. These criteria will be used to discuss the role of the plasma 
effects in the production of metastable NSs where the {\em rates} of supply and redistribution 
of material and energy are critical.

Importantly, many exotic metastable nanostructures and self-organized arrays can be produced in plasma-solid 
systems under far-from-equilibrium conditions. These conditions are very different compared to common 
physical/chemical vapor deposition systems and are related, e.g., 
to thermal non-equilibrium between the plasma species, anisotropicity and non-uniformity of distribution of 
electric field, pressure, surface stress (e.g., due to ion bombardment), and local heating. 

Some other 
factors such as strong imbalance between the rates of heating and cooling and non-uniformity of long-range 
electromagnetic forces also contribute to the departure of the system from equilibrium. 
In turn, these factors lead to a variety of unique, plasma-specific driving forces for self-organization over 
different spatial scales. 

For example, strong thermal non-equilibrium (\ref{thermnoneqpl}) of low-temperature plasmas 
leads to the enhanced production of building units via electron-impact reactions;
these reactions are not common to neutral gases.
As a result, higher rates of material supply, and hence, more non-equilibrium conditions can be 
produced on the surface. In turn, these conditions may lead to stronger 
driving forces for diffusion-based self-organization at shorter time scales, characterized by non-uniformities 
of adatom density on the surface 
(e.g., within the adatom capture zones around the nanostructures). 

On the other hand, 
longer-range non-uniform electric fields make it possible to 
control surface diffusion over the scales that are much larger than the adatom capture zones around 
the nanostructures. In this way, self-organization of large mesoscopic patterns can also be affected 
by the plasma-produced electric fields. Long-range ordering is also 
possible due to the magnetic field effects, e.g., in 3D arrays of Co magnetic nanodots  
in diamond-like matrices produced by plasma processing \cite{Meletis1,Meletis2}.

The time scales for different metastable states to reach their equilibrium states  
and the associated energy barriers determine the outcome of the nanostructure formation 
through the competition of thermodynamic and kinetic effects.  
If a metastable nanostructure with exotic properties is targeted, this 
structure has to be formed well before it can rearrange into thermodynamically-prescribed 
basic shape or another, more stable (in the thermodynamic sense) 
shape, over the characteristic time $\tau_{\rm trans}$.

Therefore, if the least stable structure is the target, the building material should be delivered and 
arranged and then kinetically stabilized in the desired shape,  fast, well before 
the processes of atom relocations become significant.
Indeed, atomic transport from one facet to another determines shape evolution  
 and may lead to metastable nanocrystals with a very strong presence of reactive 
 (typically higher-index, less equilibrium) 
facets \cite{Jensen-changing-shapes,Non-Eq1,TiO2-reactive-111-facets,Non-Eq3}. 
To minimize the possibility of relaxation of the desired 
structure to any other (more stable) structure, $\tau_{\rm trans}$ should satisfy \cite{OstrikovPOP2011}:
\begin{equation}
\label{ineq-times-kinetics}
\tau_{\rm trans} > \tau_{\rm gen} + \tau_{\rm deliv} + \tau_{\rm diff} + \tau_{\rm incorp} + 
\tau_{\rm recryst} + \tau_{\rm coat},  
\end{equation}   
where $\tau_{\rm gen}$ and $\tau_{\rm deliv}$ are the times required to produce and deliver 
the BUs, $\tau_{\rm diff}$ is the characteristic time of their diffusion/hopping over the surface
(e.g., from one facet to another), $\tau_{\rm incorp}$ is the time of BU incorporation into the facet, 
and  $\tau_{\rm recryst}$ is the recrystallization time. Additional processing (e.g., surface passivation) 
is often required to retain the structure in the desired metastable state. This process effectively 
increases the transition barrier $\Delta U$ and is characterized by the time scale 
$\tau_{\rm coat}$. 

Transition from an ordered to a less ordered structure may happen as a 
result of thermal fluctuations that are present in virtually any device. This issue is particularly 
significant for small NSs because of the small energy barriers to be overcome upon transition to 
another state. This is the reason why such small structures are eager to reshape even upon addition of small amounts of 
heat, while larger nano-objects usually ``freeze" in whatever shape they were created \cite{Non-Eq1}.
Indeed, it was estimated that the activation energy for the shape transitions may exceed 
$\sim 40 k_B T$ for nanocrystals with a radius exceeding $\sim$100 nm \cite{Non-Eq3}. 
Such high barriers are very difficult to overcome via thermal activation. 

\begin{figure}
\begin{center}
\includegraphics[width=13.5cm,height=8cm,clip]{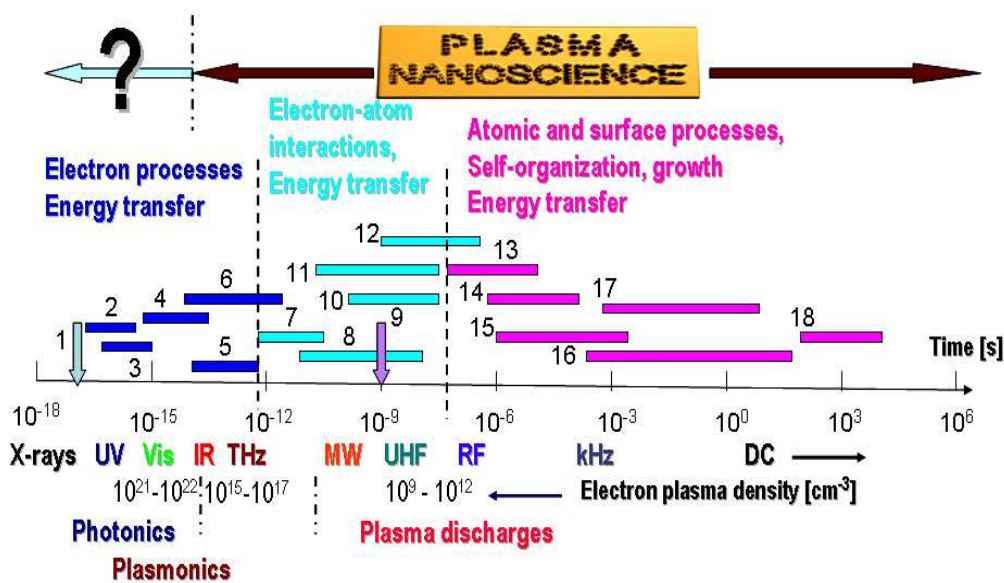} 
\caption{ \label{f6} 
Temporal scales of transfer of energy and matter at nano- and subnanometre scales \cite{OstrikovJPD2011}. 
(1) Shortest controllable time scale. (2) Electron dynamics in atoms. 
(3) Attosecond laser pulses. (4) Electron correlation, hole dynamics. (5) Femtosecond 
laser pulses. (6) Phonon processes, bond formation/breaking, atomic reorganization 
in phase transitions, electron-phonon scattering. (7) Electron-surface atom
energy transfer in femtosecond laser plasmas. (8) Energy transfer via lattice vibrations. (9) Atomic clock
precision. (10) Atomic motion, residence, and clustering on solid surfaces. 
(11) Electron collisions in low-temperature plasmas. (12) Electron transport, NP/nanofeature charging, breakdown
dynamics in low-temperature plasmas. (13) Ion collisions, transport, and residence in plasma reactors. 
(14) Neutral (e.g., radical) collisions, transport and residence in plasmas. 
(15,16) Formation of nanocrystals and self-organized patterns at high 
($\sim$1000$^{\circ}$C) and low ($\sim$10s--100$^{\circ}$C) temperatures
(17) CNT and metal oxide NW growth in a plasma. (18) Metal oxide NW growth in thermal CVD.
Copyright \copyright IOP Publishing. Reproduced from \cite{OstrikovJPD2011} by permission of IOP Publishing. All rights reserved.
}
\end{center}  
\end{figure}

Therefore, in order to be able to control the shape of small quantum dots, it is critical to produce them at low temperatures and 
increase $\Delta U$ by stabilizing their surfaces. Non-equilibrium 
plasmas show particular advantages to implement both the nanoassembly at low temperatures and the effective surface passivation
(e.g., hydrogenation) \cite{CNT-functionaliz-plasma}.         
More importantly, the plasma-based processes offer advantages at every formation 
stage of metastable NSs in (\ref{ineq-times-kinetics}) and form the basis of the 
non-equilibrium nanoarchitectronics approach \cite{OstrikovPOP2011}. Some of these 
time scales are shown in Fig.~\ref{f6} which summarizes the time scales involved in the 
transfer of energy and matter \cite{OstrikovJPD2011}.  

Indeed, electron-impact reactions in the plasma and ion-assisted reactions on the surface are 
very effective and fast channels for the BU generation ($\tau_{\rm gen}$, bars 11 and 13 in Fig.~\ref{f6} for 
the electron- and ion-assisted reactions, respectively). The ion fluxes also  
significantly shorten the material delivery time $\tau_{\rm deliv}$, compared to the typical times of 
neutral transport (bar 14). Due to the plasma heating and polarization effects, the
time scales of surface diffusion $\tau_{\rm diff}$ and other atomic processes on the surface (bar 10) 
can also be reduced. 

The time scales of BU incorporation, recrystallization, and lattice relaxation 
are also affected by the energetic states of the impinging plasma 
species and the localized surface heating through the surface 
recombination and ion bombardment. Thus, the nanostructure and  
self-organized nanopattern formation can be faster in the plasma than in equivalent thermal processes
(shift to the left in bars 15--17).  
Non-equilibrium low-temperature plasmas also show a remarkable dissociation ability and are used 
to produce reactive species (e.g., atomic hydrogen) to passivate or stabilize surfaces. This in turn may 
shorten $\tau_{\rm coat}$ and increase $\Delta U$.       

Figure~\ref{f6} also shows the time scales of other processes involved in BU  
production and nanostructure formation (``plasma-for-nano" in Fig.~\ref{f1}) such as electron transport and nanostructured surface 
charging (bar 12), and reactive radical production and transport (bar 14). 
It also shows characteristic time scales of collective oscillations in plasmas of different densities and sizes. 
Typical low-pressure plasmas used in nanoscale synthesis and processing produce plasma oscillations in the radiofrequency (RF) to microwave 
(MW) ranges. On the other hand, dense micro/nano-plasmas of hot laser plasmas are expected to produce infrared (IR) radiation.
Likewise, oscillations of cold electrons in metal nanoparticles may sustain localized surface plasmons in the UV, 
visible, and IR ranges.
For a more extensive discussion of the time scales of interest and challenges in nanoscale transfer of energy and matter 
we refer to the original publication \cite{OstrikovJPD2011}. 

Therefore, low-temperature plasmas is a {\em reactive and fast-responding non-equilibrium environment} 
 which offers a range of unique controls of self-organization in plasma-solid systems.
In particular, a range of driving forces for self-organization at different time scales 
and constructive interplay between thermodynamic and kinetic factors can be used to produce 
metastable nanostructures with exotic properties. The following section 
uses focused examples to show how to achieve these effects using low-temperature plasmas.

\section{\label{Sec3} Plasma controls in nano-solid formation: fundamentals }

The ability to control the formation of regular patterns and arrays of nanostructures 
(referred to as nanoarrays for simplicity) critically depends on the understanding of the 
plasma-specific mechanisms at the many spatial and temporal scales involved (Figs.~\ref{f3} and \ref{f6}). 
The ``building unit" approach describes the sequence of events that take place 
during the formation of self-organized patterns in non-equilibrium PSSs \cite{RMP2005}. 
This sequence includes the stages of BU generation and delivery, surface preparation,
NS nucleation and growth, and pattern self-organization.   
While the BUs are generated and delivered,
the growth surface should be suitably prepared for the nucleation. 
While the NS grows, the self-organized pattern of a large number of these 
structures also develops. 
This approach can be used for the quasi-deterministic synthesis of self-organized arrays  
which involves the sequence of steps and feedbacks that span spatial scales 
from atomic sizes ($\sim$10$^{-10}$m) to the sizes of large plasma reactors ($\sim$1 m) \cite{plasma-cond}. 
In this section, we will focus on {\em nanoscale} plasma-surface interactions, 
nucleation and growth of solid nanostructures, as well as the formation of self-organized 
patterns and arrays.

\subsection{\label{BU} Generation of building units} 

The building units that are required for nanoscale synthesis, can 
be produced in the plasma bulk and within the sheath, in addition to the 
nanostructured surface where they are typically produced in thermal CVD.   
The plasma can generate BUs in very different forms which include 
atoms, ions, radicals, molecules, and nanoclusters. 
These species can also be produced in a variety of energetic states depending 
on their electric charge and the degree of excitation.   

The generation  mechanisms of these species are very diverse and can be
split into two main groups. The first group of processes is 
based on electron-impact reactions, which stem from the strong 
thermal non-equilibrium of the plasma and a significantly higher energy of 
electrons compared to the energies of other plasma species such as ions and neutrals.  
Typical examples of electron-impact reactions that are particularly 
relevant to synthesis and processing of carbon-based nanostructures are:
\begin{equation}
\label{excitation1}
e^-  + {\rm CH}_4 \rightarrow {\rm CH}_4^* + e^-
\end{equation}    
for excitation, 
\begin{equation}
\label{ionization1}
e^-  + {\rm CH}_4 \rightarrow {\rm CH}_4^+ + 2e^-
\end{equation} 
for ionization, and 
\begin{equation}
\label{dissociation1}
e^-  + {\rm CH_4} \rightarrow {\rm CH_3} + {\rm H} + e^-
\end{equation}
for dissociation of CH$_4$ precursor molecule.  

Electron-impact reactions ((\ref{excitation1})-(\ref{dissociation1}) and many others) 
make the plasma environment unique compared to the 
neutral gas-based processes as they produce ions and lead to high rates of 
dissociation and formation of a very large number of radical species. Moreover, 
many of these species find themselves in a variety of excited 
states \cite{Bogaerts1995} which makes them more reactive 
during the interaction with the nanostructured surfaces and other plasma species \cite{BogaertsJPD2011}.

The latter interactions form the second group of reactions 
that are based on collisions of heavier species such as ions, radicals and molecules.
These reactions usually proceed through the exchange of atoms or 
electrons and also lead to the generation of a variety of reactive species in 
excited and ionized states, which in turn affect the discharge kinetics and 
plasma-surface interactions \cite{KushnerJAP1988}.   
The numerous pathways of the formation of the BUs have been 
discussed previously \cite{RMP2005}. 
For example, if a CH$_3$ radical (which is believed to play a role in the growth of single-crystalline 
carbon nanotips) is targeted, in addition to the electron-impact reaction (\ref{dissociation1}),
it also can be generated through representative neutral-neutral 
\begin{equation}
\label{n-n-r}
{\rm H}  + {\rm C_2H_5} \rightarrow 2{\rm CH_3} 
\end{equation}
and ion-neutral 
\begin{equation}
\label{i-n-r}
{\rm Ar}^+  + {\rm CH_4} \rightarrow {\rm CH_3}^+ + {\rm H} + {\rm Ar}
\end{equation}
reactions. Note that Ar$^+$ ion-assisted reactions similar to (\ref{i-n-r}) can be used to 
dissociate hydrocarbon precursors both in the gas phase and on the surface,
which is a unique feature of plasma environments compared to 
equivalent neutral gases.

Here we only stress that plasma discharges 
offer a unique environment not only for the production of a variety of reactive radical 
species but also for the clustering and polymerization, which takes place 
in the gas phase and leads to the formation of nanoclusters and nanoparticles, which in turn 
also significantly affects the discharge kinetics and species 
production \cite{PhysRepts2004,IvlevRMP}.     

\begin{figure}
\begin{center} 
\includegraphics[width=10.5cm,height=9.5cm,clip]{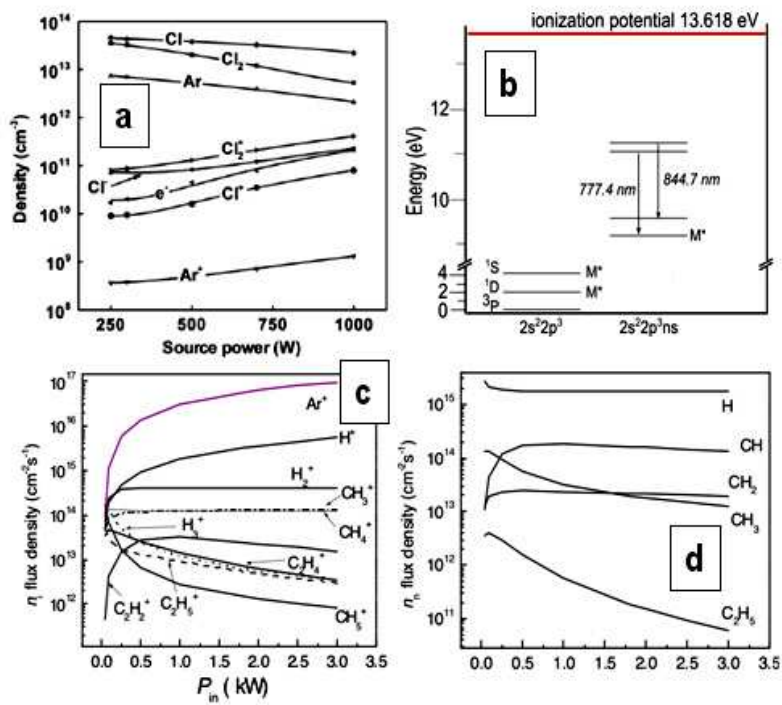} 
\caption{ \label{f8} Low-temperature, thermally non-equilibrium plasmas
sustain very high rates of dissociation and production of reactive radicals, such as Cl atoms for Si etching (a) 
[Copyright \copyright IOP Publishing. Reproduced from \cite{BogaertsJPD08} by permission of IOP Publishing. All rights reserved.]  
Long-living excited states of plasma-produced reactive species are of interest to numerous applications (b)
[Copyright \copyright IOP Publishing. Reproduced from \cite{MiranJPD2011} by permission of IOP Publishing. All rights reserved.] 
Ion fluxes may deliver more material to the surface  
compared to neutral fluxes (c,d) [Reprinted with permission from \cite{DenyJAP2004}. 
Copyright \copyright (2004), American Institute of Physics.] } 
\end{center}  
\end{figure}

Figure~\ref{f8} shows 3 more salient features of particular interest for nanoscale synthesis and processing. 
The first  feature is the ability to dissociate molecules and produce reactive radicals (Fig.~\ref{f8}(a)). 
This graph shows the computed densities of the neutral and ionized 
plasma species produced in inductively coupled plasmas in Ar$+$Cl$_2$ gas mixtures used for reactive 
chemical etching and microstructuring of Si wafers in microelectronics \cite{BogaertsJPD08}. 
Large amounts of reactive Cl atoms enable the effective, high-rate etching through the formation of silicon chloride-based volatile 
products. 

The plasma is weakly ionized, which is reflected by a low ionization degree
\begin{equation}
\label{iondegree}
\varsigma_{\rm ion} = \Sigma_j n^{i}_{j} / \Sigma_k N_k \ll 1,
\end{equation}
where $n^{i}_{j}$ is the number density of ionic species $j$ and $N_k$ is the 
number density of neutral species $k$. In the example in Fig.~\ref{f8}(a), even though 
the density of Ar$^+$ ions is 3--4 orders of magnitude lower than the density of Ar atoms,
the plasma dissociates chlorine molecules very effectively. 
Indeed, the number density of Cl atoms appears to be 
higher than the density of Cl$_2$ molecules, even at very low input powers. 

This outstanding dissociation ability of the plasma has been commonly attributed to the 
non-equilibrium (e.g., non-Maxwellian) electron energy distributions (EEDFs) \cite{SugaiEEDF}
\begin{equation}
\label{EEDF1}
f(\epsilon_e) = \xi_1 \sqrt{\epsilon_e} \exp (- \xi_2 \epsilon_e^x),
\end{equation}
where $\epsilon_e$ is the electron energy while $\xi_1$ and $\xi_2$ are energy-dependent
coefficients ($x=1$ corresponds to the Maxwellian EEDF). 
The EEDFs can be effectively controlled by the plasma parameters 
to customize electron populations in specific 
energy ranges for the production of the 
desired species. 

For example, the dissociation energy of oxygen molecule O$_2$ is lower compared to N$_2$ and if the 
EEDF peak $Max[f(\epsilon_e)]$ is adjusted above the O$_2$ dissociation threshold $\Xi_{\rm diss}^{O_2}$ yet 
below the N$_2$ dissociation threshold $\Xi_{\rm diss}^{N_2}$, 
\begin{equation}
\label{EEDFmax}
\Xi_{\rm diss}^{O_2} < Max[f(\epsilon_e)] < \Xi_{\rm diss}^{N_2} ,
\end{equation}
then the preferential generation of oxygen atoms and other reactive oxygen species can be achieved. This 
selectivity is important for the synthesis of metal oxide 
NWs (Sec.~\ref{1D-oxide}) and plasma interactions with biological objects (Sec.~\ref{bio}).
Another commonly used options to enhance selective radical production are to 
generate electron distributions with two characteristic average electron 
energies (bi-Maxwellian EEDFs) or boost electron populations in the higher-energy
tail of the distribution (Druyvesteyn EEDFs).

Plasma-produced reactive oxygen species (ROSs) are of particular interest because of their long-living 
excited states. Figure~\ref{f8}(b) shows the first three excited states of oxygen atoms which are 
metastable with radiative decay lifetimes of $\sim$1 s, $\sim$100 s, 
and 185 $\mu$s \cite{Lifetime1,Lifetime2,Lifetime3,MiranJPD2011}. These states 
are fairly stable and can be further excited to the next levels shown in Fig.~\ref{f8}(b). 
Radiative decay of these levels produces very strong IR emission with the wavelengths $\sim$777 and 845 nm,
which are commonly used as fingerprints of ROSs produced in extremely
non-equilibrium oxygen plasmas \cite{MiranJPD2011}. 

Comparing the above lifetimes with the time scales in Fig.~\ref{f6}, one can conclude that these 
metastable ROSs can very realistically take part in the 
nucleation and growth of metal oxide nanocrystals (e.g., nanowires) while retaining their 
long-living excited states. This is why plasma oxidation of nanoscale features and 
oxide nanostructure production are very effective and unique compared to thermal processes or plasma 
oxidation of bulk materials. This highly-unusual reactivity lasts 
over the entire growth process and helps explaining the  
metal oxide nanowire growth in just a couple of minutes \cite{plasma-made-nanowires} compared to days in some thermal processes
(bars 16 and 18 in Fig.~\ref{f6}).

\subsection{\label{sheath} Plasma sheath and surface conditions}

\begin{figure}
\begin{center}
\includegraphics[width=9.5cm,height=11cm,clip]{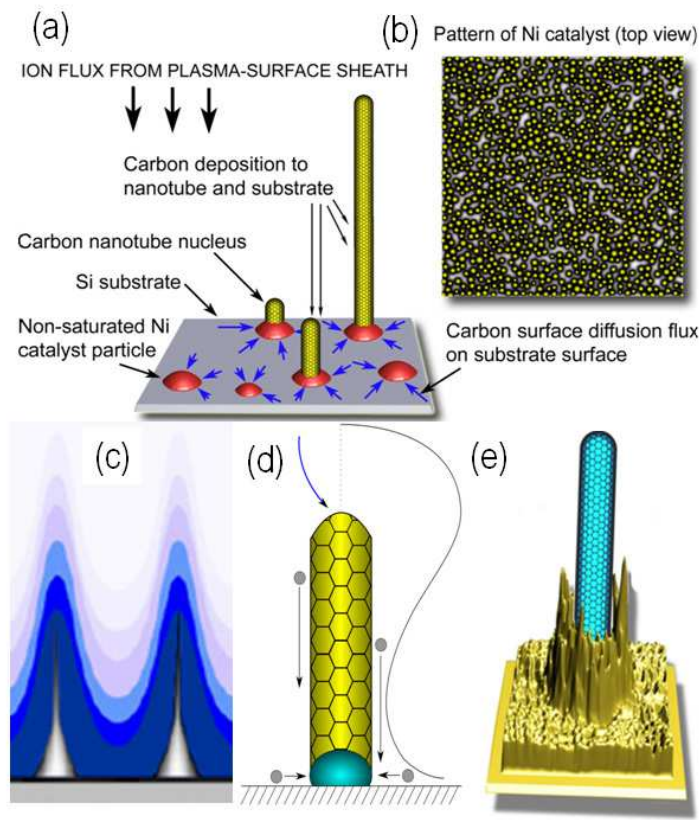} 
\caption{ \label{f9} Ion fluxes (a) and density of carbon atoms (b) in the nucleation and growth of CNTs 
on a self-organized pattern of Ni CNPs [Reprinted with permission from \cite{IgorJAP2008}. 
Copyright \copyright (2008), American Institute of Physics]. 
Topography of microscopic electric field near high-aspect-ratio 
NSs (c) [Copyright \copyright IOP Publishing. Reproduced from \cite{IgorJPD07-SpecIssue} by permission 
of IOP Publishing. All rights reserved.]  Axial (d) and azimuthal (e) profiles of ion deposition in CNT 
growth. Panel (e) reproduced from \cite{IgorIEEETPS08} with permission. Copyright \copyright IEEE (2008).} 
\end{center}   
\end{figure}

Fluxes of the neutral and charged species determine the surface conditions which are very different 
compared to other (e.g., neutral gas) environments.   
The number of neutral species deposited per unit time per unit surface can be approximated as \cite{DenyJAP2004}
\begin{equation}
\label{neutrflux}
\Phi_j^n = {1 \over 4} N_j \upsilon_j V_{Tj}^n,
\end{equation}
where $N_j$, $\upsilon_j$, and $V_{Tj}^n$ are the number density, sticking probability, and thermal 
velocity of neutral species $j$. Because of the absence of reactive dangling bonds, 
the sticking probability of non-radical species is negligibly small. 
For radical species they are typically of the order of a few percent or lower.
For example, $\upsilon_{CH} \sim \upsilon_{CH_2} \sim$ 0.025, $\upsilon_{CH_3} \sim \upsilon_{C_2H_5} \sim$ 0.01,
while $\upsilon_{H} \sim$ 0.001 \cite{31-JAP2004, 39-JAP2004, 40-JAP2004, 41-JAP2004, 42-JAP2004}. 

Within the global discharge model \cite{DenyJAP2004}, the ion flux is
\begin{equation}
\label{ionflux}
\Phi_j^k = \beta_S  n_k^i V_{Bk}^i,
\end{equation}
where $n_k^i$ and $V_{Bk}^i$ are the density and Bohm velocity of ion species $k$, and $\beta_S$ is a 
geometric factor which accounts for the discharge dimensions, ion mean free path, etc.  

The contributions from the charged species materialize through the process of 
formation of the plasma sheath, which starts from the establishment of a negative surface charge due to the 
rapid (bar 12 in Fig.~\ref{f6}) deposition of the highly-mobile plasma electrons. 
This is followed by the formation of a very unique self-organized structure with uncompensated bulk charge 
and continuous flows of ions towards the surface; this occurs over ionic time scales (bar 13). 

The energy the plasma ions acquire upon impact on the floating (i.e., electrically disconnected 
from ground) solid surface is approximately equal to the 
electric potential drop across the sheath
\begin{equation}
\label{ionensheath}
{\cal{E}}_i^k = e(\varphi_{\rm pl} - \varphi_{\rm float}) = \frac{k_B T_e}{2}\ln \left( \frac{m_i^k}{2m_e} \right), 
\end{equation}
where $\varphi_{\rm pl}$ and $\varphi_{\rm float}$ are the plasma and floating potentials, respectively, and 
$m_i^k$ is the mass of ionic species $k$, whereas $e$ and $m_e$ are the electron charge and mass, respectively. 
Importantly, potential drop (\ref{ionensheath}) is always positive; in other words, the plasma potential is always higher than 
the surface potential. This leads to continuous ion flows onto the surface and effective repulsion of
negative charges (e.g., anions or negatively charged nanoparticles or nanoclusters) away from the surface.

Panels (c) and (d) in Fig.~\ref{f8} show the fluxes of ions and neutrals in the Ar$+$H$_2+$CH$_4$ plasma-based synthesis of 
single-crystalline carbon nanotips \cite{DenyJAP2004}. Even though the 
ionization degree of the plasma (\ref{iondegree}) is very low, the
fluxes of the ions are in many cases comparable, or even exceed the fluxes of neutral radicals 
\cite{DenyJAP2004,Hash-plasma-chemistry-for-CNT-NT05}. In other words, ions play a prominent role in the delivery of 
building material to the surface in addition to surface heating. The energetics 
and kinetics of the plasma discharge can be tailored to produce 
species selectively, to achieve the favorable surface conditions.  
Figure~\ref{f9}(a,b) shows a typical situation when ion fluxes are deposited 
on the pattern of Ni CNPs on a Si substrate and the CNTs that 
nucleate and grow from these nanoparticles.  

The negative charge is a common feature of plasma-exposed surfaces which neutralize 
positive ions. It has several important implications for nanoscale processes. First, the negative electric charge 
covers even the smallest features on nanostructured surfaces and determines the microscopic 
electric fields in the vicinity of nanostructures (Fig.~\ref{f9}(c)). 
These electric fields may have both the normal and parallel (with respect to the surface)
components of the electric field.   
As a result, the electric field lines and hence, the ion fluxes converge near the sharp tips of the 
nanostructures as shown in Fig.~\ref{f9}(d) and a significant amount of ions is deposited and then 
neutralized in the top section of the nanotube. In regular patterns, the distribution of ion fluxes around the
nanostructures is typically axially-symmetric as shown in Fig.~\ref{f9}(e). This in turn leads to fairly uniform 
BU incorporation into the CNP.    

After the ions are neutralized on the surface and other neutral species are deposited, the material redistribution 
over the surface is determined by the surface conditions on and between the CNPs. The 
directions of carbon surface diffusion fluxes towards the catalysts are shown in Fig.~\ref{f9}(a). 
These fluxes are determined by 2D gradients of chemical potential over the surface, 
which act as a driving force for the self-organized 
nanotube growth from CNPs. A very non-uniform distribution of carbon adatom densities 
within the catalyst pattern is shown in Fig.~\ref{f9}(b). 

The rates of the surface processes (e.g., surface diffusion) are determined not only by the amount of material 
deposited but also by the surface temperature, as well as by the distributions of surface stresses and charges, which 
appear to be strongly interdependent. In plasmas, intrinsic surface heating mechanisms are possible 
due to ion bombardment or recombination. These effects can be controlled to adjust 
the surface temperature, e.g., to make it higher than the temperature achieved through 
external heating, or, alternatively, to completely replace the external heating by the intrinsic 
plasma heating.     

In particular, surface recombination of long-living, metastable plasma-produced radicals releases 
significant energy. In some cases, this energy is sufficient to produce solid nanostructures even without 
any external heating, e.g., metal oxide NWs of Sec.~\ref{1D-oxide}. 
The surface stress is controlled by the ion impact. Since the distribution of ion impact points 
is very non-uniform (e.g., can peak in selected areas on the NS surface) 
due to the non-uniformity of microscopic electric fields and surface electric charges, the stress distribution 
can also feature non-uniformities over comparable spatial scales.  
 
The surface conditions critically depend on the energy and fluxes of the ions upon impact on the surface, which can be controlled 
by the sheath thickness, which in turn depends on the plasma density, electron temperature 
and surface bias. It is more difficult to control the impact points 
of neutral species. However, their densities, fluxes, energetic states and lifetimes can be controlled 
by the power, pressure and gas flows in the discharge. 

As mentioned in Sec.~\ref{BU}, plasmas very effectively produce highly-reactive atomic and radical 
species that can be used to control the surface energy. 
For example, surface termination by reactive hydrogen 
atoms reduces the surface energy. If the surface is fully passivated, ion bombardment may be used 
to activate suitable dangling bonds to be used as 
anchoring points for the nucleation and growth of new nanostructures \cite{YoonPPAP2007}. 

This discussion only covers the basic conditions on plasma-exposed surfaces. 
Some of these conditions are unique even in the synthesis and processing of thin films and bulk materials. 
Importantly, nanoscale localization of energy and 
matter on the surfaces and in the vicinity of 
nanoscale objects leads to the very interesting physics of 
{\em nanoscale} plasma-surface interactions.

\subsection{\label{nano-plasma-surface} Nanoscale plasma-surface interactions} 

\begin{figure}[t]
\begin{center}
\includegraphics[width=10cm,height=11.5cm,clip]{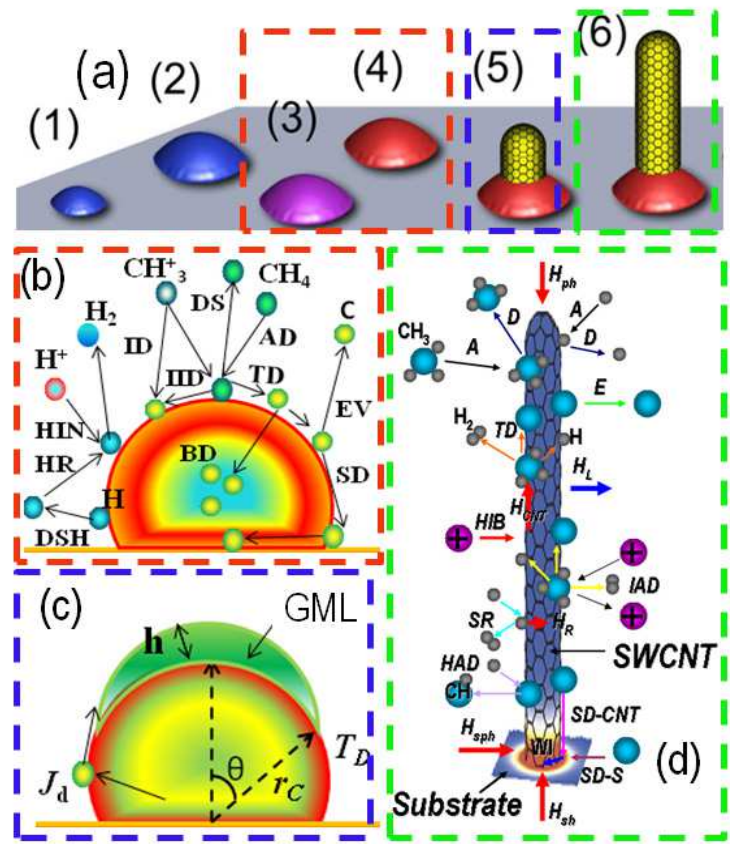} 
\caption{ \label{f10} Stages of CNT nucleation and growth from metal 
CNPs (a) [Reprinted with permission from \cite{IgorJAP2008}. Copyright \copyright (2008), American Institute of Physics.] 
Nanoscale plasma-surface interactions 
across a metal CNP and main elementary processes (b); CNP saturation, carbon extrusion, 
and monolayer nucleation (c) [Reprinted with permission from \cite{HamidACSNano2011}. 
Copyright \copyright (2011) American Chemical Society].   
Elementary processes of  
energy and matter transfer in SWCNT growth (d) \cite{OstrikovIEEE2011}.  
The most important surface processes involved
in carbon atom production and CNP heating in panel (b) 
are ion-induced dissociation (IID), ion decomposition
(ID), hydrogen recombination (HR), hydrogen-induced neutralization
(HIN), adsorption of hydrogen atoms (ADH), desorption
of hydrogen atoms (DSH), adsorption (AD) and
desorption (DS) of hydrocarbon radicals, evaporation (EV),
bulk (BD) and surface diffusion (SD) of carbon atoms, and
thermal dissociation of hydrocarbon radicals (TD).
Panel (d) reproduced from \cite{OstrikovIEEE2011} with permission. Copyright \copyright IEEE (2011).
}   
\end{center}
\end{figure}

Nanoscale localization of plasma-surface interactions arises in two typical types of situations.
In the first case, it is due to the nanometer sizes of the structures or features
being grown or processed, even in bulk plasmas. This is a very common situation 
because of the very large number of nanoscale solids synthesized or processed using plasmas. 
For example, when a small feature is produced on a Si surface using 
reactive ion etching, knowledge on the exact points of impact of the reactive ions 
is needed to study the evolution of the nanoscale etching profiles. 

The other type of situations happens when a nanoplasma is in contact with the surface.
For example, this occurs during surface electrification using 
nanoscale corona discharge generated around a conducting nanometer-sized tip of a scanning tunneling
microscope (STM) \cite{surfelectr}. However, the second situation is very rare because of 
intrinsic difficulties to generate nano-plasmas in air, primarily because of 
insufficient electron densities to reduce the Debye 
length (\ref{Debye-length}) to the nanoscales. More discussions about 
nano-plasma generation from different states of matter will follow in Sec.~\ref{Sec5}.

In another example considered in more detail below, 
carbon nanotubes nucleate and grow on metal catalyst nanoparticles following the sequence of events sketched in 
Fig.~\ref{f10}(a) \cite{CharlierRMP2007,IgorJAP2008}. 
The first 2 steps are needed to produce a pattern of catalyst nanoparticles; steps 3 and 4 denote a partial and full CNP saturation 
with carbon, respectively. After the nanoparticles are saturated with carbon, the nanotubes nucleate (step 5) and then 
grow (step 6) provided that more carbon atoms are supplied through the CNP. An example of a base-led growth of a 
SWCNT is shown. Panel (b) shows more details of the elementary processes that take place 
on the surface and within a nanometer-sized catalyst nanoparticle during its saturation 
with carbon (stages 3 and 4) \cite{HamidACSNano2011}.
Panel (c) shows the details of nucleation of a nanotube cap on the CNP surface (stage 5), while panel (d) shows a multitude 
of elementary processes on the surface of a developed SWCNT \cite{OstrikovIEEE2011}.

Importantly, all 6 stages involve interactions of the plasma-generated species with liquid (e.g., CNP, 
if the melting temperature is reached) and solid (carbon nanotube walls) surfaces with at least one dimension in the 
nanometer range. Even if a SWCNT grows to a several microns length, its thickness typically remains of the 
order of $\sim$1 nm (Fig.~\ref{f10}(d)), and nanoscale plasma-surface interactions still prevail even at 
advanced growth stages. 

Another important feature sketched in Fig.~\ref{f10}(a) is that the nanotubes are thinner than the 
CNPs. This common experimental observation suggests that the nanotube thickness is not merely determined by the 
catalyst material and sizes but also by the synthesis process. Thus, 
{\em kinetic} phemonena are crucial (Sec.~\ref{basicC}). 
On the other hand, thickness of nanotubes and  
nanowires grown in the plasma and neutral gas-based processes 
using CNPs of the same size, is also quite different  \cite{Meyyappan2009}. 

Therefore, {\em kinetics of nanoscale plasma-surface interactions} plays a major role 
in the nucleation and growth of surface-supported NSs. Why are these interactions different from 
the interactions of plasmas with bulk or thin-film materials? Basically, because of the nanoscale 
localization of the energy and matter transfer processes, such as the volume and surface of the CNP 
in Fig.~\ref{f10}(b-c). Since the surface area-to-volume ratio in NPs are much higher than in bulk materials,
elementary processes on plasma-exposed nanoscale surfaces are decisive.

The first striking observation from Fig.~\ref{f10}(b-d) is the overwhelming complexity and multitude of the 
processes involved, with only a few most significant ones shown. Below we will only discuss the 
most essential physics and refer the interested reader to the original publications for many 
other (e.g., chemical) aspects \cite{HamidACSNano2011, OstrikovIEEE2011}. After carbon precursors deposit (as ions or neutrals) 
on the CNP surface, they undergo thermal and ion-assisted dissociation. Since thermal dissociation 
of a CH$_4$ precursor requires temperatures above 800--900$^{\circ}$C, this is an important factor in the 
high temperatures required for the purely thermal SWCNT growth. In a plasma, in addition to effective production of CH$_x$ radicals via 
the gas-phase (e.g., electron-impact) dissociation, precursor dissociation also takes place directly 
on the CNP surface via a number of ion-assisted processes, similar to (\ref{i-n-r}). 

These processes are incorporated into the mass balance equation for carbon atoms on
the semi-spherical surface of a catalyst nanoparticle \cite{HamidACSNano2011}
\begin{equation}
\label{C-balance1}
\Im_C^+ + D_S \frac{1}{r_{C}^2 \sin\theta} \frac{d n_C^S}{d \theta} \left( \sin\theta \right) 
\frac{d n_C^S}{d \theta} - \Im_C^- = 0,
\end{equation}
where $n_C^S$ is the surface density of carbon species on the surface, $D_S$ is the surface diffusion coefficient, 
$r_{C}$ is the CNP radius, and $\theta$ is the azimuthal angle, see Fig.~\ref{f10}(c).
The first term $\Im_C^+$ in (\ref{C-balance1}) describes the creation of carbon atoms
on the surface of a catalyst nanoparticle due to thermal and ion-induced dissociation
of hydrocarbon radicals and ions. The second term accounts for redistribution of the 
carbon atoms due to the surface diffusion, and the third term $\Im_C^-$
quantifies carbon losses due to desorption, interactions with atomic hydrogen, and 
diffusion into the nanoparticle bulk. 

Recombination of ionic and radical species on the 
surface also leads to localized catalyst nanoparticle heating, which in turn increases the nanoparticle 
temperature $T_{\rm CNP}$ by $\Delta T$ without the need to increase the 
temperature of the whole substrate $T_H$
by $\Delta T$ via external heating. 
The nanoparticle surface temperature $T_{\rm CNP}$ is obtained from the energy balance equation which 
takes into account the heats associated with {\em localized} nanoparticle heating 
\begin{equation}
\label{heat-ch1}
{\cal{H}}_{\rm heat}^{\rm CNP} =  {\cal{H}}_{ad} + {\cal{H}}_{ibomb} + {\cal{H}}_{irec} + {\cal{H}}_{ineutr} + {\cal{H}}_{nrec},
\end{equation} 
and cooling 
\begin{equation}
\label{cool-ch1}
{\cal{C}}_{\rm cool}^{\rm CNP} =  {\cal{C}}_{dis} + {\cal{C}}_{des} + {\cal{C}}_{evap} + {\cal{C}}_{intH} + {\cal{C}}_{other},
\end{equation}
which both involve numerous channels. 
In (\ref{heat-ch1}) ${\cal{H}}_{ad}$ is due to adsorption of hydrocarbon and hydrogen species, 
${\cal{H}}_{ibomb}$ is due to kinetic energy transfer through ion bombardment, 
${\cal{H}}_{irec}$ and ${\cal{H}}_{ineutr}$ are related to ion recombination and 
neutralization, while ${\cal{H}}_{nrec}$ is due to the surface recombination of neutrals. 
The heat loss channels in (\ref{cool-ch1}) are due to species dissociation ${\cal{C}}_{dis}$, desorption ${\cal{C}}_{des}$,
evaporation ${\cal{C}}_{evap}$, interaction with hydrogen atoms ${\cal{C}}_{intH}$, and some other mechanisms 
${\cal{C}}_{other}$. 

Importantly, equations (\ref{C-balance1})-(\ref{cool-ch1}) are subject to appropriate 
boundary conditions in limited, nanometer space. The rates and energy barriers 
of several reactions also include corrections due to nanoscale size effects. 
These and several other specific factors make nanoscale plasma-surface interactions markedly 
different compared to the interactions with bulk solid materials.  

One of the consequences of localized plasma heating is that in many nanoscale synthesis 
processes {\em temperatures of external heating $T_H$ are markedly lower}  
compared to similar thermal processes. Moreover, for the effective nanoscale process to proceed, the 
catalyst nanoparticle should be predominantly and selectively heated locally (rather than the entire 
substrate). The improvement of this effect manifests good {\em energy-efficiency} of the plasma-based processes. 
Moreover, lower process temperatures lead to much lower rates of BU evaporation/desorption from the surface
(denoted EV in Fig.~\ref{f10}(b)) thereby also leading to the superior {\em matter-efficiency}.
These features represent a major opportunity for energy- and matter-efficient plasma-based nanoscale synthesis and 
processing of the future.    
     
The as-produced carbon atoms then migrate about the catalyst nanoparticle surface (denoted SD in Fig.~\ref{f10}(b); the  
associated surface diffusion flux is $J_s$) and also diffuse into the catalyst bulk (denoted BD, the 
associated flux is $J_v$). Jointly with other carbon atoms that arrive from the 
substrate surface or directly from the plasma, these atoms saturate the CNP as shown in Fig.~\ref{f10}(c). After the density of dissolved 
carbon in the catalyst nanoparticle reaches the supersaturation threshold, carbon material is extruded towards the surface, as denoted by
$J_d$ in the same panel. This extruded material joins carbon atoms redistributed on the surface by 
surface diffusion, to form a graphene monolayer (GML), which bends to form a SWCNT cap (see Sec.~\ref{nucleation}). 

Plasma-specific processes of the species production and localized heating lead to 
faster GML nucleation by relying on the surface diffusion-driven self-organization, rather than ``waiting" till 
the processes of bulk diffusion, supersaturation, and extrusion are complete. 
These conclusions are supported by the experimental measurements of activation energies of 
surface and bulk diffusion in plasma-assisted growth of various carbon nanostructures where the 
surface diffusion is the main growth rate-controlling mechanism, especially at low temperatures 
\cite{HofmannAPL2003,HofmannPRL2005}. The bulk diffusion pathway is the main option 
in many purely thermal processes \cite{DucatiJAP2002}.

However, during the nanotube nucleation (e.g., cap formation), care should be taken so as not to destroy the monolayer of carbon atoms 
by ion bombardment. Indeed, destruction of the ordered carbon network was observed at ion energies exceeding $\sim$30 eV, which is 
of the order of the energy of carbon bonding in a stable $sp^2$ hexagonal configuration \cite{Limits-PECVD-SWCNT}. 
This is why the energy distributions of the ions generated in the plasma should be adjusted to effectively 
dissociate hydrocarbon precursors on the CNP surface yet without causing any damage to the as-nucleated graphene 
monolayer.

There are several other important mechanisms of nanoscale plasma-surface interactions which 
also significantly affect the metal catalyst and the CNT nucleation on them. For example, plasma-produced 
reactive hydrogen atoms not only etch amorphous material but also prevent metal catalyst from unwanted 
oxidation; both these factors help maintaining the CNPs catalytically active throughout the nanotube
nucleation and growth process. Other known effects of the nanoscale plasma-surface interactions include but are not 
limited to ion-enhanced diffusion (mobility) of carbon atoms on the CNP surface, preventing catalyst nanoparticles from agglomeration, 
and enhancing solubility of carbon atoms in CNPs \cite{NeytsJVSTB2012}.

The unique nanoscale plasma-surface interactions considered in this section 
lead to the very interesting outcomes of 
SWCNT nucleation as will be discussed in Sec.~\ref{nucleation}. These interactions also 
continue playing a major role during the nanotube growth stage, 
see Sec.~\ref{growth} for more details.

\subsection{\label{nucleation} Nanostructure nucleation}  

Nucleation of nanostructures in a plasma involves 
several competing mechanisms. On one hand, the network-forming processes contribute to the 
stable nuclei formation while several other processes such as etching or sputtering may destroy these 
nuclei. For example, impact of ions with the energies higher than the energy of C-C bonds in a hexagonal 
$sp^2$ network, may break these bonds and produce defects in the graphene monolayer on a catalyst nanoparticle.
Therefore, care should be taken to customize the ion fluxes and energies to avoid unnecessary damage to the semi-spherical  
cap that is formed during the nucleation stage, which will be considered below.

\begin{figure}[t]
\begin{center}
\includegraphics[width=9.5cm,height=11cm,clip]{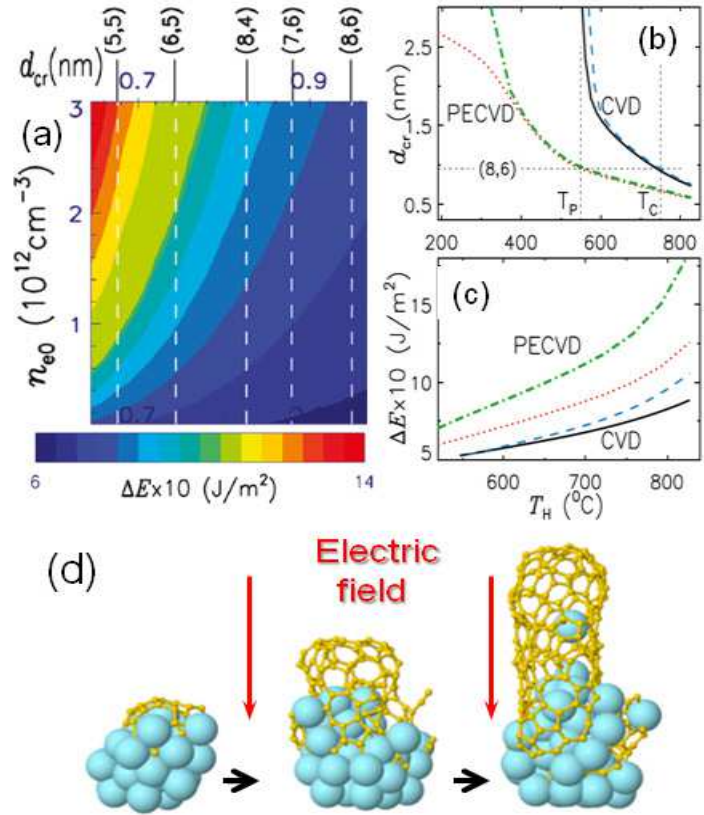} 
\caption{ \label{f11} Effects of the nanoscale plasma-surface interactions on the 
nucleation of SWCNTs (a-c) [Reprinted with permission from \cite{HamidACSNano2011}. 
Copyright \copyright (2011) American Chemical Society]. Numerical simulations 
confirm experimentally observed SWCNT nucleation at the Ni CNP summit, followed by 
the vertical alignment (d) \cite{NeytsJACS2012}.} 
\end{center}   
\end{figure} 

The effect of the plasma exposure on the nucleation of SWCNTs on Au CNPs is shown in 
Fig.~\ref{f11}(a-c) \cite{HamidACSNano2011}.
The as-formed graphene monolayer features a critical diameter $d_{cr}$ (Fig.~\ref{f11}(a,b)), 
beyond which it is stable. The minimum energy (nucleation barrier) $\Delta G_N$ should be overcome to enable the 
graphene monolayer nucleation. The critical GML diameter 
\begin{equation}
\label{GML-crit-diam}
d_{cr} = 2 B r_C / (A^2 + B^2)^{1/2}
\end{equation}
is determined by the CNP radius $r_C$ (Fig.~\ref{f10}(c)),
energies of the GML-vapor $\gamma_{GV}$, GML-CNP $\gamma_{GS}$, and 
CNP-vapor $\gamma_{SV}$ interfaces (entering $A$), energy of the strained 
CNP edge (entering $B$), and the difference in the chemical 
potentials in the solid  and liquid  phases
\begin{equation}
\label{deltamuGT}
\Delta \mu_{sl}^{GT} = \mu_{s} - \mu_{l} - \frac{2 \Delta \Omega_{ls} \gamma_{SV}}{r_C}
\end{equation}
modified by  the size-dependent Gibbs-Thomson effect. This effect is described by the 
third term on the right hand side of Eq.~(\ref{deltamuGT}) and is particularly important for 
small nanoparticles because of the $\propto r_C^{-1}$ dependence. 
Here $\mu_{s}$ and $\mu_{l}$ are the chemical potentials of the solid and liquid phase 
without the GT effect, and $\Omega_{ls}$ is the difference in elementary volumes per atom in the 
liquid and solid phases.

Importantly, $\Delta \mu_{sl}^{GT}$ enters $A$ and is the {\em driving force} for GML nucleation. 
As can be seen from (\ref{deltamuGT}), this driving force becomes weaker for smaller CNPs due to the 
Gibbs-Thomson (GT) effect \cite{Dubrovski-NW-PRB2008}, which reduces the nanoparticle supersaturation. 
This effect increases the nucleation barrier $\Delta G_N$ in thermal processes, and 
very high temperatures and pressures are usually required to nucleate SWCNTs on small catalyst nanoparticles. 

The plasma and the Gibbs-Thomson effects interplay constructively to nucleate and then bend the GML to form a 
nanotube cap. The work of graphene monolayer bending 
\begin{equation}
\label{workbending}
\Delta E = E_K - W_{ad}
\end{equation}
 is the difference between the kinetic energy $E_K$ of coordinated 
detachment of carbon atoms from the catalyst surface and the work of adhesion $W_{ad}$ of 
the as-nucleated layer \cite{HatakeyamaACSNano2010}. 
            
This interplay leads to several effects suggesting significant advantages 
of the plasma-based processes in the nucleation and growth of size- and even possibly
chirality-selective growth of thin 
SWCNTs, on small catalyst nanoparticles, at low process temperatures. First, the same critical diameter $d_{cr}$ can be achieved at temperatures 
significantly lower than in thermal CVD (Fig.~\ref{f11}(b)); this is the consequence of a dramatically reduced 
nucleation barrier $\Delta G_N$. Second, at the same temperatures, $d_{cr}$ is also smaller, which means that 
thinner SWCNTs can nucleate in a plasma, on catalysts of the same size. Third, it is much easier to bend the 
GML in a plasma as the energy of bending $\Delta E$ is typically several times higher, see Fig.~\ref{f11}(c).

A combination of these effects allows one to compute the 3D process parameter maps to determine the optimum plasma 
process parameters when selective nucleation of SWCNTs of a certain thickness (and hence, chirality) is possible; one such 
map is shown in Fig.~\ref{f11}(a). These calculations help explaining the results of recent experiments 
which suggest the possibility of effective control of SWCNT thickness and chirality distributions using 
plasmas and tailored catalysts \cite{DaiNL2008, HatakeyamaJACS2010, KeidarACSNano2010, HatakeyamaACSNano2010}. 

The fast plasma kinetics also leads to the possibility to clearly separate 
the onset of nucleation of SWCNTs of different thickness/chirality, which is characterized by the 
thickness-dependent incubation times \cite{HamidJACS2012}
\begin{equation}
\label{inctime}
\tau_{inc} = \tau_d + \tau_p = \frac{2 n_C^S}{J_v + J_s}, 
\end{equation}
which combine the times needed for carbon atoms to dissolve into the catalyst nanoparticle ($\tau_d$) and
then precipitate ($\tau_p$) at the CNP's surface, where $n_C^S$ is determined from (\ref{C-balance1})
and $J_v$ and $J_s$ are the fluxes of bulk and surface diffusion, respectively.  
This selectivity offers an interesting opportunity to
enable the as yet elusive time-programmed synthesis of SWCNTs with 
controlled thickness and possibly chirality \cite{HatakeyamaACSNano2010}. 
   
There are a few more manifestations of the benefits of 
carbon nanotube nucleation in low-temperature plasmas. 
First, the electric field in the plasma sheath promotes nucleation of SWCNTs 
at the top of the metal catalyst nanoparticle as shown by the results 
of hybrid molecular dynamics/force-biased Monte Carlo simulations 
in Fig.~\ref{f11}(d) \cite{NeytsJACS2012}. The SWCNT cap not only nucleates 
and bends at the CNP summit, but also prominently elongates along the direction of the
electric field, giving rise to the commonly observed vertical growth of carbon 
nanotubes in the plasma \cite{ZFRen-Science1998,MeyyappanPSST03}.   
Interestingly, in thermal CVD the nucleation points of many one-dimensional nanowire-like structures 
are more often observed closer to the nanoparticle edge with 
the interface \cite{Tersoff-Nucleation-Science08}.   

The physics behind this preferential nucleation is in the electric-field 
enhanced mobility of carbon atoms on the CNP surface. Under conditions of typical electric fields 
generated in the plasma sheath at low pressures used for SWCNT growth \cite{HatakeyamaJACS2010}, the 
contribution of the electric field-induced mobility turns out to be stronger 
compared to thermal hopping, which in turn leads to the upward surface fluxes that tend to converge at the 
catalyst nanoparticle summit \cite{NeytsJACS2012}.     

\begin{figure}[t]
\begin{center}
\includegraphics[width=11.5cm,clip]{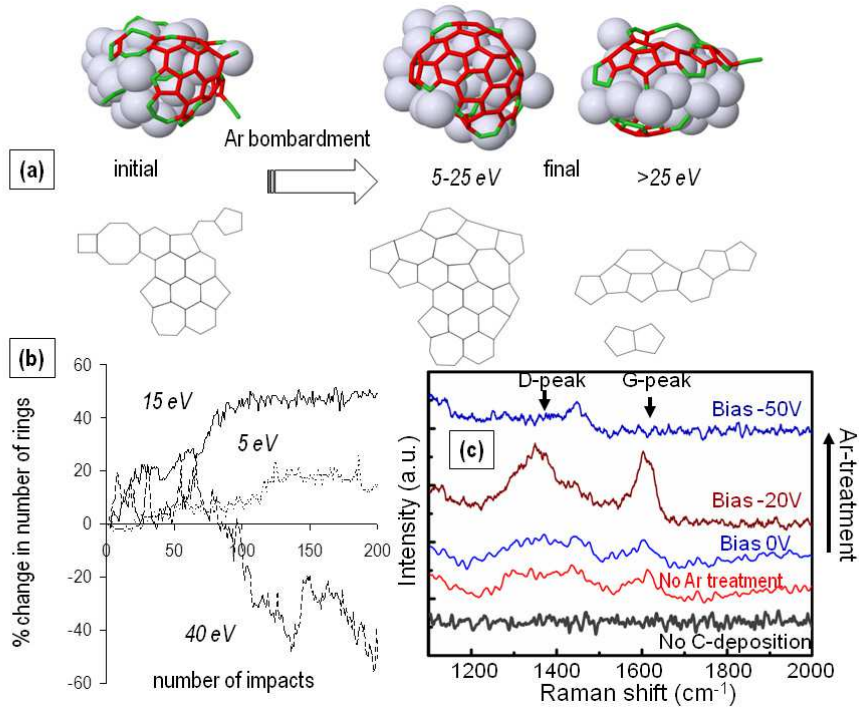} 
\caption{\label{f-ion-bomb} Enhanced nucleation of a SWCNT by low-energy 
ion bombardment [reproduced with permission from \cite{Neyts-ion-bomb-2012}. 
Copyright \copyright (2013) by the American Physical Society]: (a) Representative snapshots of the effect of Ar$^+$  
bombardment with medium (15-–25 eV) and high (30 eV) energy on the SWCNT cap; (b) average percent 
change in the number of rings in the cap over 10 simulations, each consisting of 200 consecutive ion impacts, 
for 5, 15, and 40 eV; (c) experimental Raman spectra for as-nucleated SWCNTs subjected to Ar$^+$ 
ion bombardment under various biases. } 
\end{center}   
\end{figure} 

Since the electric field can control the mobility of carbon atoms during the nucleation process,
lower temperatures of external heating may be required for nanotube 
nucleation. This is consistent with the plasma-enabled reduction of the nucleation temperature 
presented in Fig.~\ref{f11}. The most obvious benefit arises when the temperature is reduced below 
the CNP melting temperature. In this case the catalyst can be maintained in the 
solid (crystalline) state, enabling its facets to act as a template for the nanotube nucleation. 
Unfortunately, a very large number of defects inevitably appear during the nucleation and growth 
at low temperatures. 

Interestingly, the ion bombardment in a narrow energy window effectively 
enhances the nanotube cap nucleation and reduces the defect concentration, 
without increasing the growth temperature. This healing effect of ion bombardment is
rather counterintuitive, because defect creation and other damaging effects are 
commonly expected. 

This possibility was very recently confirmed by a complementary study involving 
state-of-the-art reactive molecular dynamics simulations and experiments on 
ion bombardment of as-nucleated single-walled carbon nanotubes \cite{Neyts-ion-bomb-2012}.  
These simulations involved a model of the repulsive Ni/C -– Ar$^+$ interaction 
described by a repulsive Moli$\grave{e}$re potential employing Firsov constants:
\begin{equation} 
\label{Erik-potential}
V_M = \frac{Z_1 Z_2 e^2}{4\pi \varepsilon_0 r} \sum_{i=1}^3 \eta_i\exp \left( - \frac{\delta_i }{l_{\rm scr}} r \right),
\end{equation}
where $r$ is the scalar distance between the impinging Ar$^+$ ion and the targeted C or Ni atom, 
$Z_i$ is the nuclear charge of atom $i$, $e$ is the elementary charge, $\varepsilon_0$ is 
the permittivity of space, and $l_{\rm scr}$ is the screening length, 
determining the effective interaction range of potential (\ref{Erik-potential}). 
The values for $\eta_i$ are \{0.35; 0.55; 0.1\}, and the values of $\delta_i$ are \{0.3; 1.2; 6.0\} for $i=$ \{1; 2; 3\},
respectively.   

The processes observed in the numerical experiments are clearly different in three energy 
ranges, and are found to be directly related to the maximum transferable kinetic energy
\begin{equation} 
\label{Erik-energy1}
{\cal T}_{\rm max} = {\cal E}_{\rm ion} \frac{4 m_1 m_2}{(m_1 + m_2)^2}, 
\end{equation}
where $m_1$ and $m_2$ are the masses of the impinging ion (with energy ${\cal E}_{\rm ion}$) and 
the C-atom hit. At low ion impact energies, when ${\cal T}_{\rm max}$ is below the energy required to 
displace C-atoms from their positions at the surface \cite{LehtinenPRB2010}, 
the effect of ion bombardment is negligible. In the range 10-–25 eV, however, 
the ions can displace the surface C-atoms, 
allowing more extensive and stable ring structures to form (Fig.~\ref{f-ion-bomb}(a,b)). 
At still higher energies, when ${\cal T}_{\rm max}$ is near or above the C-displacement energy, 
carbon atoms can be displaced from their stable lattice positions in existing 
ring structures, thus damaging the nucleating cap \cite{Neyts-ion-bomb-2012}. 

Experimentally, ultra-short SWCNTs were grown to 
validate the simulations. When a bias of -20 V was applied, 
a significant increase in the ($sp^2$ order-related) G-peak in the Raman spectrum was 
found, whereas a strong decrease in the G-peak is observed 
when applying a bias of -50 V (Fig.~\ref{f-ion-bomb}(c)) \cite{Neyts-ion-bomb-2012}.

It should be noted that the ion bombardment differs in two main aspects with thermal annealing. 
First, the ions act as an external agent, interacting directly with the carbon 
network, rather than acting through the metal nanoparticle. Second, ion bombardment 
is a local influence, affecting only (or mostly) the C-atom(s) directly hit, 
whereas thermal annealing affects the system as a whole. 

Therefore, the ion bombardment, often 
regarded as an ``evil", can be gainfully used 
in even so delicate processes as nucleation 
of single-walled carbon nanotubes.   
In the following subsection, we will consider the plasma-specific effects during 
the next, growth stage of the nanotubes.

\subsection{\label{growth} Nanostructure growth}

\begin{figure}[t]
\begin{center}
\includegraphics[width=9.5cm,height=11.5cm,clip]{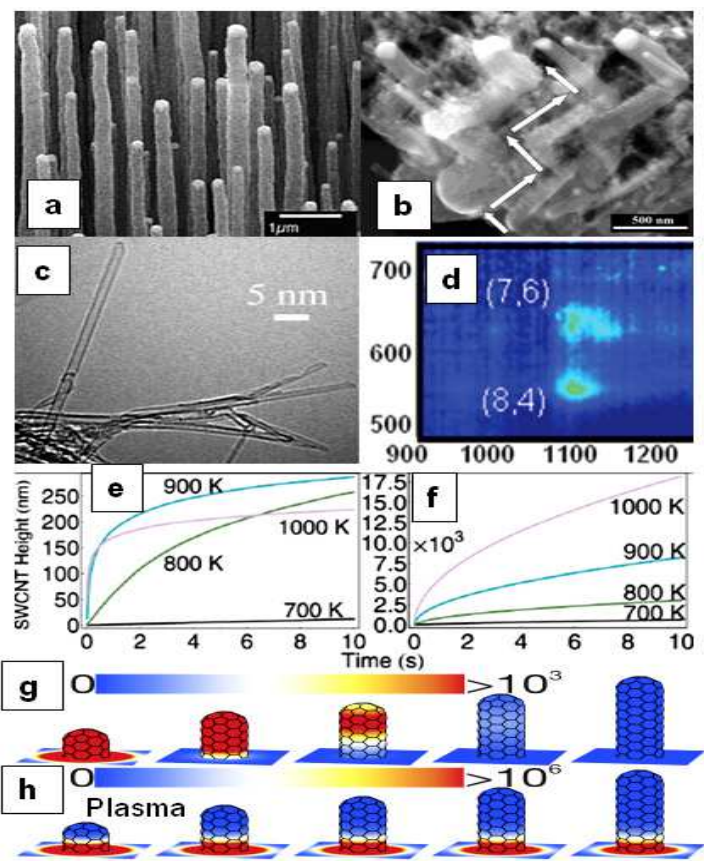} 
\caption{ \label{f12} Vertical alignment of MWCNTs in a plasma (a) 
[From \cite{ZFRen-Science1998}. Reprinted with permission from AAAS]. 
Zig-zag nanotube structures produced by 
varying the direction of DC electric field during the plasma growth (b) 
[Reprinted with permission from \cite{AuBuchonNL2004}. Copyright \copyright (2004) American Chemical Society].  
Plasma-produced SWCNTs show superior structural quality (c) 
[Reprinted with permission from \cite{DaiNL2008}. Copyright \copyright (2008) American Chemical Society]  and 
chirality distributions (d) [Reprinted with permission from \cite{HatakeyamaACSNano2010}. 
Copyright \copyright (2010) American Chemical Society] evidenced by TEM and PLE imaging, respectively. 
SWCNTs grow faster in a plasma 
than in thermal CVD (e,f), 
in part because of building unit delivery to the catalyst nanoparticle at the nanotube base (g,h), as confirmed by 
numerical simulations [Reprinted with permission 
from \cite{ET-SWCNT-APL08}. Copyright \copyright (2008), American Institute of Physics]. } 
\end{center}  
\end{figure}

The pronounced vertical alignment is perhaps one of the most frequently 
cited benefits of using plasmas in the growth of carbon nanotubes and many other 1D
nanostructures \cite{MeyyappanPSST03,my-plasma-nanoscience-book}. 
This effect was observed by a very large number of researchers, with the first reports on the 
observation and interpretation of this phenomenon (see Fig.~\ref{f12}(a))
dating back to 1998--2001 \cite{ZFRen-Science1998,Bower-vert-alignment,MerkulovAPL2001}. 

This alignment is indeed very strong and the electric field control of the nanotube growth direction is 
very effective. In the example shown in Fig.~\ref{f12}(b), the direction of the applied 
electric field was changed, and the CNTs with 
a well-defined zigzag structure were produced 
during DC plasma-enhanced chemical vapor deposition (PECVD). 
Moreover, the nanotubes maintained the same diameter before and after each 
bend while preserving the catalyst particle at the tip \cite{AuBuchonNL2004}. 

This alignment is a result of a complex interplay of 
several mechanisms. One of the earlier mechanisms \cite{MerkulovAPL2001} explains 
the vertical alignment 
by the balance of stresses in  the developing one-dimensional nanostructures. Recently, it was revealed 
that this vertical alignment may not necessarily emerge during the nucleation stage and 
proceeds through the 3 stages from randomly entangled, to partially aligned and fully aligned 
nanotubes. More importantly, the plasma etching (again, process kinetics) is an important   
factor that determines the carbon nanotube growth mode and vertical alignment \cite{ZFRen-NT2011}.

Temperatures for solid nanostructure growth using plasmas are usually markedly lower compared 
to equivalent thermal processes; in some cases this temperature difference can reach a few hundred 
degrees. For example, SWCNTs grow in a plasma at 
temperatures as low as $\sim$450$^{\circ}$C, while heating to
$\sim$800--900$^{\circ}$C is used in thermal chemical vapor deposition \cite{Robertson-SWCNT-APL08}. 
As discussed in Secs.~\ref{nano-plasma-surface} and \ref{nucleation},
these lower growth temperatures are due to the lower energy barriers for 
the NS nucleation.

Despite much lower growth temperatures, the 
plasma-produced CNTs and their arrays often feature superior quality, as evidenced in Figs.~\ref{f12}(c,d). 
The nanotubes in panel (c) have 
a clear, defect-free single-walled structure, which is also evidenced by the strong radial breathing modes
and a typically very high ratio of the intensities of the G and D bands 
in the Raman spectra \cite{DaiNL2008}. 

The SWCNT patterns synthesized in a plasma can feature 
narrow size and chirality distributions as shown in a 
photoluminescence emission (PLE) map in panel (d) \cite{HatakeyamaACSNano2010}. As a result, very large 
fractions (e.g., as high as 96\%) of semiconducting SWCNTs can be produced \cite{DaiNL2008}. 
The mechanisms of reducing the size and chirality distributions have been discussed in Sec.~\ref{nucleation}. 
However, the physics of the  observed SWCNT chirality selectivity is still unclear, in part because 
of the recently revealed changes of the nanotube chirality during the nucleation stage \cite{NeytsJACS2011}.      
  
During the advanced growth stages, the CNT growth rates can be estimated by 
\begin{equation}
\label{grrate}
{\cal R}_{\rm gr}^{\rm CNT} = \frac{m_C}{\rho_{\rm CNT} S_{\rm CNP}}(J_s + J_v),
\end{equation}
where $m_C$ is the mass of a carbon atom, $\rho_{\rm CNT}$ is the nanotube mass density, and 
$S_{\rm CNP}$ is the surface area of the catalyst nanoparticle which supports the growth
(e.g., through which carbon atoms incorporate into the developing nanostructure). 

As discussed above, several plasma-related mechanisms may increase the fluxes $J_s$ and $J_v$
and decrease $S_{\rm CNP}$ (e.g., reduce the nanotube thickness). This is why it is not surprising that 
plasma-based processes also very often feature higher growth rates 
compared to similar thermal processes in the same gases and under the same conditions, see Figs.~\ref{f12}(e,f). 
In particular, these results suggest that in the base-led SWCNT growth mode, plasma-produced carbon atoms 
reach the catalyst and incorporate into the 
nanotube walls much faster than without the plasma. 

In addition, effective etching of amorphous 
carbon by the plasma-produced reactive hydrogen atoms maintains the CNP catalytically active longer, 
which results in taller nanotubes. Panels (g) and (h) show the possibility to control the plasma parameters 
to deliver the ions as close as possible to the catalyst nanoparticle where carbon BUs are needed 
for the growth (red spots denote higher density of ion impacts) \cite{ET-SWCNT-APL08}. 

This is possible to implement by controlling the topography of the microscopic 
electric fields near the nanostructured surface 
\begin{equation}
\label{elfield}
{\bf E}({\bf r}) = \sum_i \int_{S_i} \frac{\rho_i dS_i}{4 \pi \varepsilon_0 r^3}{\bf r} + {\bf E}_{\rm sheath},  
\end{equation}
and the energy of ions that follow the electric field lines \cite{IgorJPD07-SpecIssue}. 
The first term in Eq.~(\ref{elfield}) represents the microscopic field produced by the nanotube arrays, while 
the second term ${\bf E}_{\rm sheath}$ is the field within the plasma sheath. 
Here, $\rho_i$ is the charge density on the CNT surfaces. The curvature of the resulting 
electric field near the nanotubes strongly depends on both the plasma and the nanoarray parameters
and can be optimized to enable selective delivery of the ions to the specific areas
on the nanostructure surface. 

This targeted BU delivery is complemented by the 
fast downward travel of carbon species along the nanotube length and also by the 
reduced desorption rates of these species from the surface. A combination of these factors   
leads to very high carbon nanotube growth rates in plasma-based processes.

\subsection{\label{self-org-patterns} Self-organized pattern formation}   
  
The previous sections dealt with the nucleation and growth of individual solid nanostructures. 
Let us now consider plasma-specific effects in the development of self-organized 
patterns and arrays of two types. 

The first type of patterns uses pre-formed features such as 
arrays of catalyst nanoparticles as in Fig.~\ref{f2}(a). In this type of patterns the 
positions of individual nanostructures are predetermined, e.g., by the CNP positions.    

In a plasma, one-dimensional nanostructures grow in the direction 
of the electric field, see Sec.~\ref{growth}. 
Some nanotubes may grow with different rates and vary in length. 
Since all the NSs in the pattern share the incoming flux of 
precursor species, the nucleation and growth exhibits collective behavior. 
The underlying mechanism is based on the dynamic 
flux redistribution between the substrate and surfaces of the nanostructures. 
For example, simultaneous saturation of catalyst nanoparticles is desirable for 
the formation of length-uniform arrays of carbon nanotubes and other one-dimensional 
nanostructures (e.g., inorganic nanowires). 

It was reported that the
degree of simultaneity of catalyst nanoparticle saturation with carbon in plasma-based processes is higher than 
in the equivalent thermal CVD. This effect owes to the more regular redistribution of 
carbon flux caused by the microscopic electric field effects \cite{IgorAPL08-Catalyst}. 
Surprisingly, plasma-guided self-organization thus also plays a role even in the 
arrays with pre-determined positions of individual nanostructures.

\begin{figure}[t]
\begin{center}
\includegraphics[width=9.5cm,height=12cm,clip]{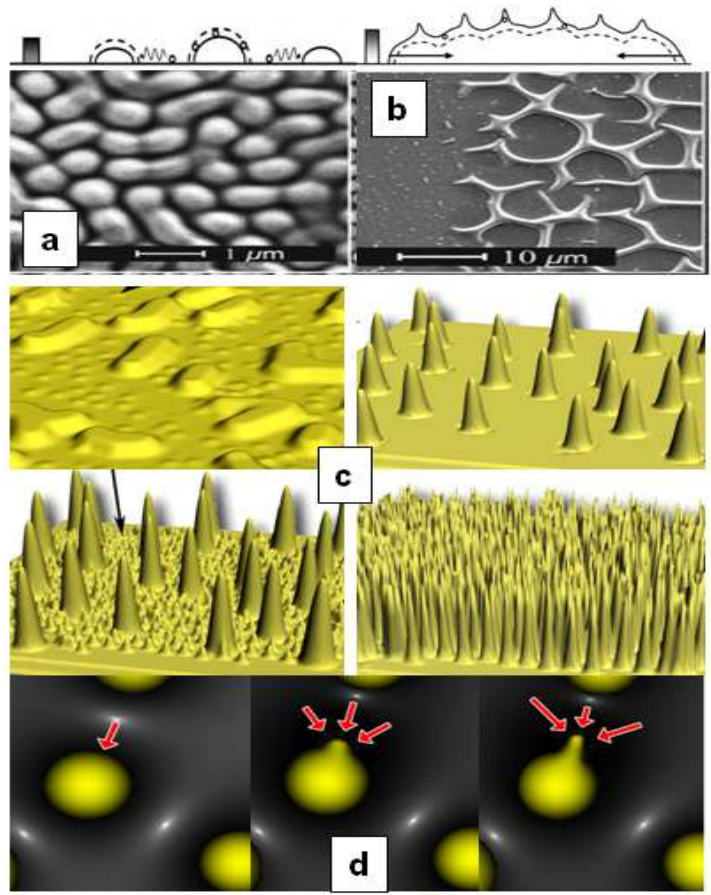} 
\caption{ \label{f13} Self-organized SiO$_2$ nanoarrays from Ar$+$O$_2$ microplasmas (a,b) 
and sketches of surface temperature gradients (bars), stress levels (arrows), and profiles formed 
[Reprinted with permission from \cite{BelmonteAPL2010}. Copyright \copyright (2010), American Institute of Physics]. 
3D self-organized growth of uniform arrays of carbon nanocones from
non-uniform catalyst patterns (c) 
[Reprinted from \cite{Carbon2007}, Copyright \copyright (2007), with permission from Elsevier]. 
Self-alignment of carbon nanowire connections between 
Ag particles on Si (d); carbon fluxes are shown by arrows 
[Reprinted from \cite{Carbon2009-Letter}, Copyright \copyright (2009), 
with permission from Elsevier]. }
\end{center}    
\end{figure}

Figure~\ref{f13} shows examples of self-organization in the second type of 
patterns where solid nanostructures nucleate randomly. This randomness leads to additional 
complexity. However, reasonable ordering persists 
even in self-organized 2D and 3D arrays  \cite{Meletis1,Meletis2}. In addition to the balance of 
material supply and consumption which controls the 
first type of patterns, the stress and 
surface energy landscapes, along with non-uniform 
temperature distributions play a vital role in the second case.   

An interplay of these factors leads to very different self-organized 
patterns of SiO$_2$ nanostructures (Fig.~\ref{f13}(a,b)) on stainless steel substrates exposed to 
remote microwave microplasmas of Ar$+$O$_2$ gas mixtures \cite{BelmonteAPL2010}. 
These patterns require different plasma conditions,  
temperature gradients, and stress levels. Weak thermal gradients under higher temperatures 
lead to the low stress levels and hence,  
the nanodot morphology (a). On the other hand, stronger thermal gradients
at lower temperatures generate much higher stress levels (arrows in (b)) 
leading to the self-organized hexagonal cells. 
Relative contents of working (Ar$^+$ ions) and 
building (O atoms) units are also different in both cases. 
A higher density of Ar and a lower density of oxygen lead to
highly stressed (e.g., due to stronger ion bombardment) and less dense (due to reduced BU delivery) patterns in Fig.~\ref{f13}(b). 
Customized oxygen-based plasmas can produce self-organized 
patterns of randomly-nucleated nanodots, nanowires, nanowalls, and other oxide 
NSs (see Sec.~\ref{1D-oxide}).

Figure~\ref{f13}(c) shows the temporal dynamics of the formation of a dense and uniform 
3D array or carbon nanocones (bottom right) on very non-uniform patterns of Ni catalyst 
nanoparticles (top left). A combined experimental and numerical 
study \cite{Carbon2007} has revealed that this self-organization proceeds through 
several stages. First, a primary rarefied array of nanocones is formed (top right), followed by 
nucleation and faster growth of smaller nanocones between them (bottom left).
Importantly, this process is due to the plasma-specific redistribution of carbon BUs 
across the 2D substrate and the surfaces of developing nanostructures and is not 
possible in the equivalent thermal process.    

An example of one such 2D rearrangement of carbon BUs on a 
plasma-exposed Si surface between Ag particles is shown in Fig.~\ref{f13}(d). 
As mentioned in Sec.~\ref{growth}, this leads to 
non-uniform microscopic electric fields with a component parallel to the 
surface. Adatoms respond to this long-range non-uniform electric field. 
Both the electric field and its 
non-uniformity are the strongest along the line connecting 
two adjacent catalyst nanoparticles. As a result, carbon wires nucleate and then grow along this
direction \cite{Carbon2009-Letter}. Physically, this effect is quite similar to the SWCNT nucleation 
in a vertical electric field (Fig.~\ref{f11}(d)). This horizontal growth 
evidences that the growth direction of 1D carbon wires can also be controlled 
using {\it self-organized} microscopic electric fields, in addition to 
external fields (Fig.~\ref{f12}(b)). 

The development of self-organized nanopatterns in Figs.~\ref{f13}(c,d) can be described 
using the following two-dimensional diffusion equation of carbon atoms \cite{Carbon2009-II} 
\begin{equation}
\label{diffeq3}
\frac{\partial \xi}{\partial t} = D_S \left( \frac{\partial^2 \xi}{\partial x^2} + \frac{\partial^2 \xi}{\partial y^2}  \right) 
+ \Psi^+ + \Psi^-,  
\end{equation}
where $\xi$ is the surface density of carbon atoms, whereas $\Psi^+$ and $\Psi^-$ are the source and sink terms of
carbon atoms that add to the surface and leave it, respectively. 
The diffusion coefficient $D_S$ is given by Eq.~(\ref{diffcoeff}), with the effective 
energy barrier 
\begin{equation}
\label{effdiffbar}
{\cal E}_d^{\rm eff} = \frac{\Sigma_k \varepsilon_k \eta_k}{\Sigma_k \eta_k}   
\end{equation}
and pre-factor $D_0$ modified to incorporate several features (e.g., imperfections) 
of surfaces or real solid materials summarized in Table~\ref{symbols}. 
In Eq.~(\ref{effdiffbar}), $\eta_k$ is the surface density of $k$-th feature expressed in m$^{-1}$, and 
$\varepsilon_k$ is the energy barrier associated with $k$-th feature. 

\begin{table}[t]
  \tbl{Activation energies for surface processes and densities of features \cite{Carbon2009-II, 30-Carbon2009-II,
34-Carbon2009-II, 35-Carbon2009-II, 41-Carbon2009-II, 42-Carbon2009-II}.}
{\begin{tabular}{@{}lll}\toprule
   Energy/Density
  & Feature & Numerical value \\
\colrule
   $\varepsilon_d$ & Surface diffusion activation energy for Si surface & $\sim$1.1 eV \\
   $\varepsilon_V$ & Adatom activation energy on surface vacancy & $\sim$1.43 eV \\
   $\varepsilon_I$ & Adatom activation energy on interstitial atom & $\sim$1.25 eV \\
   $\varepsilon_F$ & Adatom activation energy on Frenkel defect & $\sim$1.5 eV \\
   $\eta_V$ & Surface density of vacancies & $\sim 5.0 \times 10^{15}$ m$^{-2}$ \\
   $\eta_I$ & Surface density of interstitial atoms & $\sim 5.0 \times 10^{15}$ m$^{-2}$ \\
   $\eta_F$ & Surface density of Frenkel defects & $\sim 1.0 \times 10^{15}$ m$^{-2}$ \\
   $\varepsilon_D$ & Adatom activation energy on dislocation & $\sim$1.2 eV \\
   $\eta_D$ & Surface density of dislocations & $\sim 1.0 \times 10^{15}$ m$^{-2}$ \\
   $\varepsilon_S$ & Adatom activation energy at step terrace (Ehrlich–-Schwoebel barrier) & $\sim$2.2 eV \\
   $\eta_S$ & Surface density of step terrace sites & $\sim 8.5 \times 10^{15}$ m$^{-2}$ \\
   $\varepsilon_P$ & Adatom activation energy on nano-grain boundaries of Si surface & $\sim$2.2 eV \\
   $\eta_P$ & Surface density of nano-grains on Si surface & $\sim 1.0 \times 10^{15}$ m$^{-2}$ \\
   $R_C$ & Correction coefficient for surface diffusion on a real-roughness surface & $\sim 0.97$ \\
   $\varepsilon_{mS}$ & Adatom activation energy on substitutional atom & $\sim$1.3 eV \\
   $\eta_{mS}$ & Surface density of substitutional atoms & $\sim 8.5 \times 10^{15}$ m$^{-2}$ \\
   \botrule
  \end{tabular}}
\label{symbols}
\end{table}

Although the focus of the original report \cite{Carbon2009-II} was on the effect of the electric  field 
on the formation of self-organized connections between the microparticles, several other effects can drive this 
process. Similar to Fig.~\ref{f13}(a,b), surface stress between the particles 
plays an important role. Indeed, the stress magnitude is higher in the area between the two adjacent
particles. Hence, the surface stress should enhance carbon 
adatom incorporation into the micrograins. Consequently, the levels of the particle 
supersaturation with carbon should be higher at the surfaces facing the
nearest neighbor. This also helps the nucleation
of the 1D carbon structures connecting the closest micrograins.

Here we stress that during this self-organized 
growth process, the electric, adatom density,
stress, and chemical potential fields self-organize in a complex
and possibly, coordinated way to sustain
the connection growth between the two nearest micrograins \cite{Carbon2009-II}.
This interesting possibility still awaits its conclusive experimental 
verification.

Plasma-specific phenomena also affect development of many other 
self-organized nanopatterns and arrays, with further examples related to
diamond-like \cite{Golding}, Ni \cite{Ni-island-self-org-remote-plasma}, 
Ge \cite{JamesHoJAP07}, and SiO$_2$ \cite{Igor-Uros2013} nano-islands.  
Although every process is different, the physics of these  
self-organization phenomena is quite similar and based on 
the effects discussed in this section. With this knowledge, we can now review several classes
of nanoscale solids and plasma processes.

\section{\label{Sec4} Nano-solids from plasmas: plasma-specific effects and physical, chemical, and functional properties} 

The previous section focused on a sequence of events 
and elementary processes during the formation of solid 
nanostructures and nanoarrays on plasma-exposed 
surfaces. Here we use a different approach to cover a  
broad range of materials systems and nanostructures. 
These objects are split according to 
their dimensionality, similar to Fig.~\ref{f1}. 

This choice is dictated by the ability of nano-solids to 
confine electrons in a different number of dimensions, which in turn determines 
their properties and applications.   
Each of these ``dimensional" categories features 
typical materials systems, e.g., carbon- and silicon-based, 
inorganic, organic, living, etc. materials. Where appropriate, 
we also discuss the plasma-specific physical effects and 
the associated materials structure and 
performance in applications.       

This is why, as mentioned in Sec.~\ref{introE}, each section 
first introduces the envisaged applications of the  
solid materials, then any existing problems in their synthesis 
and processing, followed by a critical discussion of the 
plasma-specific effects that lead to any process improvements or enable 
some new features. The focus of these discussions is on the
final outcomes rather than the dynamics of nucleation and development 
of the structures, which were discussed in greater detail 
in Secs.~\ref{Sec2} and \ref{Sec3}. 
Nonetheless, wherever possible, we comment on these features and 
refer the reader to the appropriate sections in this review. 

The number of nano-solid systems referred to in this section is very large.
This is why we discuss only the most essential of those 
properties of these solids that were affected by any 
relevant plasma-specific effects. The focus is mainly on 
the morphological, structural, and selected functional properties. 
We also provide links to relevant solid-state physics and chemistry,
materials science and nanoscience literature which provide 
exhaustive coverage of the properties and applications for the 
main materials types considered in this section.

\subsection{\label{0D-structures} Quantum dots, nanocrystals and nanoparticles}   

Here we consider three basic options of the production and organization 
of zero-dimensional solid nanoparticles using plasmas: quantum dots 
embedded in bulk solid materials,
unsupported nanocrystals and nanoparticles, 
and surface-supported nanoarrays.  

\subsubsection{\label{QD} Quantum dots}

\begin{figure}[t]
\begin{center}
\includegraphics[width=12.5cm,clip]{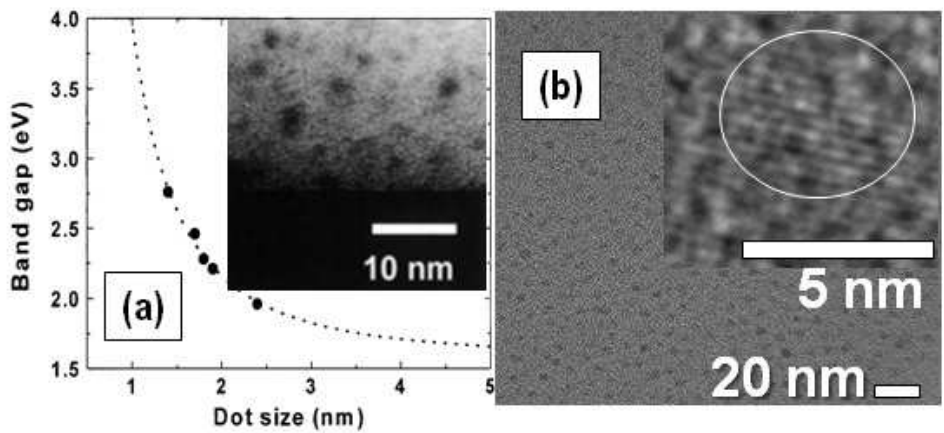} 
\caption{ \label{f14} $a$-Si QDs in a SiN film grown by PECVD and their bandgap 
{\it versus} QD size [Reprinted/adapted with permission from \cite{ParkPRL2001}. 
Copyright \copyright (2001) by the American Physical Society]. Crystalline Si QDs in an $a$-Si matrix produced 
using SiH$_4$ plasmas with very low or no dilution in H$_2$ (b) [Reprinted/adapted with permission from 
\cite{QJ-nc-Si-2}. Copyright \copyright (2009) American Chemical Society].  } 
\end{center}  
\end{figure} 

\paragraph{Semiconducting quantum dots}

Semiconducting quantum dots, which feature strong electron quantum 
confinement in all three dimensions, are usually unsupported, surface-supported, or embedded in a 
matrix of a bulk material. A large variety of unsupported quantum dots has been produced using
colloidal chemistry approaches; this is a very active and well-developed research 
field. Here we discuss the latter two options which have been 
demonstrated using low-temperature non-equilibrium 
plasmas. Figure~\ref{f14}(a) 
 shows a cross-sectional high-resolution transmission electron
microscopy (HRTEM) image of $a$-Si QDs embedded in a silicon nitride
film \cite{ParkPRL2001}. These 3D quantum dot patterns 
are produced at very high growth rates 
from 1.4 to 3.2 nm/min using reactive mixtures 
of SiH$_4$ and N$_2$ gases. The dark spots in Fig.~\ref{f14}(a) represent QDs with an
average size $\sim$1.9 nm and a dot density of $\sim$1.0$\times$10$^{19}$cm$^{-3}$. 

The quantum confinement effects are manifested by the 
wider bandgap energies for smaller quantum dot sizes. The 
experimental points fit the dashed line plotted using 
the effective mass theory for three-dimensionally 
confined Si QDs \cite{ParkPRL2001}
\begin{equation}
\label{bandgap4}
E(a) = E_{\rm bulk} + \sigma_{\rm qconf} / a^2,
\end{equation}
where $E_{\rm bulk}$ is the bandgap of bulk Si, $a$ is the quantum dot size (diameter), and 
$\sigma_{\rm qconf}$ is the electron confinement parameter. 
These results thus evidence the effective electron confinement in the $a$-Si
QDs, which leads to the observed strong photoluminescence (PL). 
The value of the achieved electron confinement parameter $\sigma_{\rm qconf} =$2.4 eV$\cdot$nm$^2$  
is more than 3 times larger compared to two-dimensional $a$-Si quantum wells \cite{LockwellPRL}.
This can be explained by noting that 
\begin{equation}
\label{effmass5}
\sigma_{\rm qconf} \propto {1 \over m^*} = {1 \over m_e^*} + {1 \over m_h^*},
\end{equation} 
where $m_e^*$ and $m_h^*$ are the effective masses of electrons and holes,   
$m^*$ is the reduced mass, and the effective masses in amorphous materials are commonly assumed to
be isotropic in all 3 directions  \cite{ParkPRL2001}.

Importantly, the degree of passivation of QD surfaces which determines the 
PL efficiency, can be effectively controlled through altering the rates 
of reactive species production (e.g., nitrogen or hydrogen dissociation) in the plasma discharge by the 
operating parameters. The degree of passivation of $a$-Si by SiN in Fig.~\ref{f14}(a) 
is very high \cite{ParkPRL2001}. This very effective 
plasma-assisted surface passivation by SiN  
is commonly used in the fabrication of photovoltaic 
solar cells \cite{XuJPD2011-SpecIssue}.   
The degree of passivation of Si QDs determines the 
difference between the luminescence energy $E_{\rm lum}$ and the absorption energy $E_{\rm abs}$, 
which is in turn determined 
by the density of deep luminescence centers such as the surface states.

The density of such states is usually higher in nanocrystalline Si 
({\it nc}-Si) made of single-crystalline Si quantum dots embedded in an $a$-Si matrix  
(Fig.~\ref{f14}(b)). To improve the surface passivation, heavy dilution of 
SiH$_4$ in H$_2$ is commonly used \cite{Roca-TSF2004,QJ-nc-Si-2}. 
However, this very significantly affects the growth rates, and hence, the material cost 
in the solar cell production. The outstanding ability of thermally non-equilibrium 
plasmas to dissociate precursor species has led to the unique possibility to avoid 
hydrogen dilution and produce high-quality {\it nc}-Si films at low process temperatures
($\sim$150$^{\circ}$C) where the quantum dot formation was observed. 

Si QDs in Fig.~\ref{f14}(b) feature the preferential (111) growth orientation and the 
crystalline fraction can be controlled from zero to 86\%. The QDs can be grown 
at very competitive growth rates of up to $\sim$2.5 nm/s using fairly low power densities of 
$\sim$40 mW/cm$^3$, and very low or no hydrogen dilution of silane precursors. 
These features are very difficult to achieve otherwise. This is why 
low-temperature plasmas are commonly used in flat display and solar cell technologies.  

\paragraph{\label{QD-simulation} QD formation and properties: numerical simulations}

The mechanism of the Si QDs formation from the plasma is not fully understood. 
Nonetheless, it is clear that the nanocrystals may form either through localized 
crystallization on the surface as the $a$-Si film grows or may form and 
be delivered from the gas phase. 
The energy of the plasma-grown Si nanocrystals upon deposition appears to be a critical 
factor which determines the structure selectivity between the amorphous and nanocrystalline Si 
films \cite{Roca-TSF2004}. This energy is determined by the interplay of  
charging mechanisms and the (electrostatic, 
ion and neutral drag, etc.) forces acting on the 
nanoparticles in the plasma \cite{Kortshagen-review,PhysRepts2004,IvlevRMP}.

\begin{figure}[t]
\begin{center}
\includegraphics[width=12.5cm,clip]{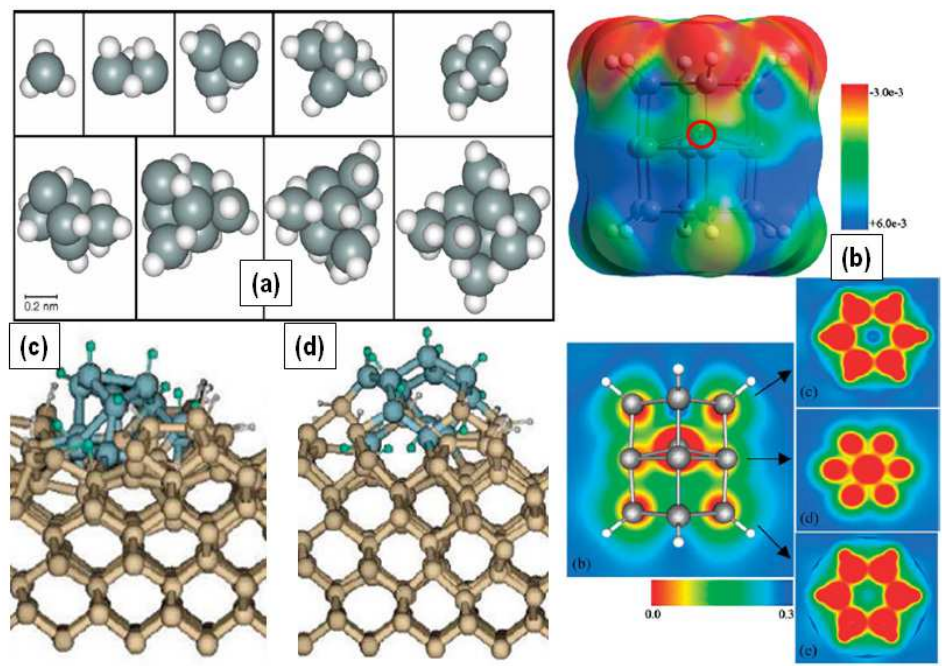} 
\caption{ \label{f-Vach} Plasma-grown silicon hydride nanoclusters, their electronic properties and role as building units 
in crystalline Si growth studied by {\it ab initio} numerical simulations \cite{Vach2005,Vach2006,Vach2010}. 
Si$_n$H$_m$ clusters grown under typical plasma conditions (a) [Reprinted from \cite{Vach2006}, 
Copyright \copyright (2006), with permission from Elsevier]. 3D profiles 
of electron density of a hydrogenated Si nanoparticle with the ``off center" Si atom (encircled); this 
non-symmetric atomic arrangement leads to a very strong permanent dipole moment visualized by graded 
colors (b) [Reprinted with permission from \cite{Vach2005}. Copyright \copyright (2005) by the American Physical Society]. 
Plane cuts of the Si$_{19}$H$_{12}$ nanoparticle in different directions produce 
2D contour plots of the electron density also shown in the lower section of panel (b) \cite{Vach2005}. 
Panels (c) and (d) show a temporal
dynamics of epitaxial recrystallization of a Si$_{12}$H$_{12}$ 
nanocluster after 0.42 and 6.0 ps after the 
impact on the Si surface; surface temperature is $T=$300 K 
and the cluster impact energy $E_{\rm imp}=$60 eV [Reprinted
with permission from \cite{Vach2010}. Copyright \copyright 
(2010) American Chemical Society].} 
\end{center}  
\end{figure} 

Using a multilevel simulation strategy, Vach et al. \cite{Vach2005,Vach2006} 
simulated the controlled growth of silicon nanoclusters in a plasma. 
Most importantly, it was found that when there was no atomic hydrogen present in the plasma, 
small Si:H nanocrystals could not be formed. In contrast, under plasma conditions leading 
to very low H-impact rates, the formation of hydrogen-rich, small nanocrystals was found. 
When higher H-impact rates were allowed, hydrogen-poor crystallites appeared. 

Representative examples of Si$_n$H$_m$ clusters studied by these numerical simulations 
under typical plasma conditions are shown in Fig.~\ref{f-Vach}(a). 
Interestingly, the hydrogen atoms always remain on the outside of the Si$_n$H$_m$ clusters. 
The variety of such structures indicates that the plasma conditions can be adjusted such that the resulting 
nanoparticle properties can be tuned. This ability is of crucial importance, in the plasma-assisted 
fabrication of polymorphous solar cells \cite{Roca-TSF2004}. 

For example, using an  intermediate flux of hydrogen atoms (which is possible to implement in real plasma experiments 
by controlling the degree of hydrogen dissociation), one can produce a unique structure with 
a Si atom shifted off the center shown in Fig.~\ref{f-Vach}(b). This asymmetry  leads to the very large
permanent dipole moment $\sim$ 1.9 D (Debye units), which is comparable to a water 
molecule \cite{Vach2005}. This dipole moment is visualized by the 3D and 2D profiles of the 
electron density calculated from the total self-consistent field density mapped with the 
corresponding electrostatic potential in Fig.~\ref{f-Vach}(b). 
Importantly, the presence of this dipole moment enables the unique possibility to align these 
Si:H nanoclusters in the plasma electric field. This alignment in turn may lead to the electrostatic 
attraction of the oppositely charged poles of different nanoclusters thereby forming chain- or wire-like
nanoassemblies. This is an interesting example of the use of tailored building units to 
assemble more complex structures. We emphasize that both stages of this process benefit from 
the plasma-specific effects \cite{Vach2005}.       

In a subsequent computational study \cite{Vach2010}, it was shown that the reaction mechanisms 
of the Si:H nanoclusters on Si-surfaces depend mostly on the impact energy, 
but are not very sensitive to the cluster size. Moreover, a good agreement with experimental 
results was obtained. To allow a comparison between impacts of various cluster sizes, 
the effect of the impact energy was quantified through the impact energy coefficient 
\begin{equation}
\label{impencoeff-vach}
\alpha_{\rm imp} = \frac{E_{\rm imp} / N_{\rm cl}}{E_{\rm sub}},
\end{equation}
where $E_{\rm imp}$ is the cluster impact energy, $N_{\rm cl}$ is the number of silicon atoms in the cluster, and 
$E_{\rm sub}$ is the cohesive energy of the substrate. 

Using the definition (\ref{impencoeff-vach}), one can discern three distinctive regimes \cite{Vach2010}. 
Low impact energies lead to reflection of 
the cluster, medium impact energy lead to soft 
landing, while higher energies lead to fragmentation of the clusters. 
Impact-induced dissociation followed by fragment migration, penetration and 
partial surface damage, on the other hand, is observed at higher cluster energies. 
The reported difference in reactivity of smaller and larger clusters 
stems from the different and size-dependent
structural stability of the clusters. These findings are consistent 
with the experimental results \cite{WatanabePRL98}.

Of particular interest is the possibility to epitaxially grow nanocrystalline Si films 
by such cluster impacts. The results of simulations in Figs.~\ref{f-Vach}(c,d)
show that such epitaxial growth is possible, as evidenced by the epitaxial-like recrystallization of 
clusters (d) after the loss of the initial cluster structure following surface impact (c) \cite{Vach2010}.  

The ability to tailor the structure and electronic 
properties of nanocluster BUs and their integration into 
nanostructured Si films is indispensable for the development 
of the third-generation photovoltaic solar cells, thermoelectric, energy storage, optoelectronic,
and several other advanced applications \cite{ordered_dense_arrays_PV,XuJPD2011-SpecIssue}.
A clear opportunity for future research exists in the understanding of the 
possible contribution of the plasma-specific effects to improve 
the formation and ordering of self-organized three-dimensional structures 
in nanocrystalline Si which may form during heteroepitaxial growth and 
chemical synthesis \cite{nc-Si-nature2000}.

\paragraph{\label{C-dots} C-dots}

The discussions above are related to more traditional examples of 
 plasma-assisted synthesis of Si-based quantum dots. In these examples, 
which have dominated the literature for about a decade, low-pressure 
(e.g., in the 10s-100s mTorr range) plasma discharges in 
highly-toxic and flammable gaseous precursors of relevant materials 
(e.g., silane, germane, etc.) have been commonly used.

\begin{figure}[t]
\begin{center}
\includegraphics[width=11.5cm,clip]{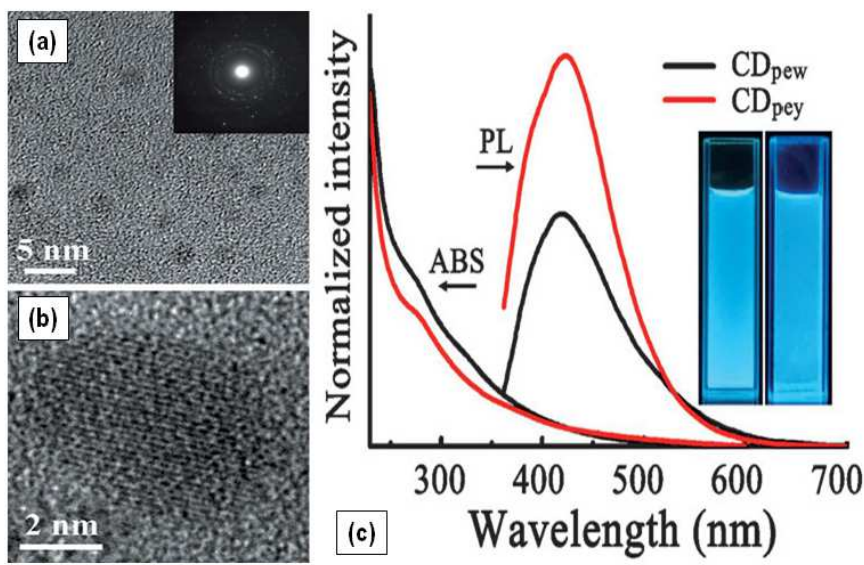} 
\caption{ \label{f-C-dot} High-resolution TEM micrographs (a,b) and optical emission/absorption spectra (c) \cite{C-dot-plasma} 
of atmospheric-pressure plasma-synthesized 
C-dots using natural egg yolk precursor in aqueous solution. Panel (b) shows a selected C-dot at higher magnification.
Inset in panel (a) shows a typical electron diffraction pattern. Spectra of photoluminescence emission 
and UV/Vis absorption of C-dots in aqueous solution (c); excitation wavelength is 360 nm.  
CDpey and CDpew denote egg yolk and white produced carbon dots. Inset in panel (c)
shows optical images of CDpew (left) and CDpey (right) solutions 
under UV light exposure. 
Reproduced with permission from \cite{C-dot-plasma}. Copyright \copyright 2012 Wiley-VCH Verlag GmbH \& Co. KGaA, Weinheim.} 
\end{center}  
\end{figure} 

Recently, several new possibilities to synthesize 
QDs of other materials systems by using different plasmas and  
precursors have been demonstrated. For instance, 
atmospheric-pressure dielectric barrier discharge (DBD) plasmas 
were used to produce luminescent carbon QDs (commonly termed C-dots \cite{C-dot-review})
from natural egg yolk and white precursors \cite{C-dot-plasma}. 
As can be seen from Fig.~\ref{f-C-dot}(a,b), plasma-produced C-dots have quite similar sizes 
and morphological appearance compared to Si QDs in Fig.~\ref{f14}(b).

The C-dots represent a new class of recently discovered carbon nanomaterials 
which typically feature spherical morphology and tunable surface functionalities. 
They are mostly composed of $sp^2$ hybridized carbon and as such may be termed 
nanocrystalline graphitic nanodots, in particular because of 
the above mentioned similarity to nanocrystalline Si. 
This makes them very different compared to diamond nanoparticles
which will be considered in Sec.~\ref{nanodiamond}.

The size and 
surface termination of the C-dots can be effectively tuned by the 
synthesis process parameters. This leads to the size- and 
excitation energy-dependent photoluminescence emission.  
Importantly, because of their biocompatibility, 
the carbon dots are considered as viable alternatives to commonly 
used metal and semiconducting quantum dots, which are often not only 
toxic but are also produced using hazardous precursors such as the 
SiH$_4$ used to synthesize $nc$-Si discussed above. 
Despite many recent achievements, the degree of control of size, 
elemental composition (e.g., oxygen content), defects, and surface 
functional groups still remains quite limited. 
Consequently, tunability of the optical luminescence
while minimizing absorbance, quantum yield of photon emission,
and the fluorescence lifetime still remain significant issues. 

Unlike many conventional methods of synthesizing C-dots, the 
atmospheric plasma-based process \cite{C-dot-plasma} relied on
natural and renewable carbon precursors, namely egg yolk and white. The 
carbon dots showed good crystalline quality as can be seen in 
HRTEM micrograph in Fig.~\ref{f-C-dot}(b); the typical size of the
C-dots produced was under $\sim$5 nm.  
X-ray Photoelectron Spectroscopy (XPS) and
Fourier Transform Infrared Spectroscopy (FTIR) analyses also 
confirmed the presence of C$=$C, C--OH, C--O--C, and C$=$O functional 
groups. 

The yolk-produced C-dots showed somewhat better crystalline structure,
which was explained by the higher presence of lipids in egg yolk
($\sim$33\%) compared to egg white ($\sim$0.01\%). It was suggested that 
the higher density and viscosity of the yolk was favorable to
enhance electron-ion recombination processes on the surface,
which in turn promotes nanoparticle crystallization. 
However, the mechanism of the plasma-enabled nanodot formation 
is presently unclear and requires a dedicated study of the different 
mechanisms of interaction of atmospheric-pressure plasmas with 
liquid media of different density and lipid content. 
Despite significant recent progress \cite{Bruggeman}, studies of plasma-liquid interactions 
are presently at the early stage.                   

The significant difference in the structural properties has led to the stronger 
photoluminescence emission from the egg yolk-derived C-dots, as can be seen in 
Fig.~\ref{f-C-dot}(c). The maxima of the absorbance and PL emission show a 
very clear spectral separation allowing for the blue emission to be 
detected easily, as can be seen in the inset. Moreover, the time-dependent fluorescence 
\begin{equation}
\label{fluoresc}
\Upsilon (t) = \alpha_1 \exp(-t/\tau_1) + \alpha_2 \exp(-t/\tau_2)
\end{equation}
showed contributions from two time-resolved decay lifetimes $\tau_1$ and $\tau_2$
with fractional contributions $\alpha_1$ and $\alpha_2$, respectively. 
By fitting the experimental spectra with (\ref{fluoresc}) provided the estimate 
of the average lifetime \cite{C-dot-plasma}
\begin{equation}
\label{lifetime}
\bar{\tau}  = [\alpha_1 \tau_1^2 + \alpha_2 \tau_2^2]/[\alpha_1 \tau_1 + \alpha_2 \tau_2] = 6.37 \pm 0.05 {\rm ns},
\end{equation}
which is comparable with the lifetimes of C-dots produced by other 
(e.g., wet chemistry, laser ablation, thermal, etc.) routes \cite{C-dot-review,C-dot-3}. 
The competitive quantum yield ($\sim$ 8\% and $\sim$ 6\% for the egg yolk and white-produced
C-dots, respectively) and the improved (by $\sim$10 \%) photostability suggests that the 
atmospheric-pressure plasma-based processes are indeed viable alternatives to the 
many existing wet chemistry-based approached to the synthesis of luminescent C-dots.  

This very interesting development opens an opportunity to develop 
cheap (e.g., vacuum-free, atmospheric-pressure) and environment-friendly plasma processes 
based on natural and renewable precursors. In this example, 
the egg yolk and white served mainly as carbon precursor,
while the non-equilibrium plasmas (equation \ref{thermnoneqpl} 
also holds for atmospheric-pressure 
DBDs) were used to effectively reform the precursor material into 
relevant hydrocarbon species. One can thus expect that 
a wider range of atmospheric plasma- and natural precursor-based 
processes will emerge for the synthesis and processing of 
a range of nanoscale solid materials using precursors with 
high natural abundance. This area is very new and offers 
numerous opportunities for the studies of the plasma-specific mechanisms
that lead to precursor conversion into solid nanostructures.

\subsubsection{\label{Nanocrystals} Nanocrystals and nanoparticles}

Nanoparticle production has been demonstrated from the plasmas in both gases and liquids.
The plasmas used also range from thermally non-equilibrium plasmas of low-pressure 
discharges to thermal plasmas at atmospheric pressure and more recently, to 
non-equilibrium atmospheric-pressure microplasmas in gases and liquids 
\cite{Kortshagen-review, Meyyappan2009, XingguoLi, Editorial_JPD2011, 
Mariotti-Sankaran, Mohan-Book-2012, Tony-Shigeta}.   
The materials composition is also very diverse ranging from more traditional 
carbon- and silicon-based nanoparticles, to oxides, nitrides, metal nanoparticles of 
binary, ternary, and quaternary elemental compositions. These nanoparticles also range 
in size thereby exhibiting very different electron confinement properties. 

In this section we will only focus on the selected types of 
crystalline nanoparticles produced by using three distinctively different types of plasmas, 
namely non-equilibrium low-pressure plasmas, gaseous microplasmas, and 
microplasma-assisted nanoparticle synthesis in liquid media.  
There is some ambiguity regarding the allocation of 
any particular nanoparticles to the specific dimensionality because 
only very small NPs qualify as zero-dimensional nanostructures. 
Therefore, any larger nanoparticles in this subsection are mentioned either 
from the earlier development perspective or have a significant potential 
for further size reduction.  

\paragraph{Nanocrystals from low-pressure plasmas}

Formation of semiconducting nanoparticles in the plasma bulk has been a subject of 
intense studies for more than two decades. 
Recently, the paradigm has shifted from the management of (shapeless, amorphous, agglomerate, 
etc.) particulate contamination \cite{PhysRepts2004} to intentional, quasi-deterministic synthesis of high-quality 
semiconducting nanocrystals with specific shapes \cite{Kortshagen-review}. 
The mechanisms of the formation of amorphous Si clusters 
in reactive SiH$_4$-based plasmas are relatively well 
understood \cite{Si-clustering, Kaitlin2004, VieraJAP02}. 

\begin{figure}[t]
\begin{center}
\includegraphics[width=12cm,clip]{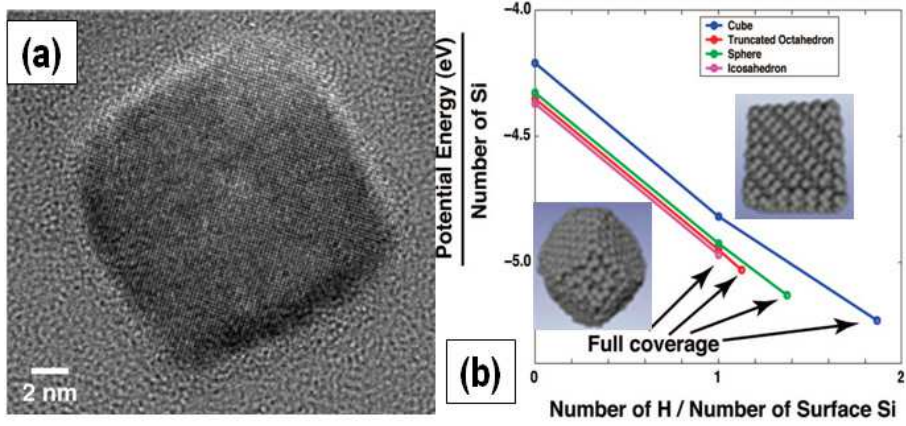} 
\caption{ \label{f15} HRTEM image of a plasma-produced Si nanocrystal of a cubic shape 
(a) [Copyright \copyright IOP Publishing. Reproduced from \cite{KortshagenJPD2011} by permission 
of IOP Publishing. All rights reserved]; surface coverage by hydrogen is a 
key factor in thermokinetic shape selection (b) [Reprinted/adapted with permission from 
\cite{Si-hydrogenation}. Copyright \copyright (2008) American Chemical Society]. } 
\end{center}  
\end{figure}

However, it only recently became possible to use plasma-grown 
Si nanocrystals in photovoltaic and nanoelectronic 
devices \cite{RocaJPD07,KortshagenJPD2011,KortshagenNL2012}. 
In addition to the effective and high-rate production of 
{\it nc}-Si films (Sec.~\ref{QD}), well-shaped and passivated Si and Ge 
nanocrystals can be produced in the plasma bulk and 
then deposited on the surface without any significant damage. 
A Si nanocrystal with a clear-cut cubic shape is 
shown in Fig.~\ref{f15}(a).

This is why the synthesis of faceted nanocrystals in a plasma 
has attracted strong interest. This process shows several interesting 
features discussed in Sec.~\ref{basicC} because it 
is conducted under far-from-equilibrium conditions when the developing 
nano-solid phase effectively exchanges energy with the plasma environment. 
At equilibrium, the fluxes of the energy received $J_{\cal E}^+$ and lost $J_{\cal E}^-$ 
by the developing nanoparticle are balanced and determine its surface temperature \cite{KerstenJPD2011}. 
The integrated incoming energy flux 
\begin{equation}
\label{enfluxin}
J_{\cal E}^+ = J_{\cal E}^e + J_{\cal E}^i + J_{\cal E}^{\rm rec} + J_{\cal E}^{\rm ass} 
\end{equation}
includes contributions from electron $J_{\cal E}^e$ and ion $J_{\cal E}^i$ fluxes, as 
well as the energy influx densities due to recombination of charged species $J_{\cal E}^{\rm rec}$ and 
dissociated molecules $J_{\cal E}^{\rm ass}$. 
The energy flux released by the particle is  
\begin{equation}
\label{enfluxout}
J_{\cal E}^- = J_{\cal E}^{\rm rad} + J_{\cal E}^{\rm cond}, 
\end{equation}
where $J_{\cal E}^{\rm rad}$ is the energy loss due to radiation and $J_{\cal E}^{\rm cond}$ is due 
to heat conduction through the interactions with neutral gas species.  
The electron and ion fluxes in (\ref{enfluxin}) are determined by the floating potential on the 
nanoparticle surface $V_{\rm float}$ which is retarding to the electrons and accelerating 
to the ions. The ion- and neutral recombination terms entering (\ref{enfluxin}) will be 
used in Sec.~\ref{1D-oxide} in the discussion of plasma-enabled inorganic nanowire 
synthesis, while all other terms can be found in the original publication \cite{KerstenJPD2011}. 

As a result, nanoscale plasma-surface interactions (which are similar to Sec.~\ref{nano-plasma-surface})
lead to highly-non-equilibrium nanoparticle heating, with their surface temperatures being
up to several hundred degrees higher than the neutral gas temperature \cite{KortshagenJPD2011}.
These particles can also reside in the plasma long enough (in part due to electric charging and 
levitation) to change their shape. This shape change happens faster when the surface 
temperature is higher, similar to Fig.~\ref{f4}. 
The shape formation is controlled by the dynamic changes in 
the surface energy through the effective surface hydrogenation 
in a plasma \cite{Si-hydrogenation}. 

The shape of the nanoparticles is determined by minimization of the temperature-dependent 
Gibbs free energy \cite{Barnard2006}
\begin{equation}
\label{GibbsEn-Barnard}
G_{\rm NP}(T) = G_{\rm NP}^{\rm bulk}(T) + G_{\rm NP}^{\rm surf}(T) + G_{\rm NP}^{\rm edge}(T) + 
G_{\rm NP}^{\rm corner}(T) + G_{\rm NP}^{\rm def}(T), 
\end{equation}
which is a sum of contributions from the NP bulk $G_{\rm NP}^{\rm bulk}(T)$, surface $G_{\rm NP}^{\rm surf}(T)$,
edges $G_{\rm NP}^{\rm edge}(T)$, corners $G_{\rm NP}^{\rm corner}(T)$, as well as planar or point defects 
$G_{\rm NP}^{\rm def}(T)$. As shown in Fig.~\ref{f15}(b), the final shape of a Si nanocrystal is determined by the 
extent of surface hydrogenation, which strongly affects $G_{\rm NP}^{\rm surf}(T)$. 
The cubic shape requires the largest surface passivation  
of almost 2 hydrogen atoms per Si atom on the surface. These results obtained 
through the study of the process kinetics by Molecular Dynamics (MD) simulations are also 
consistent with thermodynamic arguments suggesting that the equilibrium shape of a surface-hydrogenated 
Si nanocrystal is cubic while spherical shapes are ubiquitous if the Si surface is bare
\cite{Barnard-shapes1}. 

This surface passivation is easier to implement in a plasma where the 
rates of production of H atoms are typically much higher than in neutral gases,
because of very high dissociation rates (e.g., Fig.~\ref{f8}(a)), which can be achieved by adjusting the 
EEDF (\ref{EEDF1}) to maximize the number of electrons with the energy above the dissociation 
threshold of a hydrogen molecule of $\sim$4.53 eV.  

Structural stabilization is the next step in the process kinetics 
(\ref{ineq-times-kinetics}), where the plasma non-equilibrium 
also plays a critical role.  
Effective heat exchange between the nanoparticles and background gas lead to the 
heat extraction and the nanoparticles ``freeze" in the as-formed shape, 
see Sec.~\ref{basicC}. Moreover, fairly large amounts of plasma-dissociated 
atomic hydrogen help producing another plasma-specific effect of fast recrystallization of 
the developing crystalline Si network assisted by the rapid and effective 
incorporation of reactive hydrogen atoms \cite{Aydil-Nature2002}.    

This is a clear example of the very unique non-equilibrium process 
conditions enabled by the plasma environment, which in turn lead to 
the formation of exotic yet very useful (see the discussion about PL emission 
and surface passivation in Sec.~\ref{QD}) nanocrystals that do 
not normally form in neutral gases under thermal equilibrium conditions. 
It is possible that quantum dots (Sec.~\ref{QD}) also follow a
similar scenario. However, because of the very small size, 
it is presently very difficult to control their shapes 
and size distributions.  

These advances made it possible to fabricate light emitting devices 
by using small Si nanocrystals, which are produced, passivated, and then deposited 
onto glass substrates pre-coated with a transparent conducting oxide 
layer \cite{KortshagenNL2012}. Importantly, all the stages of this process are 
conducted in the plasma gas phase, while the parameters of the
resulting nanocrystal film can be effectively tuned to control the
emission characteristics. For example, films containing Si nanocrystals 
with the average diameters of 5.1, 4.1, and 3.1 nm produce 
photoluminescence peak wavelengths of 900, 817, 
and 707 nm, respectively \cite{KortshagenNL2012}. 
Importantly, bulk Si which features an indirect energy bandgap, 
is not a luminescent material. 
Intense research efforts by several research groups are underway to 
explain the mechanisms of photoluminescence from plasma-produced 
Si quantum dots and one should expect many interesting results 
in the near future.     

\paragraph{\label{microplasma} Nanocrystals from microplasmas}

Microplasma environments offer a very unique environment for the 
effective synthesis and functionalization of a large number 
of nanoparticles and nanostructures \cite{Mariotti-Sankaran}.
As mentioned in Sec.~\ref{nano-plasma-surface}, microscopic plasma localization is beneficial for the 
selective treatment (e.g., patterning or functionalization) or 
nanostructure growth within microscopic spots on the surface.  
The plasma size can be reduced by producing a discharge within thin channels or reducing the
inter-electrode gap. This leads to many interesting physical phenomena which 
make microplasmas very different from their bulk counterparts.

Most prominently, size reduction to micrometer dimensions makes it possible to maintain thermal non-equilibrium 
for continuously-operated (i.e., not pulsed or transient) plasmas even at atmospheric pressure. 
When the discharge size decreases, the relative size of the charge-non-neutral areas (e.g., sheaths) 
increases, which leads to stronger electric fields between the electrodes. Consequently, the electron energy 
distribution departs from Maxwellian, with the maximum shifting towards higher energies. The electron density 
also increases, most likely because of the higher rates $\nu_{en}$ of the electron-neutral collisions. 
Moreover, more frequent collisions of neutrals with the electrodes or reactor walls 
lead to dissipation of a substantial fraction of the energy they receive through collisions with other species. 
As a result, gas temperature $T_g$ may become much lower than $T_e$ and the
plasma is thus thermally non-equilibrium.

\begin{figure}[t]
\begin{center}
\includegraphics[width=12cm,clip]{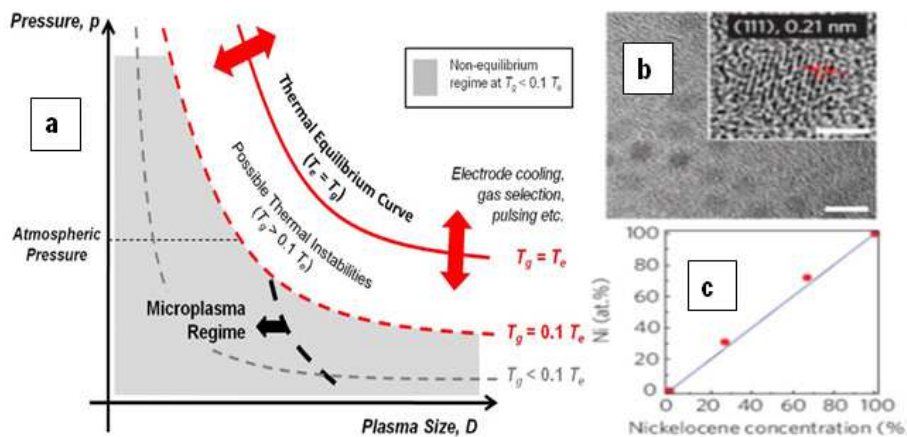} 
\caption{ \label{f31} Isotherm boundaries of plasma  
non-equilibrium and conditions for microplasma regime (a) 
[Copyright \copyright IOP Publishing. Reproduced from \cite{Mariotti-Sankaran} by permission 
of IOP Publishing. All rights reserved].  
TEM of  Ni$_{0.27}$Fe$_{0.73}$ (b) CNPs for SWCNT growth. Linear dependence of Ni 
content in CNPs and concentration of nickelocene precursor (c).  
Panels (b) and (c) are reprinted from \cite{SankaranNatureMater2010} 
by permission from Macmillan Publishers Ltd., Copyright \copyright (2009).}  
\end{center} 
\end{figure}

Conditions of the energy and species balance lead to the following basic scaling law 
\begin{equation}
\label{micropl-law}
T_g \propto (p_0 D^2 \varepsilon n_e \nu_{\rm nheat})^{2/3}
\end{equation}
which relates the gas pressure $p_0$, plasma size $D$, electron density $n_e$, 
average energy $\varepsilon$, and the effective rate of collisions leading to the heating of neutrals 
$\nu_{\rm nheat}$. Equation (\ref{micropl-law}) shows that the gas temperature can be decreased 
quite effectively ($T_g \propto D^{4/3}$) by reducing the plasma size. This effect is quantified 
in Fig.~\ref{f31}(a) which shows the thermal equilibrium ($T_g = T_e$) and non-equilibrium 
($T_g = 0.1 T_e$) isotherms \cite{Mariotti-Sankaran}. The gray area represents the thermal 
non-equilibrium regime ($T_e \gg T_g$). More importantly, this regime can be achieved 
at atmospheric pressure (a horizontal
short dashed line) when the plasma size is reduced below a certain threshold.  
The area of non-equilibrium microplasmas is also indicated by an arrow pointing to the left.

Under such conditions, the plasma is very reactive while remaining cold.
This is favorable for the effective generation of reactive species at low temperatures,
which is required in most nanoscale processes.  

A striking example of this effect is shown in Fig.~\ref{f31}(b,c) where microplasmas 
of gas mixtures of argon with metalorganic precursors nickelocene and ferrocene 
were used to synthesize small Ni$_x$Fe$_{1-x}$ catalyst nanoparticles 
for the growth of SWCNTs \cite{SankaranNatureMater2010}. 
The particle diameter and the lattice spacing were fine-tuned by varying the relative 
concentration of Ni and Fe in the alloy. This in turn leads to the ability to control 
and in particular, significantly narrow down the distributions of chiralities of 
the SWCNTs grown on these catalysts.    

More importantly, the content of Ni in the catalyst nanoparticles increases linearly with the nickelocene flow, 
see Fig.~\ref{f31}(c). This is a clear indication of the very effective 
precursor dissociation in non-equilibrium microplasmas. Moreover, this linear relationship 
makes it possible to predict the relative content of Ni and Fe in the catalyst alloy 
even before the  measurements of elemental composition.  

These non-equilibrium features of gaseous microplasmas have led to a variety 
of self-organized nanopatterns, with representative examples shown 
in Fig.~\ref{f13}(a,b,d). Strongly non-uniform electric fields, densities 
of BUs, surface stresses, and temperature distributions on microplasma-exposed surfaces 
lead to a variety of interesting self-organization phenomena under far-from-equilibrium 
conditions. These phenomena in turn  not only lead to exotic nanopatterns \cite{MariottiAPL07},
but also to the many high-quality hierarchical nanoarchitectures of particular interest to several 
applications including energy storage, optoelectronics, and sensing 
\cite{MariottiJPD2009, Nozaki-Si-nc-microplasma, Nozaki-SWCNT-microplasma}.    

\paragraph{\label{nc-from-liquid} Nanocrystals from plasmas in liquids} 

\begin{figure}[t]
\begin{center}
\includegraphics[width=12cm,clip]{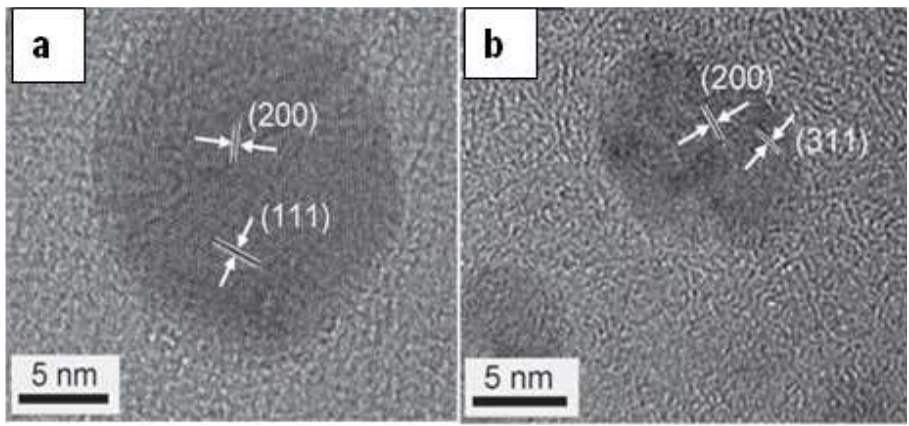} 
\caption{ \label{f32} TEM of Ag (a) and Au (b) NPs 
produced by microplasma reduction of AgNO$_3$ and HAuCl$_4$  
 precursors, respectively, in thin PVA films. 
 Reproduced with permission from \cite{MohanAdvFunctMater2011}. 
Copyright \copyright 2011 Wiley-VCH Verlag GmbH \& Co. KGaA, Weinheim.
}  
\end{center} 
\end{figure}

A variety of nanocrystals have also been synthesized and functionalized using 
non-equilibrium plasmas in liquids. 
Strong thermal and chemical non-equilibrium together with the 
very high densities makes such plasmas suitable for the liquid-phase production of a variety 
of nanoparticles and hybrid nanostructures 
\cite{Meyya-liquid-plasma,GrahamJPD2011}.

One can thus boost electrochemical reactions in the liquid precursor phase,
e.g., in AgNO$_3$ and HAuCl$_4$ solutions to produce high-quality crystalline Ag and Au 
nanoparticles, respectively \cite{MohanAdvFunctMater2011}. The liquid precursor solution 
can be used as an electrode of the discharge while a gas-phase 
plasma jet as another electrode \cite{Janek-Electrochemistry}. 
In this case the nanoparticles are produced by the direct reduction 
of the precursor in the plasma. The nanoparticle sizes 
can also be monitored {\em in situ} by analyzing their
size-dependent plasmonic responses \cite{XX-AgNP2013}. 
However, after synthesis, the nanoparticles need to be collected and then 
transferred to be used in device fabrication. 

Using DC microplasmas in contact with water or ethanol, 
it is possible to tailor the sizes, structure, and surface properties of 
Si nanocrystals \cite{MariottiAFM2011}. In this way, the Si nanocrystal 
surfaces can be passivated with oxygen- or organic-based 
functional groups. The microplasma treatment in ethanol 
drastically enhances the photoluminescence intensity and also leads to 
the very strong shift of the PL maximum towards lower photon 
energies. Importantly, the reported changes are microplasma-enabled 
and are not possible to reproduce using conventional 
electrochemistry. Moreover, as the crystal sizes are typically 
below 5 nm, and the photoluminescence exhibits a clear size-dependence,
these nanocrystals may also be reasonably considered as 
zero-dimensional Si quantum dots. This simple microplasma-based process
may therefore be considered as a potential viable alternative 
to the high-precision Si QD synthesis in low-pressure plasmas 
discussed above.  

Surface-supported nanocrystls can also be 
produced by direct plasma exposure of 
aqueous solutions of metal salts and polymers spin coated onto a substrate 
\cite{MohanAdvFunctMater2011}. An atmospheric-pressure, room-temperature microplasma 
jet in an Ar flow was used for electrochemical reduction 
of metal cations to crystalline Ag and Au nanoparticles 
shown in Fig.~\ref{f32}. The achievable spatial localization of the microplasma 
was $\sim$30 $\mu$m, which is particularly interesting for the development of new 
micropatterned devices.  

This interesting result owes to the strong non-equilibrium, 
high reactivity, and micrometer spatial localization of microplasmas. The 
plasma is cold and is suitable for polymer processing without causing any significant damage. 
The produced Au and Ag nanoparticle arrays in turn enable 
a variety of size-dependent plasmonic effects upon excitation with a suitable light source.

\subsubsection{\label{Nanoarrays} Nanoarrays}

Regular arrays of small nanodots are of a significant interest for many applications ranging from 
bio-sensing to optoelectronics and photonics. A certain degree of uniformity 
of the feature sizes and shapes as well as the pattern regularity over reasonably large surface areas
are required in most cases. This can be achieved using either the bottom-up or top-down 
approaches. 

In the top-down approach, pattern transfer using pre-formed masks 
is usually required. These masks may be prepared by direct feature writing, e.g., using focused ion beams 
or lasers. However, it is very difficult to achieve very small feature sizes to meet the demands of the 
present-day nanofabrication. 

Recently, various nanoporous structures proved very effective as 
etching masks. These structures are typically formed through the customized growth leading to the 
nanopore or nanochannel formation. In some cases, such as in self-organized porous alumina templates,  
the dimensions and mutual position of the pores can be tailored. This in turn leads to 
the possibility of customized transfer of regular patterns to the surface 
using plasma-assisted reactive ion etching (RIE). 

The pores may also be used for the formation of regular arrays 
of nanoparticles or nanorods on the surface by evaporating or sputtering solid material through the 
porous template. This approach is a combination of top-down pattern transfer and bottom-up growth 
and is very effective to produce regular arrays of catalyst nanoparticles for CNT growth 
similar to Fig.~\ref{f2}(a). 

The delivery of building material for the nanoarray 
formation commonly benefits from plasma-assisted techniques, such as ionized physical vapor 
deposition (i-PVD) \cite{AndersJPD07}, pulsed laser deposition \cite{Rosei-PLD-APL06}, 
and several others. The obvious benefit of using plasmas is in the generation of ion species
whose energies and fluxes can be optimized to pass through the nano-channels and reach the substrate. 
Reactive ions, e.g., CF$_x^+$ can be used to etch the pattern through the mask opening. On the other hand,
metal ions may condense at the substrate surface to produce regular arrays of metal nanodots or nanorods. 
In this way, using plasmas may extend the lifetime of such templates and also 
substantially improve the resulting pattern quality because of the very high rates of nanopore 
clogging when evaporation or thermal CVD are used. Plasma discharges are also effective for 
the conditioning of the nanoporous templates after the 
deposition; this can be easily arranged in the same reactor without disrupting the vacuum cycle.  

\begin{figure}[t]
\begin{center}
\includegraphics[width=9.5cm,height=11.5cm,clip]{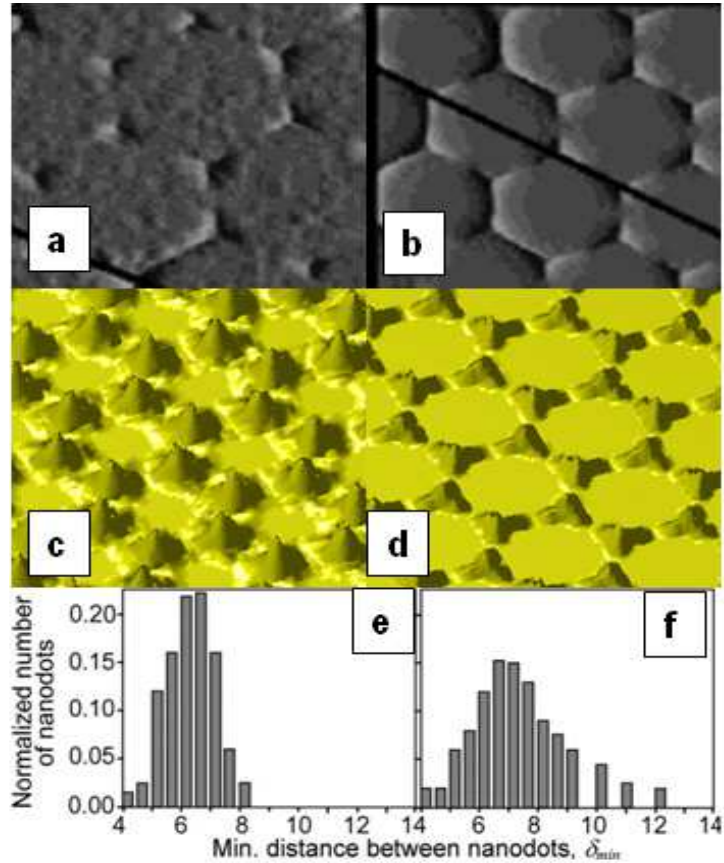} 
\caption{ \label{f16} Regular nanopore patterns on Si(111) using 
self-assembled nanosphere monolayer as a mask  for plasma etching
(a,b) 
[Reprinted with permission from \cite{nanosph-lithogr-NL2004}. 
Copyright \copyright (2004) American Chemical Society]; larger nanospheres produce pattern (a). 
Numerical simulations reveal correlation between the nanopatterns 
and micro-topography of ion fluxes (c,d) 
[Copyright \copyright IOP Publishing. Reproduced from \cite{XX-nanosphere-lithography} by permission 
of IOP Publishing. All rights reserved].   
Direct plasma exposure improves ordering in self-organized  
Ni nanodot arrays on Si(100) (e,f) 
[Reprinted with permission from \cite{our-APL-2008-Davide}. Copyright \copyright (2008), American Institute of Physics].
 } 
\end{center}  
\end{figure}

Another option for nanoscale pattern transfer is based on plasma-assisted RIE
through an etching mask made by a self-assembled monolayer of nanospheres (Fig.~\ref{f16}(a-d)). 
The plasma ions and other reactive species (in this example, mostly CF$_x^+$ and CF$_x$) 
pass through the pores between the closely packed nanospheres in a monolayer and interact 
with the underlying Si surfaces to form regular hexagonal patterns of small 
nanopores shown in Fig.~\ref{f16}(a,b) \cite{nanosph-lithogr-NL2004}. 
Numerical simulations of ion deposition  
\cite{XX-nanosphere-lithography} show a clear correlation 
between the positions of the peaks of the ion flux densities upon 
impact on the surface (Fig.~\ref{f16}(c,d)) and the triangular dips and 
linear trenches etched on the Si surface (panels (a,b)). 

Hence, the plasma-produced ion species play a major role in the 
nanoscale etching. This process can be conducted at room 
temperature and is suited for nanostructuring 
of delicate and temperature-sensitive materials.  
The ion fluxes and energies can be  controlled (e.g., to 
control the energy released through the ion collisions with the surface) 
using substrate bias.
To mitigate any adverse effects of surface charging, the bias 
can be tailored as a sequence of pulses; the pulse duration can 
be adjusted to drive the ions through the pore channels before any
significant charge build-up and let the charge dissipate between the ion pulses.  
In this way, the process throughput and precision can be improved 
\cite{XX-template-APL08}.

The second, bottom-up approach leads to the formation of reasonably 
ordered nanodot arrays through self-organization 
in randomly nucleated nanopatterns. Indeed, 
self-organization in non-equilibrium plasmas produced better size and positional uniformity of Ni nanodot
arrays on a Si(100) surface compared with the 
similar neutral gas-based process \cite{our-APL-2008-Davide}. 
The results of numerical simulations of the nanopattern development under 
conditions of neutral and ionized metal vapor deposition
are shown in Figs.~\ref{f16}(e) and (f), respectively. In case (f), the Si 
surface is in direct contact with the plasma, while the surface in panel (e) 
is uncharged. The Ni nanodot arrays on the plasma-exposed surface
are more uniform in size and in terms of the local order 
measure (LOM)
\begin{equation}
\label{locordermeas}
{\cal M} = \frac{\sigma_{\rm NND}}{\bar{\chi}_{\rm NND}},   
\end{equation}
where $\bar{\chi}_{\rm NND}$ and $\sigma_{\rm NND}$ are the mean 
nearest neighbor distance (NND), and the standard deviation of the NND. 
The LOM in (\ref{locordermeas}) can be used as a measure of positional 
 uniformity of self-organized nanoarrays.  

The observation in Figs.~\ref{f16}(e) and (f) was explained 
by introducing 2D patterns of Ni adatom capture zones
(see Sec.~\ref{basicC}). 
These areas become larger relative to
the small nanodot sizes  when the surface is charged. 
In this case, smaller dots  grow faster than the 
larger ones, eventually leading to the fairly size- and position-uniform 
arrays. These results suggest that plasma non-equilibrium and complexity  
lead to better ordering and size
uniformity in self-organized arrays of randomly 
nucleated metallic nanodots. This phenomenon is generic and applies to other  
materials systems. 

Controlled ordering and inter-particle spacing is crucial 
for the envisaged functional properties of the self-organized nanoarrays.
For example, it is imperative to estimate length scales over which 
a quantum dot interacts with its neighbors. An order of magnitude 
estimate can be obtained by analyzing the Hamiltonian \cite{Delanty}
\begin{equation}
\label{Hamilt1}
{\cal H} = \frac{{\bf p}^2}{2m_e^*} - V_0 [e^{-(r_1^2/2R^2)} + e^{-(r_2^2/2R^2)}]
\end{equation}
of a model system of an electron in 
two Gaussian potential wells, which approximates a system 
of two neighboring quantum dots. 
Here, $m_e^*$ and ${\bf p}$ are the effective mass and momentum of the electron,  
$V_0$ is the depth of the potential well, and $R$ is the radius of the both quantum dots. 
Furthermore, $r_1 = \sqrt{x^2 +y^2}$ and $r_2 = \sqrt{(x-x_0)^2 +y^2}$, where
$x_0$ is the spacing between the two Gaussian wells (quantum dots). 

The ground states of the model Hamiltonian in (\ref{Hamilt1}) show 
several important features that are applicable to most two-dimensional
arrays of quantum dots \cite{Delanty}. When the distance between the quantum dots is small, 
electrons can tunnel between the two dots, resulting in a lower
ground state energy. While the distance between the two
quantum dots increases, the system's ground state energy asymptotically
approaches the energy of a the ground state of a single quantum dot. The energy 
also increases when the QD radius becomes smaller, which is the consequence of
a stronger electron confinement.

Therefore, the electronic and optical responses in nanoarrays with 
the mean inter-dot distances much larger than the estimated 
electron tunneling length, will be mostly determined by the 
combined properties of non-interacting individual quantum dots, their 
sizes and positioning within the array. In the opposite case,
the electron tunneling should be carefully taken into account. 
 
Relating the plasma-process-specific morphological parameters of 
nanopatterns and arrays to their 
electronic, phononic etc. properties and consequently, functional performance,
is one of the key goals of the plasma nanoscience research. 
Due to the overwhelming complexity, only phenomenological or process-specific 
links presently exist. An interesting methodological approach 
to link the process parameters, morphological, and electronic
properties of silicon-germanium (Si$_{1-x}$Ge$_x$, $0 \leq x \leq 1.0$) 
nanostructures produced using plasmas has recently been proposed \cite{ValeriJPD2011}. 
In particular, the effect of the QD-morphology-controlled density of the 
electron trap states on the photoluminescence intensity 
can be approximated as \cite{ValeriJPD2011} 
\begin{equation}
\label{IPL-Valeri}
I_{\rm PL} = I_0 \exp (-V_c N_{\rm nr}), 
\end{equation}
where $I_{\rm PL}$ is the actual intensity of the PL emission line 
affected by quantum confinement and non-radiative recombination via trap states
in the quantum dot volume, $I_0$ is due to 
the radiative recombination of the electron–-hole pairs, 
and $N_{\rm nr}$ is the
volume density of non-radiative recombination centres (traps).

The effect of the surface morphology comes along through  the
contribution of $V_c$, which 
is the volume of confinement of the excited carriers. 
In other words, $V_c$ is the quantum dot volume, which can be roughly estimated
as $\sim h_{\rm rms}^3$, where $\sim h_{\rm rms}$ is the root-mean-square 
roughness of the surface, which can be obtained, e.g., through
the Atomic Force Microscopy (AFM) measurements. 
Therefore, the plasma control of surface morphology may be used 
to tailor the electronic and optoelectronic (e.g., luminescent) responses of 
self-organized nanoarrays.   

Dense 3D arrays of Si QDs may enable the as yet elusive 
multiple exciton generation in the photovoltaic photon energy conversion   
\cite{Shiratani-QD-solar-cells, NozikNL2010}. However, the 
degree of spatial uniformity of such arrays still requires  
improvement. The QD arrays can also form on plasma-exposed surfaces, e.g., Ge quantum dots produced using 
germane-based precursors \cite{Shieh-Ge-QD-plasma}. An interesting hybrid 
structure made of SiC nanoislands with dense nanocrystalline inclusions,
was also reported \cite{ChengAPL2007}.

\subsection{\label{1D-nanostructures} One-dimensional nanostructures}   

Plasmas are particularly suitable for the synthesis of 
many one-dimensional (1D) nanostructures, which commonly leads to the pronounced vertical alignment, substantially 
reduced growth temperatures, higher growth rates, and in many cases also superior structural 
properties and performance in applications compared to the equivalent thermal processes. 
Given the very large number of relevant reports, we will only use a few representative examples to show 
a few salient features and comment on the most important physics involved. 
This subsection considers 1D carbon nanostructures (Sec.~\ref{1D-Carbon}), silicon 
(Sec.~\ref{1D-Si}) and other inorganic (Sec.~\ref{1D-oxide}) nanowires.

\subsubsection{\label{1D-Carbon} 1D carbon nanostructures}

\begin{figure}[t]
\begin{center}
\includegraphics[width=9.5cm,height=12cm,clip]{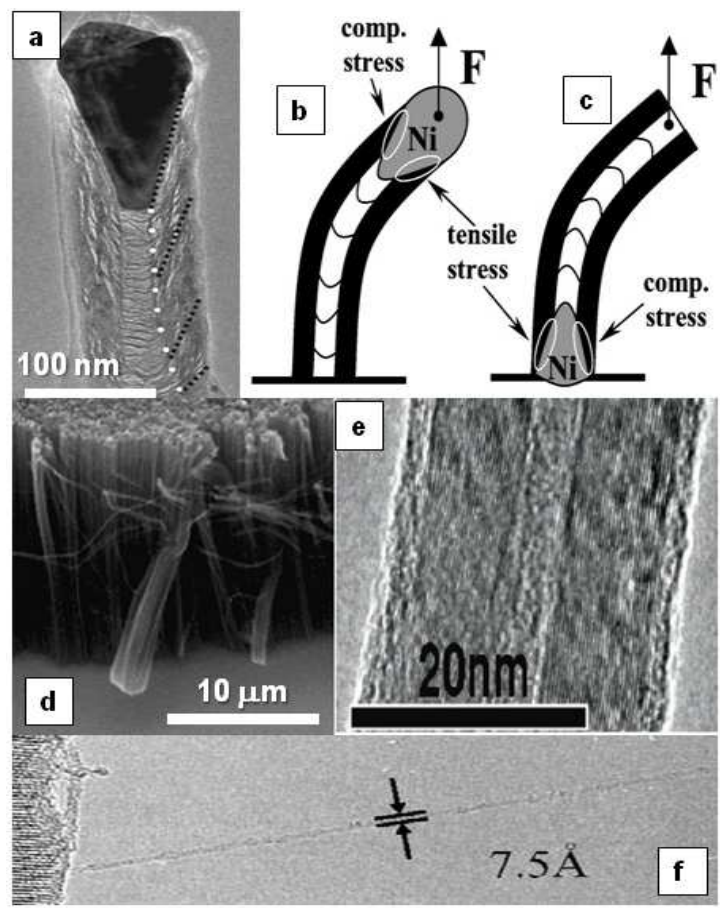} 
\caption{ \label{f17} HRTEM image of a CNF produced in a fast (5 min) plasma process shows  
cone-like catalyst and bamboo-like walls (a) 
[Reprinted with permission from \cite{CuiAPL2004}. Copyright \copyright (2004), American Institute of Physics]. 
Mechanism of vertical alignment 
of CNFs in a plasma (b,c) 
[Reprinted with permission from \cite{MerkulovAPL2001}. Copyright \copyright (2001), American Institute of Physics]. 
SEM and HRTEM images  
of MWCNTs grown by catalyzed PECVD (d,e) 
[Copyright \copyright IOP Publishing. Reproduced from \cite{Meyyappan2009} by permission 
of IOP Publishing. All rights reserved]. 
A TEM image of a SWCNT grown by PECVD on zeolite support (f) 
[Reprinted from \cite{HatakeyamaCPL2003}, 
Copyright \copyright (2003), with permission from Elsevier]. } 
\end{center}  
\end{figure}

Figure~\ref{f17} shows typical examples of 
1D carbon nanostructures produced using low-temperature plasmas, namely carbon 
nanofibers (CNFs) (a-c), multiwalled carbon nanotubes (MWCNTs) (d,e), and 
a thin ($\sim$0.75 nm) SWCNT (f).
Many plasma-specific effects related to SWCNT nucleation 
and growth have been discussed in previous 
sections. This is why the main focus here is on 
CNFs and MWCNTs. 

In a plasma, the carbon 
nanofibers and  multiwalled carbon nanotubes more commonly grow in a tip-led mode \cite{RobertsonMaterToday}. 
In the early growth stage, the initially semi-spherical metal catalyst nanoparticles reshape and detach 
from the surface as the nanotube/nanofiber walls nucleate; this is evidenced 
by {\it in situ}, real-time TEM \cite{Helveg_Nature2004}. 
The formation of MWCNTs proceeds through the formation of  
step terraces on the catalyst surface. This surface structure is metastable and has a relatively high 
surface energy. Nucleation of carbon chains on these terraces minimizes the surface energy and 
eventually leads to the formation of straight nanotube walls, which support and lift 
the catalyst nanoparticle as the growth proceeds. The number of the terraces presumably determines the 
number of the walls. The CNP elongation and
detachment from the substrate is facilitated by surface charges and is 
one of the reasons for the preferential growth 
of MWCNTs and CNFs in the tip-led mode in PECVD \cite{MeyyappanPSST03}.   

Therefore, the process kinetics and non-equilibrium, as well as 
the metastable states play the key role in the selectivity of the nanotube  
growth mode. Moreover, fast carbon material extrusion through the catalyst 
(see Sec.~\ref{nucleation}) give a stronger upward push to the 
CNP to detach from the surface and determines its reshaping. 

The location of the catalyst nanoparticle on top of the nanotubes or nanofibers 
largely determines the nature of the nanoscale plasma-surface interactions.
Although these interactions mostly follow the basic trends discussed in 
Sec.~\ref{nano-plasma-surface}, there are a few distinctive features 
of these interactions. 

The most notable feature is the effect of direct heating or even possible 
cooling of the catalyst nanoparticle through its energy exchange with the ions and neutral species in the
plasma. The ion focusing by the sharp tips of the CNTs/CNFs and recombination of the plasma species 
lead to the heating of the CNPs on their top, quite similarly to the heating effects of 
Si nanocrystals considered in Sec.~\ref{Nanocrystals}. 
Since the length of the plasma-grown nanotubes or nanofibers may be a few tens or even hundreds 
of micrometers, the difference between the temperatures of the catalyst nanoparticle $T_{\rm CNP}$, the top surface of the
supporting substrate $T_{\rm sub}$, and the substrate holding platform $T_{\rm sh}$ 
can be quite significant. 
These temperatures are related through \cite{DenyJPD2009}
\begin{equation}
\label{tempscnfgrowth} 
T_{\rm sh} = - T_{\rm sub}^0 \ln \left[ \exp \left( -\frac{T_{\rm sub}(T_{\rm CNP})}{T_{\rm sub}^0} \right) -
f(d,T_{\rm CNP},\xi_{\rm Si},\xi_{\rm C}) \right],  
\end{equation}
where $T_{\rm sub}^0$ is the temperature of the external substrate heating in the absence of the plasma,
while $\xi_{\rm Si}$ and $\xi_{\rm C}$ are the heat conductance of the Si substrate and the
carbon nanostructure, respectively. 

One important conclusion from (\ref{tempscnfgrowth}) is that the difference 
\begin{equation}
\label{tempdiffcnf}
\Delta T = T_{\rm CNP} - T_{\rm sh} 
\end{equation}
between the temperatures of the catalyst and the substrate holder can in some cases 
be as high as $\sim$100$^\circ$C.
This result is consistent with numerous experimental observations \cite{MeyyappanPSST03,Meyyappan2009}
and is valid under conditions when the effect of ion bombardment is particularly important 
while the temperature of the operating gas $T_g$ is higher than $T_{\rm CNP}$. In this case, the CNP receives heat 
 through the impacts of both ions and the hot neutral species. 
 
However, in the opposite case when $T_{\rm CNP} > T_g$ and the effect of ion collisions with the catalyst nanoparticle is weak, the 
 temperature of the substrate holder may be even higher than $T_{\rm CNP}$. 
 Consequently, some energy supplied to the substrate heater would unnecessarily be wasted to heat the cold background gas
 while the CNP heating may be ineffective. This in turn may lead to the much reduced nanostructure growth rates and even possibly
 to the growth termination, e.g., due to the catalyst poisoning because of the insufficient carbon diffusion
 at reduced temperatures. 
 
This conclusion points at the crucial importance of controlling energy consumption in nanoscale synthesis
and in particular, reducing the unnecessary energy waste. Ultimately, only the catalyst nanoparticle on top of the 
CNF should be heated to sustain the nanostructure growth, and not the entire substrate, the substrate holder or the 
reactive gas in the entire growth chamber. Implementation of this effect may be a very significant advance 
towards energy- and atom-efficient nanotechnology of the future. 

Although it is presently not clear 
how to implement this idea in practice (e.g., in large-area nanotube growth), there are some indications on the 
principal possibility of this effect. For example, by heating the nanotubes developing 
in the tip-led growth mode externally by a laser beam, it is possible to drastically reduce the 
temperature of the external substrate heating to $\sim$350$^\circ$C \cite{SRPSilva} to make the
growth process compatible with the present-day integrated circuit technologies.
Further efforts in this direction are therefore warranted in the near future.

Carbon nanofibers commonly feature a bamboo-like structure shown in Fig.~\ref{f17}(a). 
In this typical example, the catalyst
nanoparticle has an elongated near-conical shape. 
The nanostructure walls nucleate as
stacked conical layers on the catalyst as can be seen in panels (a-c). 
Kinetics-wise, it means that the rates of carbon atoms delivery, hopping and rearrangement on 
the catalyst nanoparticle surface favor the formation of fairly large graphene 
monolayers that encage the lower, elongated section of the CNP. 
Several plasma-specific effects, such as 
ion-assisted production of carbon atoms, 
localized catalyst heating through ion impact, energy exchange between the 
CNF and the environment, presence of hydrogen 
in the precursor gas mixture, and several others, play a critical 
role \cite{DenyAPL07,DenyJPD2009}.

The mechanisms of vertical alignment during the 
nucleation stage of SWCNTs in a plasma were discussed in Sec.~\ref{nucleation}. 
For larger nanostructures such as CNFs or MWCNTs, additional factors such as 
stress generation and relaxation also play a major role as sketched 
in Fig.~\ref{f17}(b,c). When a 1D nanostructure grows along the direction of the electric 
field in the plasma, the electric field creates a uniform tensile stress across 
the interface between the catalyst and the nanofiber; this stress distribution becomes 
non-uniform if the nanofiber bends \cite{MerkulovAPL2001}. 
Interestingly, in panels (b) and (c) the stresses are distributed in the opposite way 
which leads to the restored CNF alignment only in the case of the tip-led growth, which 
is most common for plasma-based processes.   

The pronounced vertical alignment of the plasma-produced carbon nanotubes, nanofibers and some other 
one-dimensional nanostructures is important for many applications \cite{Intro2}.
One of the most common examples is a strong enhancement of the 
electron field emission (EFE) from the CNT/CNF arrays 
which is commonly described by the Fowler-Nordheim equation \cite{FN}
\begin{equation}
J_{\rm em} = \frac{\gamma_1 (\beta E)^2}{\Phi} \exp \left( - \frac{\gamma_2 \Phi^{3/2}}{\beta E} \right), 
\end{equation}
where $J_{\rm em}$ is the emission current density, $\Phi$ is the work function, and $\gamma_{1,2}$ are 
geometry-related constants. Here,  
$\beta$ is the field enhancement factor which can be very high (e.g, $\sim$10$^3$ or even higher)
for the carbon nanotubes and nanofibers. Moreover, additional  ordering of the CNT/CNF arrays 
also prevents undesired destructive EFE interference due to the 
nanostructure crowding \cite{ChhowallaJAP2001}. 

Approximately a decade ago, the electron field emission was considered as one of the key applications of these
high-aspect-ratio nanostructures. However, the present-day range of applications of these 
nanostructures is much broader and involves areas as diverse as energy, health, environment, 
and security \cite{ZFRen-AdvPhys2011,Intro2,MelechkoJAP2005}.

For many years, it was believed that metal nanoparticles are 
absolutely essential to catalyze the growth of 1D carbon nanostructures. 
However, the first reports on the synthesis of SWCNTs  
 used non-metallic catalysts, such as zeolite
microporous materials   (Fig.~\ref{f17}(f)) \cite{HatakeyamaCPL2003}. 
Oxides and carbides (e.g., SiO$_2$, SiC, etc.) also 
showed catalytic activity to produce CNTs. 

However, any catalyst ultimately needs to be avoided to enable integration 
of carbon nanostructures into the presently Si-dominant nanodevice platform and also 
to minimize the unnecessary yet inevitable Ohmic losses during the electron current 
conduction in nanoelectronic circuitry. Hence, dielectric catalysts such as oxides or 
nitrides are even more detrimental than conventional metallic CNPs. Likewise, the CNT growth rates 
are usually lower when non-metallic catalysts are used. 
Moreover, the use of toxic and expensive metals raises 
continuously escalating environmental, cost-, and energy-efficiency concerns. 

Recently, it was shown that a direct exposure of features of arbitrary shape (e.g., lines and dots) written 
on a Si surface to high-density plasmas of CH$_4$-based gas mixtures leads to the high-rate, catalyst-free growth 
of vertically-aligned MWCNTs \cite{scratch,scratch2013}. This is only possible when the plasma is 
in a direct contact with the substrate; no nanotubes are produced when thermal CVD 
or remote plasmas are used. This is an interesting manifestation of a plasma-enabled phenomenon. 
The mechanism of this effect involves nucleation of carbon 
walls on semi-molten small Si features accompanied with the formation of the segregated 
SiC nanophase, which in turn catalyzes the CNT nucleation. 

Moreover, it was possible to resolve the nanotube nucleation events on the 
small Si nanofeatures \cite{scratch2013}. This is an important advance because 
of the intrinsic common difficulty of the 
plasma-based nanostructure growth experiments which require pressures that are 
orders of magnitude higher than during the real-time  
HRTEM or LEEM observations of nucleation processes under ultra-high-vacuum conditions.           

Nevertheless, it is clear that both oxide and metal catalysts can be eliminated and the plasma exposure 
enables a catalyst-free integration of carbon nanostructures into Si-based nanodevices. 
However, a significant reduction of the 
process temperature is still required to meet the nanodevice temperature 
tolerance requirements. 
This is why such plasma-enabled, catalyst-free growth processes 
offer very interesting prospects for future studies.

\subsubsection{\label{1D-Si} Silicon nanowires}  

Inorganic nanowires are among the most advanced building blocks 
of nanoelectronic, optoelectronic, energy conversion and 
sensing devices \cite{Meyyappan-Sunkara-Book}. Silicon NWs are of particular interest
to Si-based device platforms, not only because of the obvious compatibility and epitaxy considerations,
but also because they offer several highly-unusual properties which are difficult to achieve 
using bulk Si or Si nanoparticles. To materialize these benefits, it is crucial to maximize 
the quantum confinement effects in the nanowires. 

Reducing the nanowire thickness to the low-nm range 
is an obvious way to achieve this. However, a typical thickness of the majority of the reported Si NWs is 
a couple of tens of nanometers. One-dimensional electron transfer is extremely appealing for electronics.
However, the NWs should not only be single-crystalline, but also have a regular crystallographic 
structure with the pre-determined orientation. This structure and orientation 
should be selected to enable the fastest possible 
electron and hole transport. 
This is critical for the next-generation ultra-fast switches and computer logic devices. 
Unfortunately, it is very challenging to achieve 
all these properties in the
same nanowire array to materialize their promises.

Indeed, common thermal catalytic growth based on a Vapor-Liquid-Solid (VLS) 
mechanism and a tip-led mode usually results 
in fairly thick (several tens of nm) Si nanowires. These NWs often show a cubic 
structure and grow in (111) direction. A major challenge is to grow 
Si nanowires with the more non-equilibrium wurtzite structure and also to achieve 
the preferential (110) growth at low, nanodevice-benign temperatures. 

Moreover, (110) NWs are much thinner than (111) NWs and 
their VLS nucleation barriers increase due to the Gibbs-Thomson (GT) effect. 
As we have discussed in Sec.~\ref{nucleation}, the GT effect leads to higher solubility of building units as the
catalyst size decreases. Smaller catalyst nanoparticles also produce higher stress levels, 
which leads to strong outward BU fluxes. Hence, it becomes more difficult to nucleate the Si layers 
at the interface between the CNP and the substrate. 
Higher process temperatures and pressures are thus needed, which is not only unwelcome from the nanodevice integration 
perspective but also because of the unnecessary waste of energy and matter. 

A similar waste happens when Si NWs are thinned by using surface oxidation 
followed by the removal of SiO$_2$ from the surface. This process is 
also very difficult to control and there is a real danger to oxidize and hence,
completely destroy the whole nanowire \cite{NeytsChemMat2012}.

\begin{figure}[t]
\begin{center}
\includegraphics[width=13cm,clip]{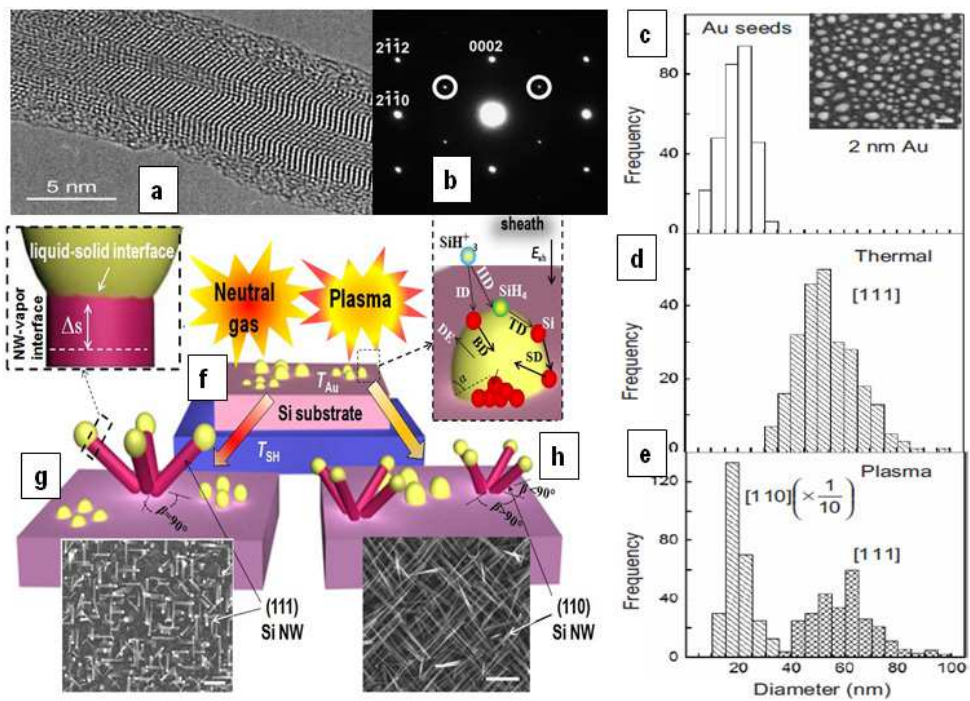} 
\caption{ \label{f18} HRTEM of a thin crystalline Si NW grown by PECVD (a) 
[Reprinted with permission from \cite{Hofmann-Si-NW-2003}. Copyright \copyright (2003), American Institute of Physics].  
Diffraction pattern of a wurtzite Si NW (b) [Reproduced from \cite{Roca-Si-NW-JMC2008} by permission of 
the Royal Society of Chemistry].
Size distributions and SEM of Au CNPs shown in the inset, scale bar is 50 nm (c); size distributions of Si NWs grown 
in thermal (d) and plasma-assisted processes (e); insets in panels (g) and (h) show SEM 
of Si(111) and Si(110) NWs grown in thermal and plasma-assisted
processes, respectively (scale bars are 500 nm); the frequency of much more abundant Si(110) NWs 
in (e) is multiplied by 0.1 \cite{AMat2007-SiNW-Plasma}.  
Elementary processes on Au CNP (top right), liquid-solid interface area (top left), 
growth experiment (f) and the outcomes in thermal (g) and plasma (h) processes \cite{Hamid-Si-NW-2012}. 
Panels (c-e) and insets in panels (g,h) are reproduced with permission from \cite{AMat2007-SiNW-Plasma}. 
Copyright \copyright 2007 Wiley-VCH Verlag GmbH \& Co. KGaA, Weinheim. Panels (g-h) 
are reproduced with permission from \cite{Hamid-Si-NW-2012}. Copyright \copyright (2013) American Chemical Society.
} 
\end{center}  
\end{figure}

Figure~\ref{f18} shows examples of the plasma processes which help solving these problems.  
Panel (a) shows a crystalline $\sim$10 nm-thin Si nanowire grown by Au catalyst-assisted PECVD
\cite{Hofmann-Si-NW-2003}. In purely thermal CVD, the NWs are much 
thicker under similar process conditions  
\cite{Iacopi-SiNW,Roca-Si-NW-JMC2008,AMat2007-SiNW-Plasma}. 
These observations are explained by multiscale, multi-phase numerical modeling which demonstrates 
that the nucleation threshold and the minimum 
thickness of the Si NWs can be substantially reduced 
(at the same external heating temperature of the substrate) 
when a plasma is used \cite{HamidAPL2011}. Nanoscale plasma-surface 
interactions and localized heating 
also mitigate the adverse GT effect and result in 
much thinner Si nanowires at much lower temperatures and 
pressures than in similar thermal CVD processes.  
 
Si nanowires with wurtzite structure can also be synthesized by PECVD, as confirmed by the 
selected area electron diffraction pattern (SAED) in Fig.~\ref{f18}(b). These NWs were produced 
through the {\it in situ} generation of indium catalyst droplets and subsequent growth on an ITO
substrate \cite{Roca-Si-NW-JMC2008}.
 
Panels (c-h) in Fig.~\ref{f18} show that even a short ($\sim$2.5 min) exposure to the plasma is very 
effective to simultaneously improve the selectivity of the NW growth direction and reduce their
thickness. Indeed, starting from the same Au catalyst nanoparticle pattern with the size distributions shown in (c),
purely thermal CVD predominantly produces Si(111) NWs with the average thickness of $\sim$60 nm (d) whereas 
thin ($\sim$20 nm) and more size-uniform Si(110) nanowires grow in the plasma-enhanced 
process (e). Moreover, the frequency of Si(110) NWs 
(see the SEM image in the inset of panel (h)) 
is almost 2 orders of magnitude higher compared 
to the much thicker ($\sim$50-60 nm) Si(111) nanowires imaged in the inset of panel (g).  

This striking experimental observation can be explained by considering the 
thermokinetic effects of nanoscale plasma-surface interactions 
sketched in Fig.~\ref{f18}(f-h) \cite{Hamid-Si-NW-2012}. 
These effects lead to the much larger difference in the Gibbs free energies between the (111) and 
(110) growth directions 
\begin{equation}
\label{deltaG-SiNW}
\Delta G = \Delta G^{(111)} - \Delta G^{(110)}
\end{equation}
in PECVD compared to purely thermal CVD. 
In other words, the
(110) becomes a clearly preferential growth direction in the plasma, which explains the
experimental observations \cite{AMat2007-SiNW-Plasma}. 

The difference in the Gibbs free energies (\ref{deltaG-SiNW}) can be used to estimate the 
diameter of the Si nanowire
\begin{equation}
\label{SiNW-diam}
d_{\rm NW} = 2 \frac{\Delta G - [\gamma^{(111)}_{\rm SV} - \gamma^{(110)}_{\rm SV}] \Delta S  }{(\gamma_{\rm LS} \delta)^{(111)} - (\gamma_{\rm LS} \delta)^{(110)}}
\end{equation}
which is assumed to be equal to the critical diameter of the Si monolayer formed at the 
interface between the catalyst and the substrate. Here, $\gamma_{\rm LS}$ and  $\gamma_{\rm VS}$ are the 
surface energies for the liquid-solid and nanowire-vapor interfaces, respectively, with the superscripts 
denoting the nanowire crystallographic orientation. Parameters $\delta^{(111)}$ and $\delta^{(110)}$ 
quantify the cross-section geometries of the (111) and (110) growth directions observed in the
experiments \cite{SiNW3,SiNW4}, and $\Delta S$ is the thickness of the nanowire surface just beneath the
interface denoted in Fig.~\ref{f18}. As mentioned above, the nanowire thickness (\ref{SiNW-diam}) is typically clearly smaller 
in PECVD compared to the equivalent neutral gas-based processes. 

These plasma-specific effects require a more thorough study, e.g., a  
combination with the thermodynamic stability arguments and extension to a larger number of 
crystallographic orientations. For more details about the various growth approaches, thermodynamics, 
and electronic properties of Si nanowires please refer to the recent review \cite{SiNW5}.
In the following subsection, more plasma-specific effects in the
synthesis of inorganic nanowires of binary solids will be considered.

\subsubsection{\label{1D-oxide} Other inorganic nanowires} 

\begin{figure}[t]
\begin{center}
\includegraphics[width=9.5cm,clip]{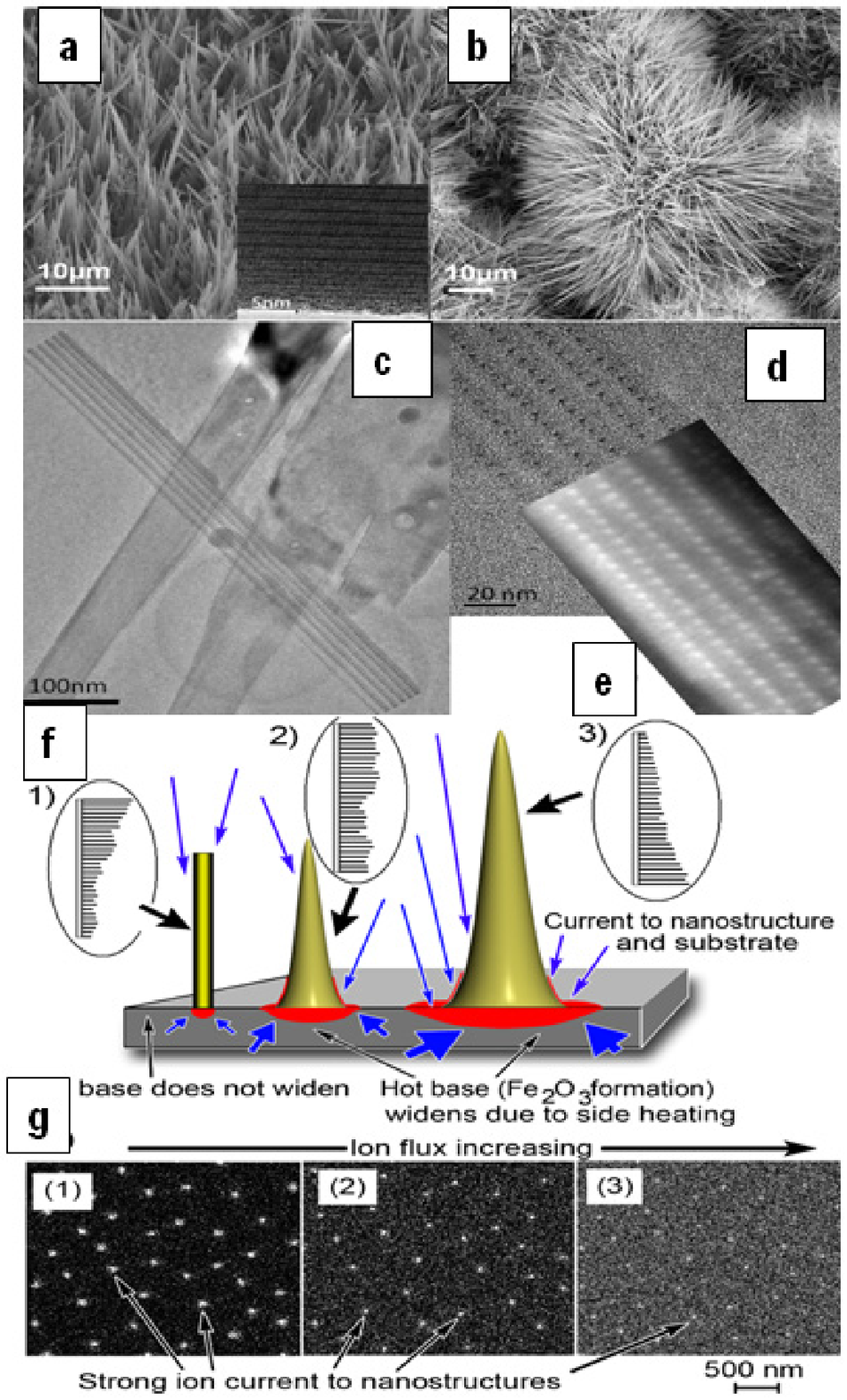} 
\caption{ \label{f19} 1D metal oxide nanowires grown  
using {\em plasmoxy nanotech} approach \cite{Editorial_JPD2011}: (a) Nb$_2$O$_2$;
(b) $\beta$-Ga$_2$O$_3$; and (c-e) MoO$_x$ imaged by SEM and HRTEM. 
Shape of Fe$_2$O$_3$ NWs is controlled by ion 
flux distributions (f) [reprinted with permission from \cite{UrosAPL09-ShapeControl}. 
Copyright \copyright (2009), American Institute of Physics]. Panels (a-e)  reproduced 
from \cite{Editorial_JPD2011} by permission of IOP Publishing. 
Copyright \copyright IOP Publishing. All rights reserved. }   
\end{center}
\end{figure}

The outstanding ability of low-temperature plasmas to dissociate and excite 
numerous molecular and radical species under conditions of strong 
thermal non-equilibrium has enabled many unique opportunities in nanoscale 
synthesis and processing. In particular, it became possible to 
dissociate reactive oxygen and nitrogen gases in the ionized gas phase while keeping the
temperature of working gas low. 
In this subsection we will mostly focus on the processes where  
interaction of oxygen plasmas with solid surfaces leads to the fast production of large 
amounts of high-quality oxide nanostructures with representative examples 
shown in Fig.~\ref{f19}(a-e).   

One of the most promising approaches towards large-scale plasma-assisted synthesis of 
oxide nanowires is based on a direct exposure of solid (e.g., metal) surfaces (e.g., thin foils) 
to reactive oxygen plasmas \cite{Uros-AdvMater2005, Uros-ChemMater2008}. 
This process has been demonstrated for a large 
number of materials systems (e.g., Fe$_2$O$_3$, Nb$_2$O$_5$, V$_2$O$_5$, CdO, CuO$_2$, SiO$_2$, WO$_2$,
WO$_3$, etc.) and is enabled by the interactions 
of ionic and (excited) neutral oxygen species with the solid surface and does not rely 
on any external heating \cite{OstrikovNanoscale2010}. The top layer of the surface 
where the nanostructures nucleate, is heated through the recombination 
of the plasma species. 

Oxygen atoms also play a role of building units and combine 
with metal atoms to form metal oxides. 
Therefore, the rates of production of oxygen atoms determine both the surface heating and 
the oxide phase formation. However, well-defined nanostructures only appear when a 
delicate balance between these two factors is reached. Indeed, too strong 
heating may easily lead to the evaporation of the entire thin metal foil. 
On the other hand, excessive (e.g., too fast) delivery of oxygen may lead to the 
uniform oxidation of the entire surface (or even the entire film) rather than 
to the nanostructure formation.

Therefore, the clue to explain the nucleation of thin nanowires on metal surfaces exposed to 
reactive oxygen plasmas is again in {\em nanoscale} plasma-surface interactions. 
The nanometer surface morphology appears to be critical for the NW nucleation.
The most delicate moment in this process is a very gentle heating of the surface layer to the
temperature which is slightly below the melting point of bulk metal, yet sufficient 
to at least partially melt small nanoscale surface features. 
This is indeed possible since the melting temperatures decrease as the nanoparticles get smaller.
For example, while the melting point of bulk Si is $\sim$1412$^{\circ}$C, 2 nm-sized nanoparticles 
can melt at only $\sim$300$^{\circ}$C \cite{SimeltT}. 

The surface heating is entirely due to the plasma and it is believed that the heat transfer is dominated 
by the surface recombination of ion and neutral species. The associated energy influx 
densities are \cite{KerstenJPD2011}
\begin{equation}
\label{Jreci-Kertsen}
J_{\cal E}^{\rm rec} = \Sigma_k j^i_k (E^{\rm ion}_k - E^{\rm diss}_k)
\end{equation}
for the surface recombination of ionic species $k$ and  
\begin{equation}
\label{Jrecn-Kertsen}
J_{\cal E}^{\rm ass} = \gamma_k N_k \sqrt{\frac{8 k_B T_{\rm gas}}{\pi m_k}} E^{\rm diss}_k
\end{equation}
for the associative recombination of neutral radicals on the surface. 
In (\ref{Jreci-Kertsen})-(\ref{Jrecn-Kertsen}), $j^i_k$ is the ion flux, $E^{\rm ion}_k$ and $E^{\rm diss}_k$ are the 
ionization and recombination energies, respectively. Here, $N_k$ is the density of recombining 
species (mostly oxygen radicals in the examples considered in this section), $m_k$ is their mass,
and $T_{\rm gas}$ is the gas temperature. The coefficient $\gamma_k$ accounts for the 
recombination probabilities of different plasma-generated species on the surface.   

This localized surface heating produces small 
nanoscale surface features where metal atoms are not so rigidly stuck in their lattices 
while oxygen atoms can penetrate easier compared to the surrounding solid material. 
Consequently, these features may be saturated with oxygen and small oxide nuclei 
may form. 

It is important that the amount of oxygen atoms matches the 
amount of metal atoms available for bonding. Otherwise, uncontrollable recombination 
of loose O-atoms may lead to the undesirable surface overheating as well as possible uniform 
and deep oxidation. This condition is a reflection of one of the basic conditions for the
successful formation of nanostructures on plasma-exposed surfaces, namely, a delicate 
balance between the demand and supply of building units and energy 
discussed in Sec.~\ref{Sec3}.  
 
After the nuclei are formed, the 1D nanostructures are shaped under strongly non-equilibrium 
conditions as shown in Fig.~\ref{f19}(f). At this stage, plasma-specific effects continue playing 
a prominent role. Many of the nanowires that form through the direct surface exposure to 
oxygen plasmas grow by basal attachment of newly-nucleated layers \cite{plasma-made-nanowires}. 
If the base does not widen, the NWs grow straight upwards (sketch 1 in panel (f)) and sharp nanoneedle-like
1D structures form. If this widening takes place, the nanowire base becomes larger during the growth process and 
high-aspect-ratio conical structures form (sketches 2 and 3 in panel (f)).
Interestingly, this effective control of the nanowire shape can be implemented by controlling the 
microscopic ion flux distributions along the surfaces of the substrate 
and the growing nanostructures. 

As was explained in Sec.~\ref{sheath}, the sheath width can be 
controlled by the plasma density and external bias; the ratio of the sheath width to the nanowire length 
determines the microscopic profile and non-uniformity of ion deposition.
In the example shown in Fig.~\ref{f19}(f) the process parameters (e.g., surface bias) were 
varied to increase the ion flux and also to vary its distribution over different areas on the 
nanostructure and substrate surfaces \cite{UrosAPL09-ShapeControl}. 
In the first case, the ion flux was predominantly deposited to the top section of the nanowire; the 
hot base did not widen and straight nanoneedle-like structures formed. When a significant fraction 
of the ion flux was deposited close to the nanostructure base, the hot basal area widened and 
nanoconical structures developed. 

Importantly, this shape selectivity is enabled by non-equilibrium process kinetics and 
nanoscale plasma-surface interactions, quite similar to the examples considered in 
Secs.~\ref{Nanocrystals}, \ref{1D-Carbon}, and \ref{1D-Si}. These strong non-equilibrium conditions lead to 
very high growth rates of the nanowires. It typically takes only up to a few minutes to 
grow a fully developed nanowire with the length of several tens of micrometers while 
fairly similar thermal or catalyst-assisted processes usually 
last several hours and even longer \cite{MahendraJACS02}. This plasma-enabled process is single-step, 
fast, energy-efficient, environment-friendly,
and does not need any catalyst or pre-formed patterns. Combined with the 
good nanowire quality, it offers exciting prospects in various 
applications such as gas sensing, energy conversion, 
etc. \cite{Uros-SensorAPL2008, UrosJPD2011}.

Other plasma-based processes, such as PECVD, catalyst-assisted growth, 
plasma flight-through synthesis, as well as hybrid approaches have also 
been successfully used for high-rate, large-quantity production 
of high-quality inorganic nanowires, not merely limited to 
metal oxides \cite{Meyyappan-Sunkara-Book, UrosJPD2011, Sunkara-nanowire-kinetics2}.  
For example, ultra-fast  kinetics and non-equilibrium conditions 
in the gas phase of dense microwave plasmas enable bulk production 
of SnO$_2$, ZnO, TiO$_2$, and Al$_2$O$_3$ nanowires \cite{MahendraPhysChemCLett2008}. 
The NWs nucleate on metal NPs while they move through the 
high (from a few Torr to atmospheric)-pressure Ar$+$O$_2$ plasma jet.
The growth process is very fast (fractions of a second) and can lead to 
nanowire production with rates $\sim$5 kg/day. This unique high-throughput process is not possible 
without the plasma. To enable partial nanoparticle melting needed 
for the NW nucleation, the heating should be very effective during  the NP flight through the 
plasma. 

A comparative analysis of the effects of heat transfer to the NPs by
conduction, convection, radiation, collisional heating, and surface reactions suggests 
that the latter mechanism is dominant. The most important reactions involve recombination 
of the plasma-produced oxygen and hydrogen radicals, as well as metal oxidation on the surface, 
similarly to the growth of surface-supported NWs discussed above. 
This similarity of the plasma-specific effects is further evidenced by the 
recent synthesis of TiO$_2$ nanowires by the direct exposure 
of Ti foils or powders to atmospheric-pressure microplasmas under 
fairly similar conditions as in the previous example \cite{MahendraCGD2011}. 
 
The nanoparticle size is also critical because the nanowire nucleation ceases above certain 
threshold when the plasma heating  
becomes less effective. The nucleation of NWs can be explained using 
thermodynamic stability analysis which relates the supersaturation levels and 
the sizes of critical nuclei to the NP temperature determined by the 
microwave input power and other discharge conditions \cite{MahendraJPCB2006}.  

There are several opportunities in this area, in particular, in deeper 
understanding of the underlying mechanisms of the energy and matter exchange 
in localized plasma-surface interactions, nucleation and thermokinetic 
selection of the growth direction, shape selectivity and control during the growth 
(e.g., maintaining 1D growth avoiding the formation of 2D nanowall-like structure) \cite{XuChemPhysChem2007},
tailoring the internal structure (e.g., vacancies and porosity), customizing the
electronic and chemical properties for targeted applications, and several others. 

Importantly, this class of nanoscale objects is a clear manifestation of a uniquely plasma-enabled 
phenomenon which certainly deserves further in-depth studies. 
For more information about the physical properties, characterization, and 
applications of inorganic nanowires please refer to the recent monograph 
\cite{Meyyappan-Sunkara-Book}.

\subsection{\label{2D-mater} Two-dimensional nanomaterials}   

As the dimensionality of nanoscale objects increases, 
the nanoscale plasma-surface interactions 
also change and are determined by the 
two-dimensional structure and morphology of these objects.
In this section, we will consider graphene, graphene oxide, and graphene nanoribbons (GNRs) (Sec.~\ref{graphene}),
vertically aligned 2D carbon structures such as carbon nanowalls and graphene nanosheets   
(Sec.~\ref{carbon-nanowalls}), and also some 2D nanostructures of other 
solid systems (Sec.~\ref{2D-other}). Generally, numerous plasma-based processes have been 
used not only to grow these 2D structures but also to modify, shape, functionalize, activate, 
decorate, and condition them. 

\subsubsection{\label{graphene} Graphene, graphene nanoribbons, and graphene oxide} 

\paragraph{\label{graphene1} Graphene} Two-dimensional graphenes are made of a single or just a few layers 
of $sp^2$-networked carbon atoms \cite{NovoselovScience2004}. 
Graphenes feature many truly unique properties such as 
exceptional electron and heat conductivity, toughness, optical transparency, 
chemical reactivity, etc., which makes them ideal for a broad range of applications 
in nanoelectronics, photonics, sensing, and several other areas. 
However, single-layer graphene shows semi-metallic properties and bandgap opening is one 
of the major present-day challenges on the way of its  
widespread applications.   

The ultra-fine, atomic-thin structure of graphene makes its 
synthesis and processing very delicate. To synthesize such a material on the 
(e.g., catalyst) surface, one needs to assemble a monolayer (or a just a few layers) of carbon atoms. 
Moreover, the monolayer should be stable and defect-free. 
Hence, the amount of material that needs to be delivered to the surface to produce graphenes, 
should be small and the monolayer nucleation should be fast and take place over the entire surface area.

The first requirement is similar to the SWCNT nucleation (Sec.~\ref{nucleation}), whereas the 
second one is different and applies to flat catalyst surfaces. 
This is why very high temperatures (typically well above 800--900$^{\circ}$C) and highly-active 
metal catalysts such as Pt, Co, Ni, Cu, Ru, Ir, and some others are usually required. 
Graphene growth using a suitable crystalline face has been demonstrated. However, 
significant difficulties (e.g., uniformity of the number of layers in different areas) still 
remain while using    
polycrystalline catalysts, which are very attractive because of the 
obvious lower material cost. 

Mechanism-wise, there are two common types of graphene growth on 
catalysts with high (e.g., Ni) and low (e.g., Cu) carbon solubility. 
In the first case, carbon atoms first saturate within the catalyst 
layer and then segregate towards the top surface to nucleate a 
graphene monolayer. When the solubility is low, the graphene nucleation 
takes place entirely on the surface. In thermal CVD, both processes 
require very high heating temperatures. The nucleated graphene 
then needs to be separated from the supporting catalyst, which 
is often commonly achieved by completely etching the catalyst,
or alternatively, by using specific chemicals that unbind 
the graphene from the growth substrate. Achieving large-area, single-layer graphene
with a small number of large 2D crystalline grains, and ultimately, the 
single-crystalline structure, is a significant challenge \cite{AgoCPL2012,Hamid-ACSNano2012}.

\begin{figure}[t]
\begin{center}
\includegraphics[width=13.5cm,clip]{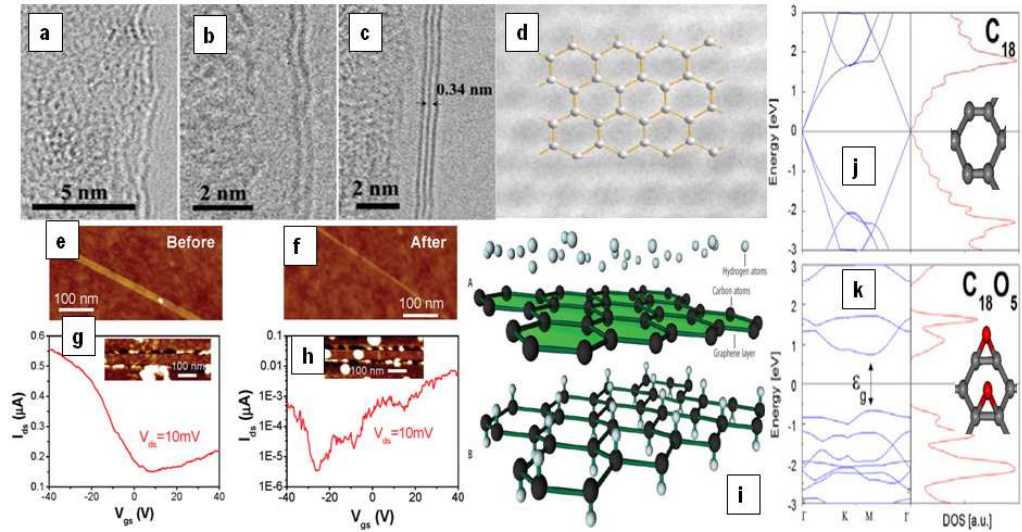} 
\caption{ \label{f20} Single-layer (a), bi-layer (b), and tri-layer (c) graphenes on polycrystalline 
Co film grown by PECVD [Copyright \copyright IOP Publishing. 
Reproduced from  \cite{Graphene-plasma-JPD2010} by permission 
of IOP Publishing. All rights reserved]. 
HRTEM image of graphene (d) synthesized by 
remote PECVD on Ni catalyst [reprinted with permission from \cite{RouvimovAPL2010}.  
Copyright \copyright (2010), American Institute of Physics]
AFM images of GNRs before (e) and after (f) etching 
in H$_2$ plasmas; nanodevice performance of these GNRs (g,h) 
[Reprinted with permission from \cite{XieJACS2010}. 
Copyright \copyright (2010) American Chemical Society].
Reversible hydrogenation of graphene surfaces  leads to non-conducting, bent 
graphane \cite{graphane-Science} (i)  (sketch reprinted from \cite{Savchenko-Science} 
with permission from AAAS). 
O$_2$ plasma exposure leads to attachment of O-atoms to carbon network 
and bandgap opening of graphene (j,k) [Copyright \copyright IOP Publishing. 
Reproduced from \cite{NourbakhshNT2010} by permission of IOP Publishing. All rights reserved]. } 
\end{center}  
\end{figure}

Plasma-assisted growth and processing of graphenes is a relatively new area with 
a relatively small number of relevant publications. The reason is that it is usually 
quite challenging to simultaneously deliver small amounts of material and enable fast and uniform 
nucleation, while avoiding any damage due to ion bombardment.     
Nevertheless, there are several examples of successful plasma-based growth of graphenes, see 
Fig.~\ref{f20}(a-d). TEM images in panels (a-c) show 1-3 layers of graphene synthesized 
on a {\em polycrystalline} Co film by a RF PECVD in a mixture of CH$_4$, H$_2$, and Ar gases.
This process is extremely fast and requires only 40 s to produce the graphenes at a relatively 
low temperature of 800$^{\circ}$C \cite{Graphene-plasma-JPD2010}. 

In another example in panel (d), the amount of carbon material involved in the synthesis was
reduced by using a remote plasma configuration \cite{RouvimovAPL2010}. 
This approach is quite similar to the synthesis of SWCNTs (Sec.~\ref{growth}). 
In this example, single- and few-layer graphenes grow on Ni(111) single crystals and also  
polycrystalline nickel films. A special advantage of using remote plasmas in the growth of graphenes 
is to eliminate the effect of the plasma electric field, which on one hand facilitates the formation 
of a nanotube cap (Fig.~\ref{f11}(d)) and accelerates the plasma ions on the other. 
Even though the plasma was remote to the growth substrate, a major  
reduction (from $\sim$1000$^{\circ}$C in thermal 
CVD to only $\sim$650$^{\circ}$C in PECVD) of the process 
temperature was reported \cite{RouvimovAPL2010}. Moreover, the growth process was very fast lasting 
only $\sim$1 min. These and several other results suggest that the 
plasma-based graphene growth is both versatile and scalable.

Significant caution towards using plasmas in the synthesis and processing of 
a so delicate 2D nanomaterial as graphene naturally stemmed from the possible 
adverse effects of ion bombardment. However, it was recently demonstrated that 
minute traces of polymer impurities can be removed from graphene surface 
by exposing it to low-density ($\sim$8.5$\times$10$^8$ cm$^{-3}$) remote Ar plasmas
\cite{ACSNano-graphene-cleaning}. Importantly, this room-temperature 
treatment does not damage the graphene structure and, moreover, is 
capable to restore its pristine defect-free state. This process is
compatible with Si-based nanoelectronics and can be regarded as a viable 
alternative to thermal vacuum or electric current annealing or wet chemical processing.    

On the other hand, graphene doping or defect incorporation may be intentional to enable 
certain functionality, such as controlling electrical conductivity type (e.g, n- or p-type) or chemical 
reactivity.  Examples in Fig.~\ref{f20} show the plasma effects on 
the bandgap of graphene. Surface hydrogenation was used 
to bend a free-standing graphene layer to form non-conducting graphane (i) 
\cite{graphane-Science}. By oxidizing the graphene surface
in a plasma, it became possible to enable a quite significant bandgap 
opening, see Fig.~\ref{f20}(j,k) \cite{GokusACSNano2009,NourbakhshNT2010}. 
Importantly, these processes are reversible,
which may lead to high-throughput, low-temperature
interconvertible graphane-to-graphene growth and processing \cite{WangACSNano2010}. 
Successful doping of graphene with nitrogen from reactive plasmas in ammonia has also 
been demonstrated \cite{LinAPL2010-GrdopN}. Opening and etching of carbon dangling bonds and creation of 
multiple vacancies is also possible using plasmas. Some examples related to graphene nanoribbons will be
discussed below. 

Other possibilities in the rapidly developing area of surface and defect engineering 
of graphene are discussed elsewhere \cite{Dresselhaus-review2012}. Raman spectroscopy 
can be used to identify the nature of defects in graphene, for example, $sp^3$-type or vacancy-type 
defects created \cite{Raman-graphdefects}. Recently, the extreme 
selectivity of the outcomes of the remote hydrogen plasma reactions with graphene of 
different number of layers was demonstrated \cite{DiankovACSNano2013}.
Indeed, fairly uniformly distributed holes are formed on basal planes of 
monolayers, while the edges are etched. On the other hand, size-uniform 
hexagonal features are etched on bi-layer and thicker graphene sheets. 
These features develop from the existing defects through highly-anisotropic etching.
The etch rates are also very sensitive to temperature and the best results are achieved 
at 400$^{\circ}$C, although room-temperature etching is also quite 
effective \cite{DiankovACSNano2013}. 

The mechanisms of these highly-promising 
plasma effects are presently unclear. Importantly, these mechanisms are more likely to 
be due to the properties of graphene layers and their interactions with the plasma species 
rather than the substrate effects because very similar observations were made on both 
relatively rough SiO$_2$ and nearly atomically-smooth mica. These results complement 
several relevant observations by other authors which include faster hydrogenation of bi-layer
and thicker sheets compared to monolayer graphene, highly-anisotropic etching along specific 
crystallographic directions, and some others \cite{Graphene-Plasma-26,Graphene-Plasma-27,Graphene-Plasma-28}. 
Identifying and better understanding of the specific roles of relevant plasma species 
and the mechanisms of their interactions with different atomic arrangements in graphenes 
of different number of atomic layers may shed some light on these highly-promising recently discovered control 
possibilities. Initial studies \cite{DiankovACSNano2013} draw parallels with erosion of graphitic 
materials by low-energy hydrogen ions \cite{graphite-erosion1,graphite-erosion2,graphite-erosion3} 
and suggest the important role of hydrogen radicals and 
H$^+$, H$^{2+}$, and H$^{3+}$ ions. However, finding the specific contributions of these 
reactive species requires precise measurements and modeling of their relative abundance 
in the plasma discharge and offers an interesting opportunity for future studies.

Reduction of graphene growth temperature still remains a significant issue where plasma-based approaches 
are expected to produce a significant contribution in the near future. A promising result on the synthesis 
of a few-layer graphenes on Cu catalyst in Ar$+$C$_2$H$_2$ microwave plasmas at temperatures as 
low as 240$^{\circ}$C has recently been reported \cite{Kalita}. However, limited details of the 
temperature measurements (which as discussed in Sec.~\ref{1D-Carbon} can be $\sim$100$^{\circ}$C or even higher
than supplied through external heating)
prevent us from commenting on this issue.

\paragraph{Graphene nanoribbons}

Low-temperature plasmas were also used for high-precision, ultra-fine selective etching of edges of 
graphene nanoribbons (thin strips of graphene) shown in Fig.~\ref{f20}(e-h). The initially 14 nm-thick GNRs (e) were thinned 
to only 5 nm (f) after a short exposure to hydrogen plasmas \cite{XieJACS2010}. This delicate 
etching did not produce any structural damage to the main body of the nanoribbon.
More importantly, this precise thinning of the GNR in one dimension has led to the bandgap opening 
of the originally metallic GNR. This in turn resulted in a dramatic increase of the 
source-drain current {\it versus} gate voltage ($I_{\rm ds}(V_{\rm gs})$ in panels (c) and (d))
nanodevice characteristics.  

Several other examples confirm the utility of low-temperature plasmas 
to fine-tune the properties of graphenes and graphene 
nanoribbons. A combination of nano-masks (e.g., nanospheres or nanowires) and
nanoscale reactive etching in low-density, low-temperature O$_2$ plasmas 
was used to fine-tune the bandgap of ultra-narrow GNRs \cite{GNR-Lithogr-AdvMat2011} 
and enable their applications in field effect transistor devices \cite{BaiNL2009}.  

Precise interactions of low-density plasmas with partially masked MWCNTs  also enabled a new class of 
graphene-based materials, namely surface-supported bent GNRs. 
In particular, by adjusting the 
plasma etch time, it became possible to produce single-layer, bilayer, and trilayer 
graphene nanoribbons, the bent analogues of flat 
graphenes \cite{GNR-Nature, JiaoNanoRes2010, HDai-GNR-2011}. 

In a recent example, narrow graphene nanoribbons ($\sim$23 nm) were grown 
directly between the source and drain electrodes of a field-effect transistor 
using plasmas of CH$_4+$H$_2$ gas mixtures and narrow (less than 30 nm) Ni nanobars
with a typical thickness below 50 nm, and a rapid heating \cite{Hatakeyama-NatureNT}.  
Under certain conditions, Ni catalyst evaporates from underneath the nanoribbon
thus creating a graphene nano-bridge between the electrodes. 

The GNRs exhibit a very clear bandgap of $\sim$58.5 meV and a very high  ($\sim$10$^4$)
on/off ratio without hysteresis during the transistor operation. 
At high temperatures, the electron transport in these nanoribbons followed 
\begin{equation}
\label{eltransp11}
{\cal G}_e \sim \exp (-\epsilon_a / 2k_B T),
\end{equation} 
where the transport activation energy $\epsilon_a \sim$16.2 meV was obtained from a linear fit to the Arrhenius plot.
At lower temperatures, the electron behavior in the nanoribbons 
followed the variable-range hoppinig approximation 
\begin{equation}
\label{eltransp22}
{\cal G}_e \sim \exp [ -(T_0/T)^{\ell} ],
\end{equation} 
where $\ell \sim$ 0.5. Importantly, these electron transport features 
are competitive with graphene nanoribbons synthesized using 
sophisticated lithographic procedures \cite{hanbrantkim}.

A combination of a quite similar nickel nanobars, plasmas, and rapid heating/cooling opened 
a highly-unusual possibility to nucleate single- and few-layer graphene
nanoribbons at the interface between the nanobar and 
the SiO$_2$ substrate \cite{Hatakeyama-ACSNano2012}. 
This advance contributes to the solution of a significant 
problem of direct, transfer-free synthesis of graphene on 
insulators such as silica. Because of very high optical transmittance 
of both silica and graphene, this combination also offers several interesting opportunities
for  applications in nano-optics. 

However, the Ni catalyst still needs to be removed
after graphene growth, e.g., by chemical etching. This is why direct, etch-free 
growth of graphene directly on SiO$_2$, BN, or similar dielectric substrates 
is highly-warranted in the near future and plasma-based approaches are expected
to contribute to the solution of this problem. 

Moreover, the mechanism of selectivity of nucleation of graphene layers 
either on top or at the bottom of Ni nanobars still remains unclear.
These phenomena may be explained by comparing the rates of carbon atom 
delivery $\Upsilon_{\rm del}$, diffusion to the bottom interface $\Upsilon_{\rm difb}$, their segregation towards the 
upper surface $\Upsilon_{\rm seg}$, graphene nucleation at the top $\aleph_t$ and bottom $\aleph_b$ interfaces, 
and Ni evaporation $\Re_{\rm evap}^{\rm Ni}$.
For example, for the formation of graphene nanobridges, it is important that
\begin{equation}
\label{ineq23}
\Upsilon_{\rm del} \sim \Upsilon_{\rm seg} \sim \aleph_t > \Re_{\rm evap}^{\rm Ni} \gg \aleph_b, 
\end{equation}
which means that graphene nucleation on the plasma-exposed Ni surface should be faster than 
the nanobar evaporation and much faster than the nucleation of graphene at the bottom interface. 
Alternatively, nucleation conditions at the bottom interface may not be met during the 
process \cite{Hatakeyama-NatureNT}. However, only detailed numerical modeling studies may 
validate this phenomenological argument.   
    
The above non-exhaustive examples suggest that despite  
significant caution, plasma-specific effects can and should be used 
for the growth and ultra-fine processing of even so delicate 
nanoscale objects as graphenes. Moreover, these effects should be explored 
to solve the persistent problem of GNR chirality control, similarly to 
the SWCNT case of Sec.~\ref{nucleation}. This may lead to better selectivity 
between the antiferromagnetic and non-magnetic responses of zig-zag and armchair GNRs 
actively pursued in nanomagnetism research \cite{nanomagnetism}.

\paragraph{Graphene oxide}

As we have seen from Fig.~\ref{f20}(j,k), plasmas are effective at oxidizing graphene, thus forming graphene oxide (GO). 
The opposite, reduction of GO is also possible, especially because of the plasma reactivity, for instance in producing reactive 
hydrogen atoms and ions. These species effectively reduce oxygen content in graphene oxide, ultimately 
leading to pure graphene. The presence of oxygen determines the GO bandgap, being wider in stoichiometric graphene oxide
and much smaller when oxygen content is low. 

\begin{figure}[t]
\begin{center}
\includegraphics[width=12cm,clip]{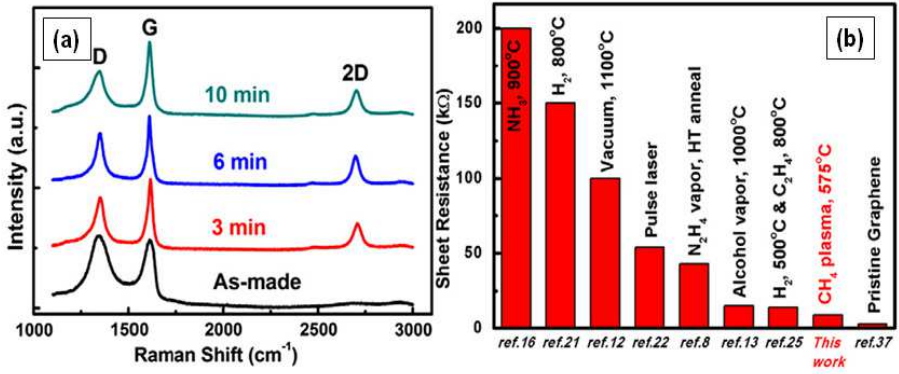} 
\caption{ \label{Fig-GO} Raman spectra of as-made and plasma-reduced (after 3, 6, and 10 min plasma treatment) graphene oxide (a).
Comparison of sheet resistances of pristine graphene, CH$_4$ plasma-reduced graphene oxide (``This work" means Ref.~\cite{GO1}), 
and other reports on GO reduction (reference numbering as in the original publication). 
Both figures are reprinted from \cite{GO1}, Copyright \copyright (2012), with permission from Elsevier.} 
\end{center}  
\end{figure}

The reduction process of GO in a low-temperature remote CH$_4$-based plasmas is also accompanied by a 
very clear defect healing as can be seen in the Raman spectra in Fig.~\ref{Fig-GO}(a) \cite{GO1}. Indeed, as the 
plasma exposure duration increases, the graphene signature 2D peak increases, while the 
ratio of the defect-related D and graphitic structure quality-related G peaks decreases 
in nearly 2 times, from 1.03 to 0.53. Therefore, 
relaxation of C--C bonds can be considered as a driving force 
for the GO reduction. The plasma-based method is very fast ($\sim$ 10 min) and requires lower temperatures 
($\sim$600$^{\circ}$C) compared to other methods labeled in Fig.~\ref{Fig-GO}(b) \cite{GO1}.

More importantly, the recovered low sheet resistance of the graphene oxide ($\sim$9 k$\Omega$/square)
appears to be reasonably close to that of pristine graphene \cite{GO1}. 
The larger sheet resistance of the reduced graphene oxide (r-GO) compared to pristine graphene
is mostly due to its rougher surface and defect presence. In this case, the carrier transport 
is described by the 2D variable hopping model with the 
conductance 
\begin{equation}
\label{conduct-rGO}
{\cal G}_e^{\rm rGO} \sim \exp \left( - \frac{\rho_{\rm hop}}{T^{1/3}} \right),
\end{equation}
where 
$$
\rho_{\rm hop} = [3 / k_B N(E_F) \lambda_L]^{1/3}
$$
is the hopping parameter, $N(E_F)$ is the density of mobile carriers,  $E_F$ is the Fermi energy, and $\lambda_L$ 
is the localization length \cite{GO1}. Larger conductance implies larger $N(E_F)$ and $\lambda_L$ which is
attributed to larger graphene crystalline domains. This is another manifestation of the importance 
of the ultimate achievement of large-area single-crystalline graphene mentioned at the beginning of Sec.~\ref{graphene1}. 

Effective reduction of graphene oxide was also demonstrated using 
atmospheric-pressure plasmas in Ar$+$H$_2$ gas mixtures, at significantly lower temepratures 
(down to $\sim$70$^{\circ}$C \cite{GO2}. The achieved sheet resistances of r-GO were also 
competitive, for example 4.77$\times$10$^4$ $\Omega$/square at 70$^{\circ}$C. 
Therefore, plasma-based reduction of graphene oxide can be regarded as a viable alternative 
to solution-based chemical methods which use strong and hazardous reactants such as 
hydrazine (N$_2$H$_4$) or sodium borohydride (NaBH$_4$) and often introduce significant 
impurities and defects \cite{GO2}. Selectivity of the plasma-assisted 
GO reduction in the presence of reactive chemicals is an interesting area to explore.   

To end this section, we stress that graphene-related research is a very rapidly expanding field with a huge 
number of publications. This is why we refer the reader to the selected reviews which cover the 
interesting physics and applications of graphene and related structures and materials 
\cite{Dresselhaus-review2012,CastroNetoRMP,AdvPhysGraph,AMat2012}.

\subsubsection{\label{carbon-nanowalls} Graphene nanosheets} 

 Let us consider vertically standing and 
unsupported graphenes (VSGs and USGs, respectively) as alternatives to 
horizontal graphenes supported by solid substrates.
 In both cases, the growth is essentially catalyst-free and is enabled by the plasma. 
 Such graphenes have an obvious advantage over the flat surface-bound counterparts.
Indeed, both surfaces and at least 3 open reactive edges of the vertically standing or unsupported graphenes
can be utilized in applications (e.g., biosensing) while surface-bound counterparts 
effectively use only one surface. 

\begin{figure}[t]
\begin{center}
\includegraphics[width=8.5cm,height=10cm,clip]{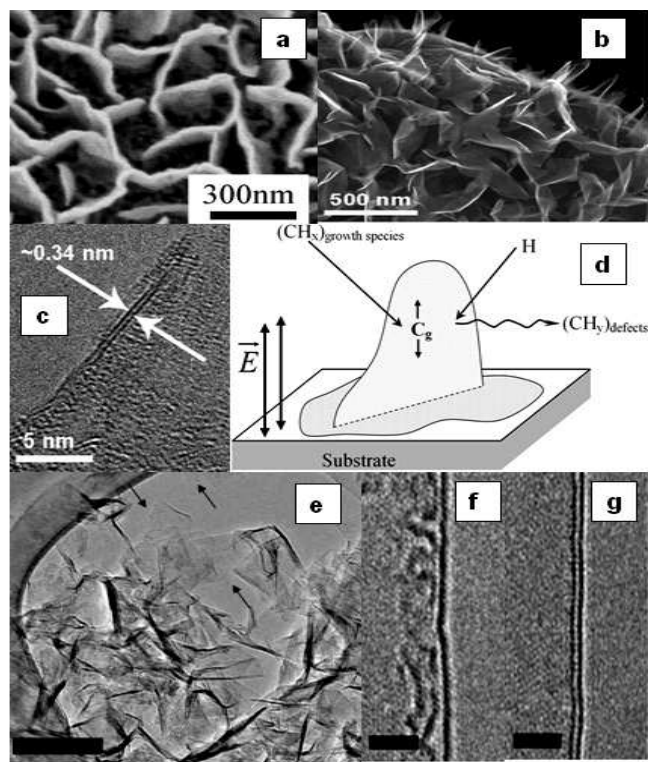} 
\caption{ \label{f21} Carbon nanowalls from plasmas 
with H radical injection (a) 
[reprinted with permission from 
\cite{Hori-nanowalls}. Copyright \copyright (2004), American Institute of Physics].
Vertical GNSs grown by RF PECVD on 
a Ni wire (b); 
HRTEM of bi-layer graphene (c); and GNS growth model 
(d) [reprinted from \cite{GNS-Carbon2007}, Copyright \copyright (2007), with permission from Elsevier]. 
TEM of GNSs grown by passing ethanol droplets through Ar plasma, arrows 
point at monolayer graphene (e); TEM images of (f) single-layer and (g) bi-layer graphene        
\cite{Substrate-free-graphene}. Scale bar is 100 nm in (e) and 2 nm in (f,g). 
Panels (e-g) reprinted with permission from \cite{Substrate-free-graphene}. 
Copyright \copyright (2008) American Chemical Society].}
\end{center}    
\end{figure}

However, producing the VSG and USG structures is quite challenging. 
Indeed, stability of straight few-atomic-layer-thin vertically standing graphenes is an issue. 
On the other hand, synthesis of USGs requires effective control of nucleation 
in the gas phase, which is very difficult because of the very fast process kinetics. 
Figure~\ref{f21} evidences solutions to these problems using 
low-temperature plasmas.

\paragraph{Surface-supported vertical graphenes}

Figure~\ref{f21}(a) shows vertically standing carbon nanowalls (CNWs) grown 
by hydrogen radical-assisted PECVD  on a Si surface without any catalyst, 
using fluorocarbon precursor \cite{Hori-nanowalls}. Similar CNWs  were
produced by other groups \cite{LongIEEE05,Robertson-nanowalls}. 
Interestingly, such catalyst-free CNW growth is only possible 
in a plasma. This plasma-enabled phenomenon was explained through the study of 
kinetics of the plasma-assisted reorganization of underlying thin oxide layers followed by the 
formation of self-organized chains of small carbon clusters on which the nanowall 
nucleation takes place \cite{SeoCarbon2011}.

As the plasma-based growth techniques matured, it became possible to produce 
very thin carbon nanosheets of only a few atomic layers thickness \cite{Cai-Outlaw}. One example of such 
vertical graphene structures supported by the curved surface of a Ni wire is shown 
in Fig.~\ref{f21}(b) \cite{GNS-Carbon2007}. Some of these structures can be as thin as 
only two atomic layers, see TEM image in panel (c). 

The growth model of these
vertical 2D NSs in panel (d) includes adsorption of hydrocarbon BUs (CH$_x$), 
diffusion of carbon species (C$_g$), etching of defects and amorphous carbon 
by reactive H atoms (CH$_y$), as well as the vertical NS orientation  
due to the anisotropic polarizability effects in the 
vertically oriented electric field $\vec{E}$ in the plasma sheath \cite{GNS-Carbon2007}.  
However, this model cannot quantify the number of vertical 
graphene walls that can nucleate and detailed numerical modeling or {\it ab initio} 
simulations are needed. It is also unclear how 
the nanostructures nucleate to  simultaneously extend in the horizontal and 
vertical directions. Physically, the reason for the vertical alignment 
may be very similar to the SWCNTs in Fig.~\ref{f11}(d).

This result was achieved using Ni catalytic support 
and the issue of direct growth of atomically thin VGS structures directly on Si still 
remains open. Recently, it became possible to produce 2--6 atomic layers-thick 
vertical graphenes using CNWs \cite{SeoJMC2011}
or silicon nanograss \cite{Shailesh-Nanoscale2011} as 
supporting and growth guiding structures. Combined with the unique 
plasma-enabled activation of the thin top reactive edges of the VSGs and their 
decoration with even thinner catalyst nanoparticles opens an opportunity 
to achieve vertically-standing single-layer graphenes.  

These and some other modifications of vertical few-layer graphenes have several interesting 
properties, such as a variety of morphologies, a very high length of open 
reactive edges (e.g., up to $\sim$10$^3$--10$^4$ m per gram of material),
as well as good electric conductivity and structural stability. 
Moreover, they may form three-dimensional networks with a high-density 
of Y- and T-shaped junctions between vertically standing 
nanosheets. This combination of properties makes such structures 
suitable for a number of applications such as electric double layer 
capacitors (supercapacitors) \cite{Cai-Outlaw}, gas and bio-sensors \cite{ChemComm2012}, and several others. 

Recently, uniquely plasma-enabled, catalyst-free synthesis of 
a few-layer vertical graphenes directly on Si/SiO$_2$ substrates using 
natural precursors has been demonstrated \cite{Seo-Honey}. 
For example, rapid reforming of natural honey exposed to reactive low-temperature
Ar$+$H$_2$ plasmas produced high-quality, ultrathin
vertical graphenes, typically 3-8 atomic layers thin. 
Importantly, this transformation is only possible in the plasma
and fails in similar thermal processes. 

The plasma-enabled process is
energy-efficient, environmentally benign, and is much
cheaper than common synthesis methods based on purified
hydrocarbon precursors. The graphenes produced also retain the essential
minerals of natural honey, feature ultra-long, open reactive edges.
Enhanced with suitable metal nanoparticles these vertical graphene structures show
reliable gas- and bio-sensing performance. Interestingly, very similar structures 
can be produced in a plasma from very different precursors in all 3 states 
of matter, such as methane (gas), milk and honey (liquid), butter (soft solid) and crystalline 
sugar (hard solid). These interesting experimental 
results will soon be reported and will require appropriate 
theoretical interpretation.

\paragraph{Unsupported graphenes from gas phase}

Using plasmas, it is also possible to produce very thin flakes of unsupported graphenes 
directly in the gas phase, without using any catalysts. The results in 
Fig.~\ref{f21}(e-g) show very thin graphene flakes synthesized in the gas phase 
of atmospheric-pressure microwave plasmas \cite{Substrate-free-graphene}. 
Some of these flakes are only 1--2 atomic layers 
thick as seen in panels (f) and (g), respectively. The graphene sheets 
were synthesized by passing liquid ethanol droplets into an argon plasma; these 
droplets have a residence time of the order of 0.1 s inside the discharge. 
During this time, ethanol rapidly evaporates, dissociates and then condenses 
under strongly-non-equilibrium conditions, which also include rapid cooling.   
This simple approach features high rates of nanoflake collection ($\sim$2 mg/min) and 
is promising for large-scale graphene production, which can also be achieved 
in high-pressure (e.g., thermal) plasma discharges. 

Arcs and other high-pressure discharges feature high plasma densities (e.g., up to 
10$^{17}$ cm$^{-3}$ in arcs) and very high rates of collisions. Likewise, the minimum spatial localization 
of the plasma (the Debye sphere) becomes much smaller compared to the low-pressure case. This is why high-pressure plasmas 
show very different effects at nanoscales. Indeed, high plasma and radical densities lead to very strong fluxes of energy and matter 
onto the growth surface, which leads to very high nanostructure nucleation and growth rates.  
This is why high-pressure plasmas are deemed viable for industrial-scale 
nanoparticle production \cite{Tony-Shigeta,Graphene-ArcJPD2009}. 

On the other hand, very high rates of collisions lead to the effective and rapid nucleation 
in the gas phase. The sheaths in high-pressure plasmas are typically
strongly-collisional and the ions lose most of their energy ($\nu_{in} \sim \nu_{nn}$) 
while traversing the sheath. In this case, the ion damage to delicate nanostructures such as 
SWCNTs is minimized. This is why the quality of carbon nanotubes, fullerenes (e.g., C$_{60}$), and graphenes 
produced in high-pressure plasmas is usually very high, which is evidenced by a very small number of
ion-bombardment-induced structural defects \cite{Nozaki-SWCNT-microplasma}.

\begin{figure}[t]
\begin{center}
\includegraphics[width=11cm,clip]{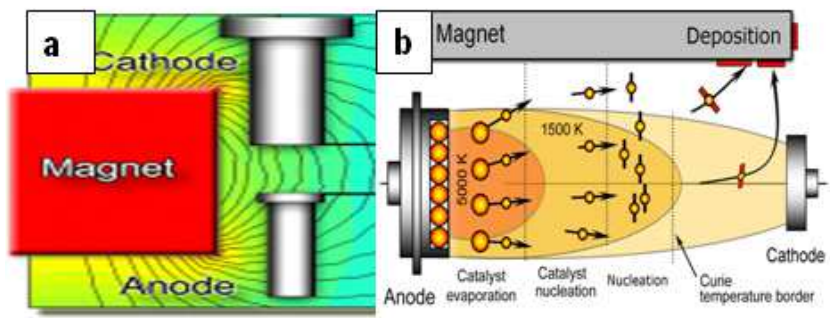} 
\caption{ \label{f30} Magnetically enhanced arc discharge for production and 
separation of graphene flakes and SWCNTs (a) [reproduced from \cite{Graphene-Nanoscale2010} 
by permission of the Royal Society of Chemistry]. 
 Non-uniformity of magnetic fields and temperature create nonequilibrium 
environment for fast moving CNPs on which graphenes 
and nanotubes nucleate (b) [reprinted from \cite{Graphene-Carbon2010}, Copyright \copyright (2010), 
with permission from Elsevier].  }  
\end{center} 
\end{figure}

These salient features lead to the nucleation of small (e.g., graphitic) nanofragments  without 
any catalyst or nucleate on fast-moving catalyst nanoparticles and produce large nanostructures 
even during the very short times needed for these particles to cross the gap between the 
points of injection and collection. In many cases the CNPs are produced by metal evaporation 
from the anode (Fig.~\ref{f30}). The gas temperature near the anode can be very high;
for example, at least several thousand Kelvin in thermal plasmas of atmospheric-pressure arcs 
\cite{MurphyPRL2002}. The nanostructures grown on these nanoparticles are usually collected at 
the opposite electrode (cathode) or at special collecting surfaces such as the surfaces of a cubic magnet 
in panel (a). The precursor material is also commonly evaporated from one or both electrodes.
 
For example,  conducting graphite electrodes are used to evaporate carbon BUs 
for the growth of graphene flakes or carbon nanotubes. 
The customized Ni-Y catalyst is embedded into the anode. 
The distributions of the density of the ionized BU vapor and the gas temperature are very non-uniform
as can be seen in Fig.~\ref{f30}. Both these features enable a   
deterministic, single-step production and magnetic separation of graphene
flakes and carbon nanotubes in an arc discharge \cite{Graphene-Nanoscale2010}.  
To achieve this, the high-temperature growth and low-temperature
separation zones are separated and the CNTs and graphenes are captured 
in different areas on the magnet.

The nanotubes and graphenes are simultaneously produced on the catalyst nanoparticles 
when they enter the high-temperature, high-plasma-density growth zone
(Fig.~\ref{f30}(b)) \cite{Graphene-Carbon2010}. 
Despite a very short residence time, the CNTs grow to 
a significant length (e.g., tens of $\mu$m); this process 
is enhanced by adding Y to the catalyst
which easily forms carbides. When the catalyst-supported 
nanostructures enter the area with much lower temperature, the 
growth process is inhibited. Moreover, Ni transits to the ferromagnetic state 
near the Curie point ($\sim$ 350$^{\circ}$C).   
This lower-temperature zone thus separates lighter CNTs and heavier graphene flakes. 
Because of this mass difference, the magnetic force is also different, and the 
nanotubes and graphenes deposit in different areas of the magnet.

Despite thermal equilibrium of thermal plasmas, the nanostructure growth process 
is strongly non-equilibrium, very fast, and is kinetically driven. The balance between 
the delivery of BUs to the catalyst surfaces (mostly determined by the density of ionized 
carbon vapor) and their gainful use in the growth (controlled by the 
gas temperature) changes dramatically while 
the CNP crosses several different zones in the discharge. This rapidly changing
kinetic process is enabled both by the non-uniformities of the magnetic field,
plasma density, and gas temperature as well as by the rapid catalyst nanoparticle transport 
across these non-uniform areas. 
More examples of the nanoparticle production 
in high-pressure plasmas, their properties and applications can be found 
elsewhere \cite{Tony-Shigeta,Editorial_JPD2011}.

\subsubsection{\label{2D-other} Other 2D nanomaterials} 

Although a large number of other 2D nanomaterials is known, the question remains if they 
can be made atomically-thin and also put in a heterolayered 
``sandwiched" stack \cite{CastroNeto2}. Of special interest is whether plasma effects can be  
gainfully used similar to Secs.~\ref{graphene} and \ref{carbon-nanowalls}.  

One typical example is heterolayered structures and superlattices, e.g., 
alternating thin layers of GaN and InGaN semiconductors used 
in optoelectronic devices. If the thickness of these layers is in the nanometer range, quantum wells 
are formed near the hetero-interfaces, which in turn offers unique quantum confinement effects. 
Plasmas have been successfully used to grow or facilitate the growth of GaN nano-layers with 
much low growth temperatures and high growth rates \cite{Sugai-GaN-PEMBE-APL07}. 

In particular, plasmas are commonly used 
in the growth of ultra-thin epitaxial semiconducting layers for the 
generation of atomic nitrogen through the effective dissociation of nitrogen molecules. 
One example is the plasma-assisted molecular-beam epitaxy of complex 
InGaGdN/GaN magnetic semiconductor superlattice structures \cite{InGaGdN-GaN}.

Low-temperature plasma treatment 
can also be used to improve the properties of heterostructured 2D nanomaterials.
For example, delicate treatment of InGaZnO periodic superlattice structures with Ar plasmas 
dramatically reduces the electrical resistivity without affecting the microstructure or  
decreasing the thermal conductivity. As a result, the thermoelectric properties 
of these two-dimensional nanomaterials are much improved \cite{InGaZnO-Thermoelectr}.

This interesting result may be related to the improvement of crystalline structure (e.g., 
compactifaction) and defect removal upon impact of low-energy Ar$^+$ ions. 
This is a typical of the many manifestations that the ion bombardment can be 
beneficial rather than commonly asserted detrimental, see other examples in
Sec.~\ref{ioneffdiff} and \ref{nucleation}. 
  
Another example is the PECVD of 
2D diamond superlattices made of heterolayers 
of pure $^{12}$C and $^{13}$C carbon isotopes \cite{diam-superlatt}. 
In this case, the plasma-specific effects are similar to the 
nano-diamond synthesis of Sec.~\ref{nanodiamond}.

Another example is the formation of ultra-thin nanolayers of magnetic alloys 
for spintronics and data storage produced using various modifications of the 
ionized physical vapor deposition.         
Functional 2D layers (e.g., transparent nanometric organic luminescent films) for organic photovoltaic and light emitting devices 
produced by plasma polymerization \cite{Angel_Barranco} are discussed in Sec.~\ref{soft-matter}. 

Vertically oriented 2D nanomaterials synthesized or processed using plasmas 
are also numerous  
\cite{MariottiNT08,CHSow-NiO-nanowalls}. 
However, their thickness is usually much larger compared 
to the thinnest vertical graphenes of Sec.~\ref{carbon-nanowalls} and a substantial effort is 
required to ultimately achieve atomically thin, vertically standing 2D nanostructures.  

2D nanomaterials produce unique quantum states of matter such as topological insulators (TIs) 
which simultaneously show insulating and conducting properties \cite{WangPRL2008}. 
Typical TIs arise in 2D quantum wells (e.g., HgTe/CdTe) or in 3D materials with 2D topological 
conducting surface states (e.g., HgTe, BiSb, Bi$_2$Te$_3$, Bi$_2$Se$_3$) which may be similar to the 
bandgap-less states of graphene \cite{FuPRL2007}. These materials enable effective 
electron spin control (e.g., spin-polarized currents along 
different interfaces in quantum-well TIs) actively pursued in spintronics. 
Exploration of plasma-based synthesis of these and many other 2D and topological materials is 
therefore warranted. 

\subsection{\label{hybrid} Hybrid nanomaterials} 

Nanomaterials with hybrid dimensionality feature many unique properties that are not common to their constituent elements 
of different dimensionalities taken separately. The constituents usually have different elemental 
composition and structure, which leads to even more exotic physical properties when combined in a hybrid 
structure. For example, combining 1D CNTs with small 0D metal or semiconducting nanoparticles leads to highly-unusual 
electron transport through the CNT-NP interface \cite{BanhartNanoscale2009,GohierAMat2012}. 
This in turn enables new functionalities in molecular-level sensing and energy storage \cite{AjayanAMat2012}. 

Moreover, the electron transport strongly depends on the 
precise NP placement. The adhesion, bonding, interface stability, and electron transport are 
very different for nanoparticles at the base, tip, or lateral surface of a SWCNT.
Indeed, when the NP is at the SWCNT base, the electrons mostly pass through 
the reactive edges of the rolled graphene sheet.
When the NP is placed at the tip of the nanotube, the electron transport is determined by the 
interface with the carbon atomic network which includes pentagonal carbon cells. The surface 
of the SWCNT cap is less reactive compared to the open edges but more reactive compared to
the lateral surface made of a rolled $sp^2$ hexagonal sheet of carbon atoms. 

Therefore, nanometer-scale selectivity of the placement and firm, defect-free attachment of small 
metal NPs to the nanotube surface, has recently become highly topical.  
This problem also applies to a large number of 1D/0D hybrid nanoarchitectures as well to other possible 
combinations of nanostructures with different dimensionalities, e.g., 2D nanowalls decorated with 0D NPs, 
3D nanocones decorated with 0D NPs which in turn catalyze and support 1D nanowires, etc.      

Let us now discuss how plasma-specific effects enable high-quality hybrid structures with the
unique properties desired for the envisaged applications 
\cite{XingguoLi,ZFRen-AdvPhys2011}. First,  
plasma-enhanced methods enable controllable synthesis and post-processing of 
the broad range of nanomaterials of different dimensionalities. 

Second,  
many examples evidence that decoration of one nanostructure with another can be 
implemented in the same reactor in a continuous, uninterrupted process flow, e.g., in the same 
vacuum cycle. 

Third, plasmas are very effective for 
the highly-controlled surface conditioning and functionalization. If special 
care is taken to avoid undesirable surface damage (e.g., the ion energy adjusted to be below the 
threshold for the destruction of a hexagonal carbon network, see Sec.~\ref{nucleation}), 
this can be used to suitably prepare the surface of the supporting (e.g., 1D) nanostructure 
for decoration by other (e.g., 0D) NSs. Importantly, surface dangling bonds may be activated 
without any significant structural damage. 

Fourth, quite differently from neutral gas- and wet-chemistry 
approaches, plasmas offer selectivity and effective control over the placement of BUs in 
specific areas with nanometer and even better precision. For example, 
the ion focusing effect (Sec.~\ref{growth}) 
allows one to deposit metal atoms directly at the tip of a SWCNT, 
with the size as small as $\sim$0.5 nm. 

Fifth, the plasma-specific selective heating 
around the point of nucleation of the decorating nanoparticles on the surface of the supporting nanostructure
is expected to lead to fewer defects and hence, a better-quality interface between the 
nanostructures of different dimensionalities.

For example, using the same plasma reactor and vacuum cycle, two-layer vertical graphene structures 
with highly-unusual electrical properties were produced \cite{SeoJMC2011}. 
Initially, well-graphitized self-organized carbon nanowalls were grown in a plasma-enabled, catalyst-free process,
without any pre-patterning. 
The top edges of the nanowalls were then activated by ion bombardment, to open 
dangling bonds. Subsequently, Cu nanoparticles were deposited on the edges, in a highly-selective fashion.
Finally, few-carbon-layer-thin vertical graphenes were grown on the Cu nanoparticle-decorated edges to produce 
a two-layer self-organized nanoarchitecture with unique properties; it was not possible 
to synthesize them without plasmas. This architecture showed an exotic 
temperature dependence of the electrical conductivity partially 
recovering the semi-metallic property of a single-layer horizontal graphene \cite{SeoJMC2011}.

\begin{figure}[t]
\begin{center}
\includegraphics[width=9cm,height=11cm,clip]{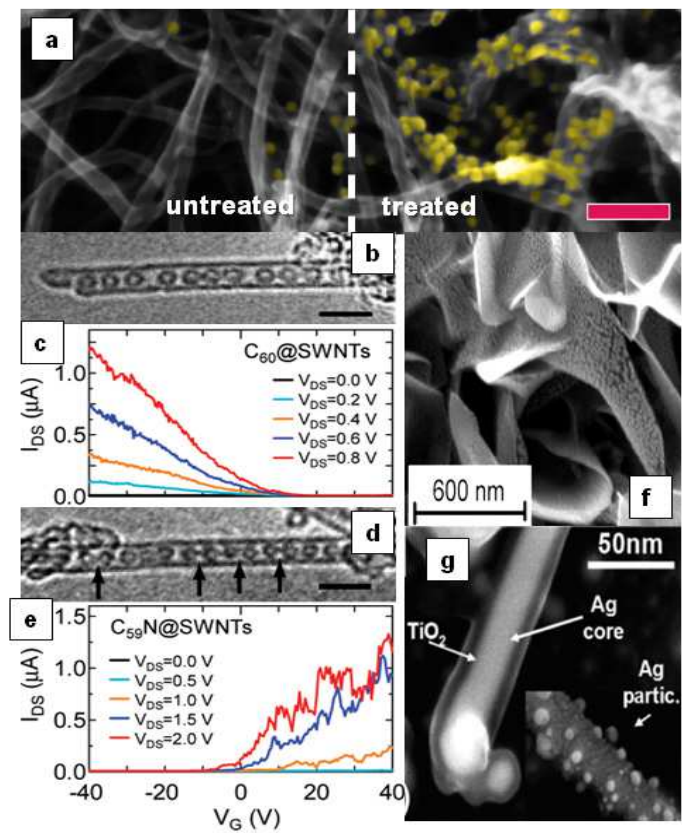} 
\caption{ \label{f22} Atmospheric plasma jet treatment improves decoration 
of MWCNTs with Au NPs (a); scale bar is 200 nm [reproduced from \cite{SamChemComm2013} 
by permission of the Royal Society of Chemistry].
C$_{60}$ encapsulated SWCNTs (b) show enhanced p-type characteristics (c) compared with pristine SWNTs,
whereas N-doped C$_{59}$N encapsulated SWCNTs (d) show n-type behavior (e) 
[reprinted with permission from \cite{Hatakeyama-C60-SWCNT}. 
Copyright \copyright (2008) American Chemical Society]. 
Vertical graphenes decorated with ordered self-organized arrays of Au NPs (f) 
[reproduced from \cite{ChemComm2012}  
by permission of the Royal Society of Chemistry]. 
Ag-TiO$_2$ core-shell nanofibers by PECVD (g) [reproduced with permission from \cite{BorrasPPAP07}. 
Copyright \copyright 2007 Wiley-VCH Verlag GmbH \& Co. KGaA, Weinheim].

} 
\end{center}  
\end{figure}

Potential damage to the supporting NSs due to the ion bombardment may be quite 
significant. This is why it is critical to select suitable plasmas and process parameter range.
For example, atmospheric-pressure plasma jet with a sub-mm spot size was used to 
treat multi-walled CNTs before depositing Au nanoparticles on them (Fig.~\ref{f22}(a)) \cite{SamChemComm2013}. 
This choice was dictated not only by the need of the 
selected-area decoration, but also by the fact that ion energies in a collisional 
sheath of atmospheric-pressure plasmas are much lower than 
under low-pressure conditions.

Importantly, the atmospheric microplasma 
jet produces a three-dimensional
microfluidic channel on dense arrays of vertically
aligned carbon nanotubes, which confines Au nanodot
aqueous solution. As can be seen in panel (a), the Au nanoparticles 
only attached to the surfaces of MWCNTs that were 
exposed to the plasma jet \cite{SamChemComm2013}. A white 
dashed line shows the rough boundary of the plasma-exposed area. 
The resulting hybrid 3D nano-structure is then used 
as an effective microscopic area-selective sensing
platform based on surface-enhanced Raman scattering (SERS). 

Using hot-filament plasma CVD, it also recently became possible to 
decorate vertically aligned MWCNTs with dense patterns of small Si nanodots. 
These hybrid structures are very stable in reactive chemical 
environments and significantly improve performance of Li-ion batteries
by combining excellent electrochemical response of Si nanoparticles and electronic conduction 
of MWCNTs \cite{GohierAMat2012}.

Figure~\ref{f22}(b-e) shows exotic hybrid structures where 0D  
fullerenes C$_{60}$ (b) and azafullerenes C$_{59}$N (d) are encapsulated by 1D SWCNTs 
\cite{Hatakeyama-C60-SWCNT}. Substitution of a N atom 
in C$_{60}$ was achieved by a mild exposure of the fullerenes to low-density nitrogen plasmas.
The plasmas have been chosen deliberately to have a relatively wide sheath 
to generate a small number of N$^+$ ions with the energies above the threshold of 
substitutional incorporation into the C$_{60}$ network. 
The SWCNTs were also synthesized using plasmas (Sec.~\ref{growth}).
 
Remarkably, the plasma ion-irradiation method was also 
used to encapsulate the C$_{60}$ and C$_{59}$N structures into the SWCNTs.
As a result, while the C$_{60}$ - SWCNT structure showed the typical p-type
characteristics (panel (c)), the C$_{59}$N - SWCNT hybrid clearly revealed the 
n-type properties as shown in panel (e).  
Both classical and density functional theory MD (DFT MD) simulations elucidated the mechanism for this process, albeit for Ni incorporation 
to produce metallofullerenes \cite{NeytsMetallofull1,NeytsMetallofull2}. 

Figure~\ref{f22}(f) shows the possibility to decorate the basal surfaces of vertical graphenes 
by Au NPs very uniformly, from the bottom to the top 
\cite{ChemComm2012}. The plasma exposure also resulted in the formation of regular, self-aligned 
linear arrays of Au nanoparticles within a certain distance from the edges. No masking or pre-patterning 
were required. The mechanism of this self-alignment is presently 
unclear. It may be enabled by the interplay of the nanoscale plasma interactions 
(Sec.~\ref{nano-plasma-surface}), non-uniform distributions of surface 
temperature and stress (similar to Fig.~\ref{f13}(a,b)), and plasma-modified adatom 
capture zones near the nanosheet edges. 

Figure~\ref{f22}(g) shows two different 
arrangements of hybrid Ag-TiO$_2$ core-shell inorganic nano-architectures 
produced by the plasma oxidation of Ag and the 
plasma deposition of TiO$_2$. The growth mechanism of these  
hybrid fiber-like architectures \cite{BorrasPPAP07} suggests the importance of 
several plasma-specific effects discussed above.  
There are many other examples of the plasma-assisted 
production and processing of multidimensional nanoarchitectures, 
e.g., using various combinations of top-down and bottom-up 
approaches. 

Physical properties (e.g., electron transport and interface phenomena) 
and applications (e.g., in electronic, energy storage, and sensing devices) of hybrid nanoarchitectures 
are discussed elsewhere (see, e.g., \cite{BanhartNanoscale2009,AjayanAMat2012} and references 
therein).

\subsection{\label{3D-nanostructures} Three-dimensional nanomaterials}   

Let us now consider nano- and micrometer objects with three-dimensional features and the relevant 
plasma-based approaches to produce and process them.  
The first two groups of examples in Sec.~\ref{etching} and \ref{Si-nanotips} are related 
to nanoscale plasma etching, one of the most 
traditional plasma applications scaled down to the nanoscale. 
This will be followed by two more common groups of nano-carbons, namely 3D cone-like
structures and nanodiamond in Sec.~\ref{carbon-nanotips} and \ref{nanodiamond}.
This subsection will conclude with the discussion of soft organic matter 
in Sec.~\ref{soft-matter} and biological objects in Sec.~\ref{bio}.

\subsubsection{\label{etching} Nanoscale plasma etching}

Plasma etching is one of the most established plasma technologies 
and the basic etching mechanisms of 
macroscopic (e.g., large-area surfaces) and microscopic 
(e.g., semiconductor microfabrication) features are well 
understood \cite{Coburn-etching}. 
Research in this area has resulted in an 
enormous number of research publications, reviews, patents, etc. which are outside the
scope of this review. 

Continuous miniaturization of integrated circuitry lead to 
 inevitable shrinking of the feature sizes. 
As the features entered the nanometer domain, new challenges emerged. 
In particular, it became critical to produce suitable 
masks with nanometer dimensions and to {\it precisely} transfer these nanoscale patterns 
to the underlying substrate. Since the feature sizes are already nearing the 10-nm 
domain, the precision of the pattern production and transfer needs to be 
in the sub-nm, and ultimately, atomic range. 
This is why we focus on the mechanisms 
of the {\it nanoscale} and {\it atomic-level} plasma etching.

\begin{figure}[t]
\begin{center}
\includegraphics[width=7.5cm,height=9.5cm,clip]{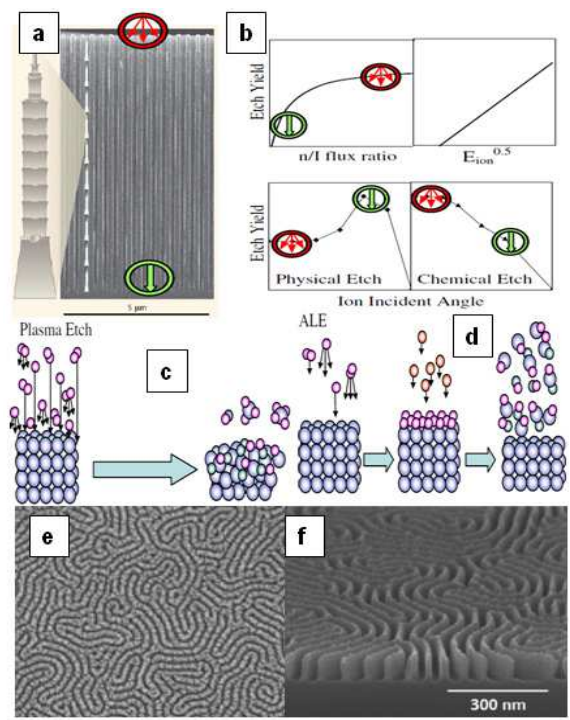} 
\caption{ \label{f23} High-aspect-ratio trenches etched in Si (a) 
\cite{Joubert,JaneChangJPD2011}; Taipei 101 building is shown for 
comparison of the aspect ratio. Effect of the flux ratio of neutral and ion species and ion energy 
on the etch yield (top row) and competition of the physical and chemical etch mechanisms (bottom row) (b); 
mechanisms of traditional (c) and atomic-layer (d) plasma etching \cite{JaneChangJPD2011}. 
Self-assembled polymer nanopatterns (e) transferred to Si (f) 
by nanoscale plasma etching \cite{Morris}.  
Copyright \copyright IOP Publishing. Reproduced from \cite{JaneChangJPD2011} 
(panels (a-d)) and from \cite{Morris} (panels (e,f)) 
by permission of IOP Publishing. All rights reserved. }
\end{center}   
\end{figure}

To better understand the fundamental differences between the conventional and the
nanoscale plasma etching, let us review Fig.~\ref{f23}(a-d). 
The conventional  mechanism of the plasma etching (c) involves a combination 
of interaction of ionic, etching, and depositing reactive species (e.g., F, Cl, H, CF$_x$, ions, etc.) 
with the surface. The rate of surface etching ${\cal R}_{\rm etch}$ is determined by a complex interplay of 
several factors such as fluxes of multiple species, reaction probabilities, energy thresholds, etc.
Under conditions when the effect of depositing species is insignificant, 
the etching rate is \cite{JaneChangJPD2011}:
\begin{equation}
\label{etchrate}
{\cal R}_{\rm etch} = \frac{J_i [A_s(\sqrt{E_{\rm ion}} - \sqrt{E_{\rm th,s}}) + B_s(\sqrt{E_{\rm ion}} - \sqrt{E_{\rm tr,s}})]}
{1 + J_i [A_s(\sqrt{E_{\rm ion}} - \sqrt{E_{\rm th,s}}) + B_s(\sqrt{E_{\rm ion}} - \sqrt{E_{\rm tr,s}})] /J_e \nu_{es}S_{es}},
\end{equation}
where $J_i$ and $J_e$ are the fluxes of ionic and etching species, $E_{\rm th,s}$ and $E_{\rm tr,s}$ are transition and 
threshold energies, while $A_s/B_s$ and $\nu_{es}$ refer to volume of substrate/polymer removed as a function of ion energy and etching species flux,
respectively. In Eq.~(\ref{etchrate}), $S_{es}$ denotes the sticking probability of the etching species to the surface.   

The reactive etching species bond with the substrate 
atoms and form volatile compounds thereby removing matter from the surface.  
It is presently possible to produce very deep trenches in Si wafers, with the aspect 
ratios compared to some of the tallest 
buildings in the world (Fig.~\ref{f23}(a)). The thickness of these channels is expected
to reduce into the low-10 nm domain, which is much less than the mean free path 
of the reactive species used for the etching. 
This creates a major problem of {\em nanoscale localization} of the plasma etching, which is 
quite similar to the effects of localization of the plasma-surface interactions 
involved in the nucleation of SWCNTs in Sec.~\ref{nucleation}. 

With regard to the high-aspect-ratio features, several important questions arise:
how to ensure that the trenches are straight (i.e., highly-anisotropic etching), how to 
deliver the etching species deeper and deeper as the features become longer and longer, 
how to ensure that these species will 
only react with the bottom surface of the trench and will not cause any damage to its side-walls, 
and how to extract the etching products out of the trench without them being redeposited on the channel walls?
More importantly, how to control the interactions of the plasma species with the 
surface in the nanometer {\em confined} space?  
  
The effect of the spatial confinement in this case may appear even more 
critical compared to Fig.~\ref{f10} because of the rigid vertical walls of the trench 
which strongly impede the mobility of the species in the gas phase.
However, the walls of the trench may provide the effective species 
delivery and escape ``slides", in a sense similar to  
evacuation of passengers from an aircraft after emergency landing.          
This in turn creates a very interesting physical similarity to the 
growth of SWCNTs in Fig.~\ref{f10}(d) where the growth rates are largely 
determined by rates of surface diffusion of carbon atoms.   
A quite similar diffusion of metal atoms effects  
metallization of micrometer-sized features \cite{AndersJPD07}.  

The mechanisms of the nanoscale plasma etching, especially in 
long trenches with the thickness of $\sim$10 nm, are not  fully understood. 
However, there is a consensus on the crucial role of the 
reactive plasma ions which is explained in Fig.~\ref{f23}(a,b). 
First, the ions can penetrate much deeper into the trenches 
because of the higher energies, better 
transport directionality and longer free paths compared to 
the neutral radical species. The ion incidence angles are crucial and they 
are determined (similar to the ion interactions with CNT arrays 
in Fig.~\ref{f9}(e)) by the plasma sheath width (hence, the ion energy), the 
microscopic topology of the electric field near the channel mouth (similar to
 Fig.~\ref{f9}(c)), and the gas pressure.  

For virtually any pattern of the etching features, the 
plasma parameters can be adjusted to maximize the ion incidence along the channel axis thereby 
enabling deep penetration into the features.    
Near the feature mouth, the impact of neutral species usually leads to the more 
isotropic etching, while the ion impact leads to the highly-anisotropic etching 
at the bottom. This is also consistent with the dependence of the 
etch yield on the ion energy in panel (b). However, since the density of neutrals in 
weakly-ionized plasmas is much higher than the ion density, the increase 
of the ion-to-neutral density ratio leads to the better etch anisotropy, yet at the 
expense of the etching rates.        

Therefore, the ions should be delivered straight to the bottom of the 
channel, without impacting on the side-walls. This leads to the 
yet unsolved problem of long-range ``ion shooting" with nanometer 
precision including the long flight 
through the nanometer-thick channel. 
One of the obvious complications is that the features get charged during the plasma exposure and the
dynamically varying charge distribution is very difficult 
to measure or compute. These charges distort the ion trajectories from straight downward 
paths thus compromising the etch anisotropy. 

This is why one of the most popular modern approaches 
to simultaneously improve the etch anisotropy and quality is to use a pulsed DC bias customized to 
let the surface charge dissipate between the ion flux pulses \cite{pulsed-etching}. 
This strategy was used to fine-tune the ion flux distribution along the walls and bottom of 
long dielectric nanopores to enhance the process precision and throughput in 
nanopore processing and nano-template-assisted nanoarray production \cite{XX-template-APL08}. 

If necessary, the ions can be removed from the etching process by placing a
porous carbon plate (aperture) between the plasma and process chambers,
where both positive and negative ions can effectively neutralize.
This process is known as neutral-beam etching and in many cases 
shows very high etching precision, such as nearly defect-free atomic-precision 
sidewall etching \cite{Samukawa1}. The neutral beam etching was recently used 
to fabricate two-dimensional arrays of silicon nanodisks for quantum dot solar cells
mentioned in Sec.~\ref{QD} \cite{Samukawa2}.

An important feature of the plasma etching is the  
synergism between the effects of neutral, ions, and photons, which 
leads to the much higher etching rates compared to the cases when these species 
are used separately.  
For example, Si etching by fluorine-based compounds may be dramatically 
enhanced by ion-assisted effects \cite{JaneChangJPD2011}, which 
are quite similar to those in carbon nanotube 
nucleation and growth (Secs.~\ref{nucleation} and \ref{growth}). 
A similar synergism is expected to play the key role in the 
plasma-enhanced atomic layer etching (ALE) sketched in Fig.~\ref{f23}(d). 

The ALE is based on highly-reproducible removal of 
atomic monolayers using a sequence of self-limiting 
etching steps \cite{JaneChangJPD2011}. Initially, species B
are deposited to fully terminate the surface to form a self-limiting 
monolayer on the surface made of etching material A. 
Species B merely form bonds with species A without binding to each other. 
The strength of the B-A bonds \cite{CQSun} should ideally be higher compared to the loosened 
bonds between the first and the second layers of atoms A.

The plasma ions may be used to energize the B-A bonded species and possibly combine with them 
to form volatile products and gently etch only one atomic layer of atoms A.
The ion energy should be adjusted to be less than the sputtering energy, yet should be enough 
to facilitate the formation of the volatile compounds and then detach them from the 
surface after breaking the already loosened bonds between the first and the second layers of the 
etched material. 

This process should also be self-limiting. The energy of the plasma ions 
and their affinity to atoms A should also be insufficient to cause 
any damage to the second layer. The plasma-assisted ALE 
was studied both experimentally and 
theoretically for a relatively small number of materials \cite{ALD1,ALD2}.

A new trend is using self-organized 
masks, e.g., made of a self-assembled monolayer of polymer nanospheres, 
followed by the precise pattern transfer using highly-selective, 
highly-anisotropic plasma etching (Fig.~\ref{f16}(a,b)). 
Another example in Fig.~\ref{f23}(e,f) evidences  
precise transfer of  self-assembled patterns of 
block PS-PMMA copolymers onto a Si substrate \cite{Morris}. 
The plasma etching takes place in nanoscale gaps  
between the polymer nanoribbons and deepens into the Si wafer by $\sim$100 nm.

Despite excellent precision in pattern transfer, this process requires mask preparation 
before the plasma etching. This raises the obvious 
question if the first step can be avoided and the self-organized 
masks could be formed during the plasma process.  
The following subsection gives a positive answer to this question.

\subsubsection{\label{Si-nanotips} Self-organized arrays of inorganic nanotips } 

\begin{figure}[t]
\begin{center}
\includegraphics[width=7.5cm,clip]{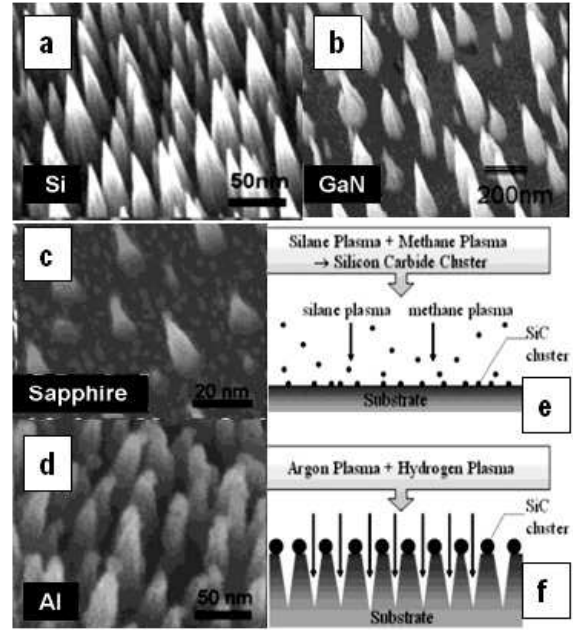} 
\caption{ \label{f24} 
Self-organized arrays of inorganic  nanotips fabricated by self-masked plasma etching on: 
single-crystalline Si (a);  epitaxial GaN on sapphire (b); 
 sapphire (c); and  Al (d). Two main process steps (e,f) 
[reprinted with permission from \cite{LCChenNanoLett2004}. 
Copyright \copyright (2004) American Chemical Society].  } 
\end{center}  
\end{figure}

Fabrication of self-organized 
arrays of Si nanotips (Fig.~\ref{f2}(b)) by self-masked plasma etching has  been 
mentioned in Sec.~\ref{basicA}. This effect is fairly generic and was also demonstrated for a large number 
of materials systems including Si, GaN, Sapphire, and Al (Fig.~\ref{f24}(a-d)).  
This self-masked dry etching technique can be implemented over large surface areas and at low process temperatures.
The etching leads to good uniformity of 3D solid nanostructures, in terms of their sizes, shapes, and 
positioning on the substrate \cite{LCChenNanoLett2004}. 

The growth mechanism of these nanoarrays is explained in Fig.~\ref{f24}(e,f). 
To produce self-organized SiC nanomasks, plasmas in a mixture of silane and methane 
gases are used as sketched in panel (e). As a result, a pattern of small SiC nanoparticles forms 
on the Si surface. This process is quite similar to the growth of self-organized arrays 
of SiC QDs mentioned in Sec.~\ref{Nanoarrays}. The size of these NPs determine 
the size of the apex of the conical structures that form during the second stage as shown 
in Fig.~\ref{f24}(f). During this stage, plasmas of Ar$+$H$_2$ mixtures are 
used. The degree of size and positional uniformity of the SiC NPs over the surface 
determine the quality of the resulting Si nanoarrays.  

According to the results of several groups, the quality and effectiveness of the plasma etching is the 
best when both Ar and H$_2$ gases are used 
\cite{LCChenNanoLett2004,LCChenNatureNT2007,ShiehJPD07}.   
This further evidences the synergistic effects of the ionic (Ar$^+$ in this case) 
and reactive radical (H in this case) plasma species during the
etching process (Sec.~\ref{etching}). 

This also implies the competition between the
physical sputtering, chemical etching and material redeposition on the surface during the 
nanoarray formation. For example, in panels (b), (c), and (g), one can 
spot small cones between the large ones. They may be formed 
through the ion shadowing effects \cite{ShiehJPD07}.
Another possibility is redeposition of material followed by the 
nucleation of the secondary nanoarray between the large cones similar to Fig.~\ref{f13}(c). 
Formation of fairly rare arrays with clear-cut faceted structures suggests that 
nucleation followed by the crystal reshaping under non-equilibrium conditions 
(similar to Fig.~\ref{f4}) may dominate over the etching 
effects \cite{APL2009-Si-Shuyan,IgorNT2010-Si-cones}. 
Conversely, very dense arrays of Si nanograss or nanotips with 
very high aspect ratios are believed to form predominantly through the
Ar$^+$ ion-assisted reactive etching of Si surface by reactive H radicals which form 
volatile SiH$_x$ products upon interaction with the surface.

Compared to the approach of Sec.~\ref{etching}, the self-masking etching 
relies on self-organized patterns of small clusters that serve as masks. The smallest 
tips of the sharp conical Si structures could be $\sim$1 nm \cite{LCChenNanoLett2004}.
This means that the SiC clusters of a comparable size can be very effective as 
etching masks. On the other hand, a dynamic interplay of physical sputtering and chemical etching 
effects may also lead to the nanotip sharpening and the formation of the structures with 
different apex angles. This ability is of particular interest for customized, bio-inspired  
biophotonic arrays \cite{LCChenMatSciEng2010}.  

These arrays represent regular conical structures where the refractive index ($n(z)$) effectively varies 
from the substrate towards the tip of the nanocones. This makes it possible to create 
sub-wavelength anti-reflection surfaces where non-uniformity of $n(z)$ can be approximated 
by several layers with slightly different refractive indexes $n_i$ \cite{LCChenMatSciEng2010}. 
The effective refractive index of such anti-reflection arrays $n_{\rm eff}$ can be fairly 
accurately calculated from 
\begin{equation}
\label{EMA}
\sum_{i=1}^k f_i \frac{n_i^2 - n_{\rm eff}^2}{n_i^2 + 2n_{\rm eff}^2} =0
\end{equation} 
using Bruggeman's effective medium approximation \cite{BruggemanEMA}. In Eq.~(\ref{EMA}),
$f_i$ denotes the corresponding volume fraction of $i$-th layer,  which decreases 
towards the summit of the nanotips.  

However, the features produced through the self-masked etching are usually not as straight as 
in Fig.~\ref{f23}(a), which suggests that the anisotropy of this etching process 
requires improvement. Nevertheless, this etching can also be considered as an example of 
nanoscale plasma etching because of the localization of the plasma-surface interactions 
within the open spaces between the small clusters that form a self-organized mask.     
This undoubtedly one of the most interesting plasma-enabled 
nanofabrication processes still warrants a significant theoretical and numerical modeling effort. 

It was recently demonstrated that a p-type Si wafer 
attains a high-quality {\it p-n} junction and a very strong photovoltaic response 
only when nanoconical structures are formed on its surface \cite{AdvEnMater2011}. 
The available interpretations involve several effects that may lead to the {\it p}-to-{\it n} type conductivity 
conversion \cite{XuJPD2011-SpecIssue}. However, the question why this effective 
conversion only takes place when a self-organized nano-texture on the Si surface is formed, 
remains open. Nevertheless, this result is a clear demonstration that 
surface-supported self-organized Si nanoarrays produced through the effects of nanoscale 
plasma-surface interactions, may also lead to the unusual and promising ways of achieving 
specific materials properties for the pursued applications. 

Indeed, the formation 
of {\it p-n} junctions for crystalline Si-based solar cells is usually implemented 
as a separate thermal diffusion process which raises significant environmental 
concerns \cite{XuJPD2011-SpecIssue}. In this example, the high-quality {\it p-n} 
junction was formed, and subsequently, the effective photovoltaic response 
of $\sim$12 \% was achieved without the seemingly unavoidable diffusion process.  
The photoconversion efficiency of this type of solar cells was subsequently 
improved to above $\sim$18.5\% by precise (and also plasma-based) surface 
passivation by SiN layers \cite{XuJPD2011-SpecIssue}
(please also see discussion on the role of surface passivation 
of nanocrystals in Sec.~\ref{QD}).

\subsubsection{\label{carbon-nanotips} Carbon nanotips }

\begin{figure}[t]
\begin{center}
\includegraphics[width=8.5cm,clip]{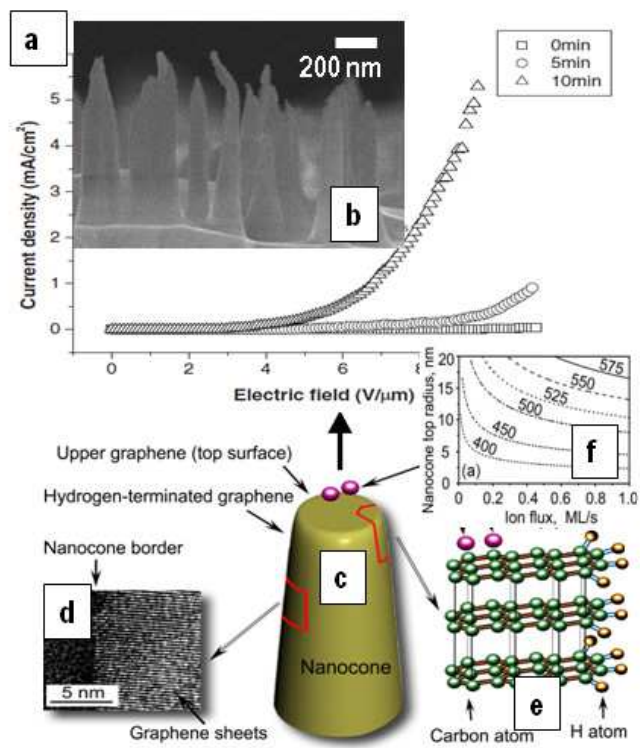} 
\caption{ \label{f25} Field emission (a) from diamond nanotips produced by PECVD (b) 
improved by N$_2$ plasma immersion ion implantation 
[Copyright \copyright IOP Publishing. Reproduced from \cite{Diam-Nanotip-FE} 
by permission of IOP Publishing. All rights reserved].
Structure and mechanism of self-sharpening of single-crystalline carbon nanocones (c) made of a vertical 
stack of graphene layers terminated by hydrogen at the edges (HRTEM image (d) and atomic sketch (e)); 
nanocone top radius versus ion flux at different surface temperatures (f) 
[reprinted with permission from 
\cite{APL07-Nanocones} Copyright \copyright (2007), American Institute of Physics]. } 
\end{center}  
\end{figure}

Carbon nanotips (CNTPs) is a typical example of 3D nanostructures. 
Several plasma-based approaches have produced CNTPs in various shapes (e.g., cone-like, needle-like, pyramid-like, etc.), 
aspect ratios, and bonding states. Typical examples of carbon nanotips are shown in Fig.~\ref{f25}. These structures differ 
by the structural stacking of carbon atoms and show examples of 
$sp^3$ diamond (a,b) and $sp^2$ single-crystalline (c-f) nanotips. 
Morphologically, these nanostructures are quite similar to the Si nanotips of 
Sec.~\ref{Si-nanotips} and in some cases feature comparable aspect ratios and array
densities (see, e.g., Fig.~\ref{f13}(c)). 

However, owing to the much higher bonding strength, chemical stability, 
and melting temperatures compared to Si, it is very difficult to produce 
such nanostructures by the direct plasma etching. 
This is why common plasma-based approaches to synthesize CNTPs involve various 
combinations of catalytic growth, chemical and physical vapor deposition, 
physical sputtering, and chemical etching effects. In most cases, the 
growth temperatures are significantly lower compared to similar thermal processes. 
Plasma-assisted techniques have also been used to customize  
the nanotip properties for specific applications.           
     
For example, a combination of hot-filament and plasma CVD, complemented with ion flux 
control using DC bias was used to produce arrays of amorphous carbon nanotips
\cite{BBWang-Carbon06}.  
Graphitic carbon films are commonly used to facilitate the nanotip formation and also 
serve as conducting support in electron field emission (EFE) devices. Historically, 
the field emission display technology has been perhaps the most common targeted application 
of the carbon nanotips. This application demands high-aspect ratio and sharp tips 
for better electron emission, conducting channels in the nanostructures to sustain electric current, 
and also excellent structural stability to prolong the device lifetime.

From the structural stability perspective, diamond nanotips are very attractive. 
However, it is challenging to synthesize diamond nanostructures with high aspect ratios 
and acceptable conductivity.  
Both these problems were solved by using plasmas   
\cite{Diam-Nanotip-FE}. First, high-quality $\sim$1 $\mu$m-tall diamond nanotips with the base 
width of $\sim$150--200 nm, firmly standing on the diamond/Si substrate were 
produced (Fig.~\ref{f25}(a,b)). This carbon nano-phase is thermodynamically less stable than graphitic carbons 
and usually requires very high temperatures and pressures. 

These conditions were partially avoided by using 
relatively high-power microwave plasmas of Ar$+$CH$_4+$H$_2$ gas mixtures
which created a strongly non-equilibrium environment even at pressures 
$\sim$0.14 Torr, which are much lower than typically used in 
PECVD of diamond films ($\sim$1--10 Torr or even higher). 
The diamond nanotips showed a good $sp^3$ structural quality and phase purity.
It is thus not surprising to see a nearly zero electron emission current in Fig.~\ref{f25}(a).  

This problem was also solved by implanting nitrogen ions produced in N$_2$ plasmas. 
This relatively low-dose ion implantation has led to the effective substitutional doping by N atoms, 
rearrangements of $sp^3$ atomic bonds into $sp^2$ in localized areas, and presumably, the 
formation of conductive paths along the nanotips. As a result, the electron emission 
was not only enabled (5 min treatment in panel (a)), but also 
 dramatically improved by precisely dosing the exposure to nitrogen ions (10 min).  
Very competitive EFE characteristics were achieved without any 
significant structural damage to the nanotip exterior structure.     

The CNTP growth and shaping is a kinetics-driven process under non-equilibrium 
conditions where several plasma-specific effects play a key role. 
Figure~\ref{f25}(c-f) shows single-crystalline CNTPs made of horizontally stacked
 graphene sheets (panel (d)) which were synthesized in low-pressure 
Ar$+$CH$_4+$H$_2$ plasmas. High rates of dissociation of 
H$_2$ in the plasma enable effective termination of the open egdes of the 
graphene layers (panel (e)), which is an important pre-requisite of structural stability 
of this type of single-crystalline nanostructures. 

These structures also self-sharpen during the growth to form high-aspect-ratio 
tapered needle-like structures \cite{APL07-Nanocones}. This 
self-sharpening is related to the plasma-specific effects, most prominently, 
the ion fluxes to the flat top of the nanocone (panels (c,f)). 
This is yet another manifestation of the major  role of the
{\em nanoscale} plasma-surface interactions.   
 
The variation of the carbon atom
density  on the flat top (a few 
nm across) of the nanotip in Fig.~\ref{f25}(c) 
\begin{equation}
\label{C-nanotip-eq}
dn/dt = \Psi_i - \Psi_d - \Psi_e - \Psi_c 
\end{equation}
is determined by the dynamic balance of the incoming $\Psi_i$ and outgoing diffusion $\Psi_d$, thermal $\Psi_e$,  
and ion-induced $\Psi_c$ detachment fluxes of carbon material.  
The critical radius for graphene nucleus formation is derived using (\ref{C-nanotip-eq}) under 
the steady-state, quasi-equilibrium conditions, and is shown in Fig.~\ref{f25}(f).
The top radius of the nanotips is in a very good agreement with the experimental results  
 and can also be controlled by the microscopic ion flux 
and the surface temperature \cite{APL07-Nanocones}.   

Importantly, ion bombardment plays a critical role. 
Indeed, ion sputtering determines the {\it a}-CNTP shapes \cite{BBWang-Carbon06} 
consistently with Sigmund's incidence-angle-dependent 
sputtering theory \cite{Sigmund-sputtering}. 
In Fig.~\ref{f25}, a moderate negative bias improves the
structural quality of diamond and single-crystalline $sp^2$ nanotips,
possibly due to the ion-assisted crystal 
densification \cite{KHMuller1986}.   
Additionally, a high-voltage bias leads to effective nitrogen 
ion implantation which enables the field emission from the diamond nanotips 
(b) without any significant structural damage. 

In addition to the strong electron field emission, amorphous carbon nanotips 
have also recently proved effective as sources of spectrally tunable 
photoluminescence (PL) emission \cite{BBWang-Carbon2012}. Importantly, plasma-specific effects 
may help adjusting the prevailing factors that determine the PL
wavelength and intensity. 

For instance, one common possibility 
to generate PL emission is through the radiative recombination 
of the electrons and holes in the band tail of $sp^2$ carbon clusters,
which are created by photo-excitation of $\pi$ and $\pi^*$ states \cite{Robertson96}.   
However, the size of spatial confinement of luminescence centers 
in $sp^2$ carbon clusters should be in the 0.94-1.15 nm range to enable 
direct $\pi - \pi^*$ band transitions \cite{Fuele}.
The size $L_{\rm cl}^{sp^2}$ of these clusters can be estimated using 
\begin{equation}
\label{clustsize5}
L_{\rm cl}^{sp^2} = (\alpha_1 + \alpha_2 \lambda) \frac{I_G}{I_D},
\end{equation}
where $\lambda$ is the excitation wavelength, and $\alpha_{1,2}$ are constants. 
Here, $I_G$ and $I_D$ are the intensities of the G and D peaks in 
Raman spectra, respectively. Here we recall that the G peak determines the degree of graphitic order while the 
D peak quantifies the degree of disorder and owes to the presence of defects, clusters etc. in 
the structure. 

From Eq.~(\ref{clustsize5}), one can see that a larger number of smaller structural 
defects (which corresponds to the higher intensity of the D peak) may be advantageous to 
generate PL due to the direct $\pi - \pi^*$ band transitions. 
The smaller $sp^2$ cluster sizes also lead to shorter bonds between carbon atoms and 
consequently, a wider bandgap. The increased separation between the $\pi$ and $\pi^*$ states
leads to stronger PL emission at higher photon energies 
(e.g., shorter wavelengths, typically in the UV range). 
The reduced spatial localization of emission centers also leads to broader band tail 
distribution, which increases the PL quenching temperature $T_q$ thereby increasing the 
PL intensity according to 
\begin{equation}
\label{PLint6}
\eta_{\rm PL} \propto \exp \left( - \frac{T}{T_q} \right), 
\end{equation}
where $T$ is the measurement temperature. In this way it is possible to tune the emission 
wavelength and intensity to achieve stable room-temperature luminescence,
which is particularly important for optoelectronic applications.  

There are other mechanisms to tune the photoluminescence from carbon nanotips, e.g., by 
controlling the hydrogen and nitrogen content. Some of these mechanisms are similar to
the mechanisms of PL from C-dots briefly discussed in Sec.~\ref{C-dots}. Addition of small amounts of nitrogen
for example lead to $sp^3$ C-N bonds which enable transitions between the $\pi^*$ and 
lone pair (LP) valance bands. These transitions produce optical emission with a longer wavelength 
compared to the direct $\pi - \pi^*$ band transitions, typically in the green range of the optical spectrum.   

As discussed earlier in this review,
plasma effects are particularly suited to generate a significant number of tolerable defects 
and also in specific locations. Moreover, high rates of dissociation of nitrogen 
can be used to control the degree of carbon nanotip doping, and hence, the presence of nitrogen-related 
radiative sites. This is why it is not surprising that plasma-produced and/or processed 
carbon nanotips are quite promising for the development of next-generation 
optoelectronic devices with tunable spectral responses \cite{BBWang-Carbon2012}.

\subsubsection{\label{nanodiamond} Nanodiamond}

Let us consider three common types of nanodiamond materials, namely nanocrystalline
diamond (NCD), ultrananocrystalline diamond (UNCD), and diamond nanoparticles. 
NCD films contain clearly-faceted diamond nanocrystals, typically 
of a columnar shape and a size of several hundred nanometers. Morphologically, NCD 
is quite similar to columnar microcrystalline Si films 
with a large percentage of the crystalline phase. 
UNCD is a morphological analogue of {\it nc}-Si  
of Sec.~\ref{QD} and features small diamond nanocrystals (typically 3--5 nm) 
randomly dispersed in an amorphous carbon matrix. The grain boundaries are typically less than 0.5 nm. 
Diamond nanoparticles can in turn be surface-supported or freestanding. 
Plasma processes were used in all 3 cases and  
in many cases led to superior outcomes.    

UNCD is optically transparent and is pursued for highly-conformal 
coatings of micro- and nanometer features. Excellent biocompatibility also makes UNCD ideal  
for microelectrode arrays in ``bionic eye" retinal implants. 
Relatively high-power-density microwave plasmas in hydrocarbon gas mixtures are typically used to synthesize UNCD.
Relatively high (e.g., $\sim$700--800$^{\circ}$C) temperatures and 
pressures (e.g., $\sim$1--10 Torr) are typical; yet they are typically much lower compared to thermal CVD.       
Detailed studies of optical emission during  the growth process have identified carbon 
dimers C$_2$ as possible BUs of ultrananocrystalline diamond \cite{UNCD-2001-Auciello}.
This emission is dominated by the $d^3\Pi$--$a^3\Pi(0,0)$ Swan band of C$_2$ species 
and is characteristic to UNCD of both laboratory and astrophysical origins \cite{GruenMRSBull2001}. 
Recent numerical studies reveal a range and specific roles of the plasma species involved in ultrananocrystalline diamond growth 
\cite{NeytsUNCD1,NeytsUNCD2}.

\begin{figure}[t]
\begin{center}
\includegraphics[width=10.5cm,clip]{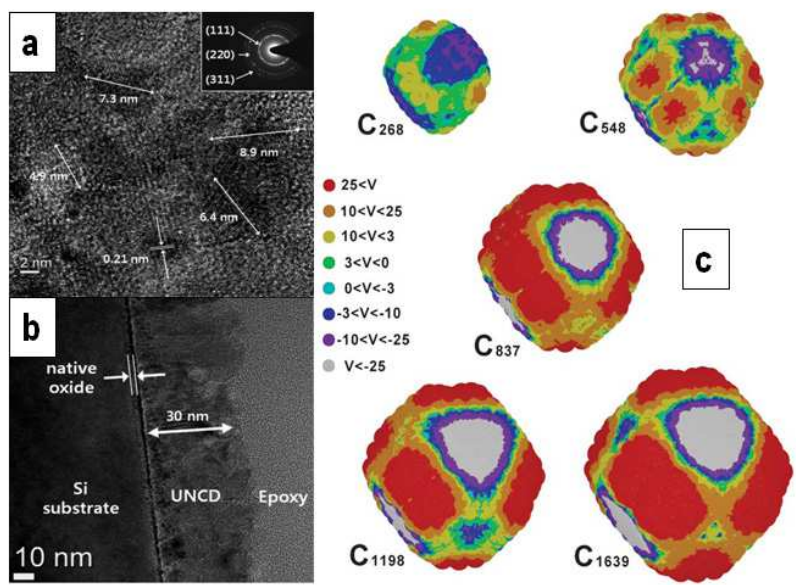} 
\caption{ \label{f26} 
Plan-view HRTEM (a) with diffraction pattern (inset) and (b) cross-sectional HRTEM of 
ultra-thin UNCD film grown by PECVD 
[reprinted with permission from \cite{UNCD-JAP2011} Copyright \copyright (2011), 
American Institute of Physics]. 3D distributions 
of the normalized surface electrostatic potential over the facets of small 
nanodiamond crystals of different sizes and shapes computed using the density functional based tight 
binding method with self-consistent charges (c) [Reproduced from \cite{BarnardJMC2007} by 
permission of the Royal Society of Chemistry]. }
\end{center}    
\end{figure}

The film thickness is typically in the micrometer range, and while 
the density of $sp^3$ nanocrystallites in an $a$-C matrix can be quite high, the 
material quality has often been an issue. Among them, achieving void-free, dense nucleation of 
small diamond nanocrystallites of fairly uniform size has been challenging.   
This issue has escalated with the continuously increasing demand for nm-thick UNCD films. 

To increase the density of nucleation, small nanodiamond seeds are used. 
These seeds help achieving continuous films and void-free interfaces. 
Figure~\ref{f26}(a,b) shows an example of an ultra-thin, mirror-smooth, and void-free ($\sim$30 nm) 
ultrananocrystalline diamond film \cite{UNCD-JAP2011} grown in CH$_4+$H$_2$ DC plasmas. The density of ultrasonically dispersed small (intentionally chosen 
in the typical UNCD size range of 3--5 nm) nanodiamond powder was in the 10$^9$--10$^{11}$cm$^{-2}$ range,
which is a typical surface density of QDs of Sec.~\ref{QD}. 
The plasma exposure has also led to markedly increased density of 
the diamond nanocrystals, presumably due to the secondary nucleation. 

This observation is supported by the reports on the effective control of the nucleation 
density and the nanocrystal sizes by the surface bias. 
In particular, it is commonly observed that the grain size decreases when a larger negative bias is used. 
This effect is believed to be due to the continuous secondary nucleation within the developing 
film \cite{Lifshits-Science-2002}, which is quite similar to the previous example. 

As mentioned above, small diamond nanocrystals nucleate and grow within an amorphous carbon 
matrix; this process was studied using MD simulations 
\cite{NeytsUNCD1,Grain-Shape-Bias}. These simulations 
explained the experimental observations and, in 
particular, showed that the plasma-produced C$_x$H$_y^+$ and H$_x^+$ ions penetrate into the 
carbon matrix and cause complementary effects. C$_x$H$_y^+$ ions 
induce nucleation of $sp^3$ clusters, which leads to the 
higher density of diamond nanocrystals. When the electric field is strong enough 
(e.g., bias exceeds -100 V), fairly deep incorporation of H$^+$ ions leads to 
the smaller crystal sizes. Moreover, stronger electric fields lead to 
the formation of $sp^2$ bonds in the nanocrystalline phase 
and a larger number of 3-coordinated C atoms in the amorphous phase.

Concurrent plasma-assisted nitrogen doping   
has dramatically (nearly 5-order-of-magnitude) 
improved the UNCD {\it n}-type electrical conductivity
\cite{UNCD-2001-Auciello}. 
This phenomenon was explained by the increased grain sizes (up to $\sim$12 nm) and grain 
boundaries (up to $\sim$1.5 nm), where the content of $sp^2$ bonds also increased. 
The  conduction path most likely formed along the $sp^2$-rich grain boundaries and involved 
$\pi$ states of carbon \cite{UNCD-2001-Auciello}. 

Free-standing nanodiamond crystals also  benefit from plasma effects.
Indeed, applications of nanodiamonds in quantum computing, medical diagnostics, 
drug delivery, and luminescent imaging require uniform surface  
functionalization. For example, surface termination of diamond nanocrystals 
with hydrogen eliminates dangling $sp^2$ bonds from the surface 
\cite{Nanodiamond-review-2012} and stabilizes the nanocrystal, similar to 
Fig.~\ref{f15}(a). Thermal hydrogenation usually requires temperatures 
$\sim$900$^{\circ}$C and even higher. These very high temperatures are typically well above 
the melting temperatures of small nanoparticles, which may be detrimental 
to the crystalline order in nanodiamond. 

Plasma hydrogenation has recently been 
shown as an effective remedy to this problem. 
This treatment leads to the formation of C--H bonds on the surface of 
diamond nanocrystals, which also showed the unusual hydrophobic properties
\cite{ND-plasma-hydrogenation}. Plasma exposure also induces  
hydrophilic response from diamond surfaces functionalized with ketone-related groups or 
carboxylic acids \cite{Nanodiamond-review-2012}.

An interesting point to stress is the synergistic effects of the small 
nanodiamond seeds and the plasma-specific effects in the ultrananocrystalline diamond 
nucleation and growth. In a sense, these 
seeds serve as catalytic supports for the void-free nucleation of the 
new diamond nanocrystals. As we have seen in Sec.~\ref{1D-Carbon}, 
it is possible to avoid using catalyst in the plasma-based nanotube growth.  
This sparks a similar question if it is possible to completely avoid seeding in the 
plasma-assisted UNCD growth. 

The most important consideration is how to 
produce or deliver similar seeds otherwise. One option would be to 
very carefully balance the supply of carbon atoms and the process parameters 
to ensure a sufficient density of ``natural" $sp^3$ cluster 
nucleation sites on the surface. As we have seen from the above examples, 
the best results in terms of UNCD film quality (Fig.~\ref{f26}(a,b)) are achieved when the 
nanodiamond seed density is in the 10$^9$--10$^{11}$cm$^{-2}$ range. 

Interestingly, plasmas in methane-hydrogen mixtures were shown to    
reproducibly generate very dense arrays ($\sim$10$^{12}$cm$^{-2}$), 
of size-uniform ($\sim$ 10 nm) orientationally-ordered (001) 
diamond nanocrystallites on Ir substrates \cite{Golding}.
These arrays emerge under far-from-equilibrium conditions when the surface bias is abruptly terminated 
followed by the rapid thermal quenching. 
It is thus worth exploring these ultra-dense arrays for UNCD seeding. 
However, it appears challenging to stabilize these metastable 
arrays and prevent a rapid island coalescence when more material 
is added during  the growth. This work is also very relevant to the 
earlier discussion of self-organized kinetic processes and emergent metastable patterns 
on plasma-exposed surfaces in Sec.~\ref{basicC} and \ref{self-org-patterns}.  

Another possibility is to assemble small amorphous nanoparticles in the ionized gas phase 
 and then transform them into crystalline UNCD seeds upon impact 
onto the growth surface. The possibility of transformation of carbon nano-onions (which can realistically form 
in the gas phase of both space and laboratory plasmas) into nanodiamonds upon surface impact 
has recently been demonstrated by {\it ab initio} numerical simulations \cite{MarksPRL2012}.
The transformation is kinetically driven over picosecond time 
scales. This strongly-non-equilibrium process is potentially a new approach for the UNCD growth. 
     
Plasma methods of nanodiamond synthesis are also energy efficient. 
For example, high-quality nanocrystalline diamond films were produced at a low temperature 
of $\sim$450$^{\circ}$C while the microwave plasma power was reduced in 50 times   
\cite{Nanodiamond-DRM2011}. This process is converse to the
UNCD case and requires suppression of the secondary nucleation, which promotes 
columnar growth.  

An interesting opportunity exists in the highly-controlled synthesis of 
size- and facet-specific nanodiamond crystals. {\it Ab initio} numerical simulations \cite{BarnardJMC2007}
show that by tuning the nanocrystal size and facet expression it is possible to very significantly redistribute 
electric charge over different facets as shown in Fig.~\ref{f26}(c). Consequently, facets with the higher 
electrostatic potential (e.g., red) will attract to the facets with the lower potential (e.g., gray).
This effect creates an opportunity to use these nanocrystals as building blocks and the arising 
polarization effects in the controlled assembly of wire-like, patterned, or compact nanodiamond 
assemblies. A quite similar effect was discussed for the plasma-generated silicon nanoclusters 
in Sec.~\ref{QD-simulation}. A plasma-based approach towards the synthesis of 
made-to-order diamond-like nano-carbons is discussed elsewhere \cite{RYP-Amanda}.

\subsubsection{\label{soft-matter} Soft organic matter}

Plasma-specific effects also enable many interesting features in the synthesis,  
surface structuring, and processing of soft organic nanomaterials such as polymers. 
This class of materials is  
highly-promising in many fields such as  
health care (e.g., biomedical implants or drug/gene delivery systems), 
organic optoelectronics (organic light emitting diodes), photovoltaics, and 
nanoelectronics. 

The common issues in the synthesis, processing, and device 
integration of such materials are high temperature sensitivity (due to
low melting temperatures), structural and morphological control at micro- and 
nanoscales, modification of surface energy to enable a certain functionality,  
gas-phase control of cross-linking and macromolecular building units, 
conformity of ultra-thin polymer layers to nanometer-sized surface features, 
and several others. Importantly, unique physical and chemical effects 
due to low-temperature plasmas
in many cases help resolving these issues.       

\begin{figure}[t]
\begin{center}
\includegraphics[width=9.5cm,clip]{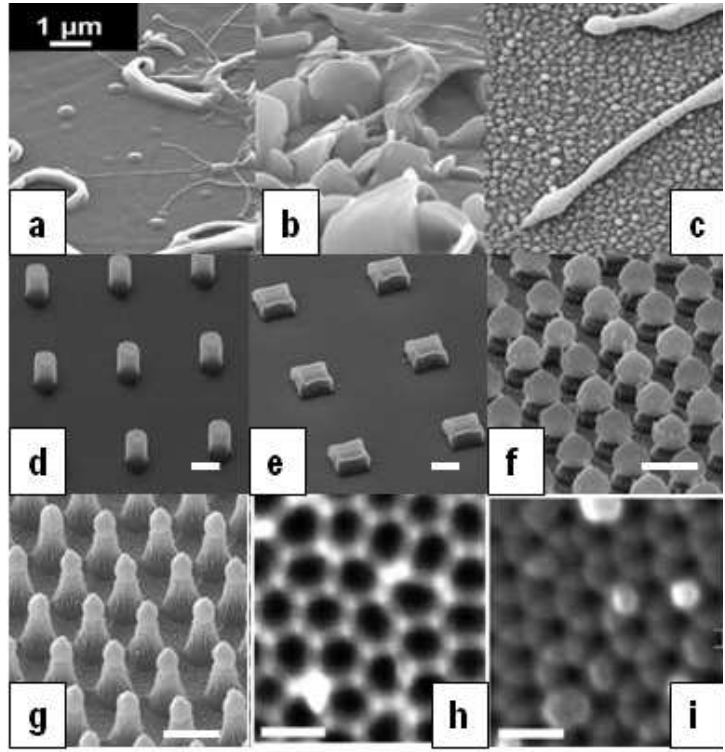} 
\caption{ \label{f27} Plasma nano-polymers. 
Osteoblast cells on ribbon-like (a), petal-like
(b), and nanodotted (c) polymers nanostructured in CF$_x$ plasmas 
[reproduced with permission from \cite{DiMundoPPAP2010}. 
Copyright \copyright 2010 Wiley-VCH Verlag GmbH \& Co. KGaA, Weinheim].
Ordered nanoarrays [panels (d,e) reproduced with permission from \cite{RossiAMat2007}
Copyright \copyright 2007 Wiley-VCH Verlag GmbH \& Co. KGaA, Weinheim; panels   
(f,g) reproduced from \cite{RossiJPD2007} by permission of IOP Publishing. 
Copyright \copyright IOP Publishing. All rights reserved] produced by plasma etching.
Conformal coating with nanometer precision allows tuning nanopore openings in 
nanoporous templates (h,i) [reproduced from \cite{Vasilev2010} with permission 
of the Royal Society of Chemistry]. 
Scale bar of 1$\mu$m in panel (a) applies to all panels (a-c). Other scale bars:  200 nm (d,e); 
1$\mu$m (f,g); and 200 nm (h,i). }
\end{center}   
\end{figure}

The two most common examples include 
plasma-produced polymers (commonly referred to as {\em plasma polymers}) for 
surface coatings with nanometer dimensions as well as precise control of 
surface energy by nanoscale surface texturing and/or functionalization. 
Owing to the very high reactivity of 
the plasma, these effects can be achieved even at room temperature. 
Figure~\ref{f27} shows typical examples 
of low-temperature nano-structuring of polymer surfaces (panels (a-g)) and conformal deposition 
of ultra-thin plasma polymers on nanometer-sized features (panels (h-i)).

The  plasma polymerization is unique because of the large
variety of reactive species produced  through the plasma-assisted fragmentation 
and remodeling of monomer precursors. As a result, a
cocktail of original precursors, reactive radicals, non-radical neutrals, 
macromolecules, ions, electrons and photons is generated. 
The large species produced have different structures (e.g., linear or aromatic) and charging states
(cations, anions, or neutral). For many years it was commonly assumed that 
merely neutral radical and molecular species play a role in the plasma polymerization \cite{Yasuda}.

However, recent advances have revealed a crucial role of the plasma ions which was 
commonly overlooked. Specifically, this role is evidenced by  
the recent demonstration of very large ions whose masses are several times larger than the masses of 
original precursors. Moreover, the plasma ions can supply as much as 50\% of the mass of the deposited 
polymeric films \cite{ShortLangmuir2011}. This is supported by calculations and direct measurements showing that 
the ion fluxes during the synthesis of plasma polymers 
can be comparable to or even larger than the fluxes of neutrals. This conclusion is very similar to 
the results of other authors \cite{DenyJAP2004,Hash-plasma-chemistry-for-CNT-NT05}
discussed in Sec.~\ref{BU}. 

The numerical modeling results also confirm that ion-neutral reactions trigger plasma 
polymerization and lead to the production of macromolecules and 
nanoclusters of various structures including chains and aromatic rings \cite{Kaitlin2006}.  
For example, generation of higher hydrocarbon species in acetylene plasmas proceeds via
different pathways, triggered by the positive C$_2$H$_2^+$ 
\begin{equation}
\label{pos-ion-pathway}
{\rm C_2H_2^+} \rightarrow {\rm C_4H_2^+} \rightarrow {\rm C_6H_4^+} \rightarrow {\rm C_8H_6^+} \rightarrow {\rm C_{10}H_6^+}
\rightarrow ... , 
\end{equation}
and negative C$_2$H$^-$ 
\begin{equation}
\label{neg-ion-pathway}
{\rm C_2H^-} \rightarrow {\rm C_4H^-} \rightarrow {\rm C_6H^-} \rightarrow {\rm C_8H^-} \rightarrow ... ,
\end{equation}
molecular ions. Both channels (\ref{pos-ion-pathway}) 
and (\ref{neg-ion-pathway}) produce 
comparable contributions \cite{Kaitlin2006}. 

This is another manifestation of the complexity of plasma polymerization in organic precursors 
compared to silane plasmas where negative ion (anion) - triggered nanocluster formation
pathway is dominant \cite{Kaitlin2004} (see also Sec.~\ref{Nanocrystals}).  
These uniquely plasma-specific ionic polymerization pathways have been largely overlooked in the 
earlier classical models \cite{Yasuda}.

These effects  lead to high-quality, conformal and defect (e.g., pinhole, pore, etc.)-free
nanometer-thick polymer films. One such example is shown in Fig.~\ref{f27}(h,i) where highly-conformal 
coating with ultra-thin films of allylamine plasma polymer was used to reduce the thickness 
of nano-channel openings in an alumina nanoporous template.  
This thickness gradually reduced as the time of the plasma treatment increased. 
As a result, it was possible to control the rates of release of a vancomycin drug from the 
plasma polymer-coated channels \cite{Vasilev2010}.  

A very important feature of the plasma polymer nanofilms is their outstanding 
compatibility and adhesion with most of the commonly used inorganic substrates (e.g., silicon, gold, etc.).
A likely reason is the ability of the plasma to suitably prepare the deposition surface 
(including surface areas with nanometer dimensions) as discussed in Sec.~\ref{nano-plasma-surface}. 
This feature makes plasma polymers suitable for biomedical implants, {\it in vivo} 
sensing, and drug delivery systems in nanomedicine.  

Low-temperature reactive plasmas are also very effective for the precise control of 
the surface energy through the nanoscale surface texturing and/or functionalization. 
The surface texturing can be implemented in two ways, namely by using the 
bottom-up and top-down approaches. The top-down approach 
is most commonly implemented by using the plasma etching (e.g., in oxygen plasmas) and 
nanopattern transfer by using pre-fabricated or self-organized masks to produce a variety of 
ordered nanopatterns and arrays, see examples in Fig.~\ref{f27}(d-g) 
\cite{RossiAMat2007, RossiJPD2007}. 
This approach is very similar to Figs.~\ref{f23}(e,f) and \ref{f16}(a-d). 

Nanoscale etching in oxygen plasmas was also 
effective in processing and branching of single-crystalline organic (e.g., metalloporphyrin, 
metallophtalocyanine, and perylene) nanowires \cite{BorrasNanoscale2011}. This 
nanoscale plasma effect still requires adequate theoretical interpretation.

Recently, it also became possible to produce self-organized polymer nanopatterns without using 
any etching masks (Fig.~\ref{f27}(a-c)) \cite{DiMundoPPAP2010}. 
Mechanisms of the plasma-specific controls of the surface roughness 
in the bottom-up self-organization-led processes 
have been proposed \cite{GogolidesJPD2011}. 
Some of the relevant effects such as the 
surface stresses, heating, ion bombardment, and charging are quite similar to 
Fig.~\ref{f13}(a,b). However, the major difference between the 
hard and soft matter cases is in the overwhelming importance of the structural and compositional 
polymer surface transformations even under relatively mild plasma exposures.

This is why it is very difficult to identify the prevailing driving forces that lead to the 
formation of the large variety of the observed self-organized patterns on plasma-exposed 
polymer surfaces. The ease of localized surface melting accompanied with possible structural transformations of polymeric 
chains is one of the critical factors which makes the self-organization processes 
even more non-equilibrium, kinetics-controlled compared to the cases involving hard matter. 
Nevertheless, the clue to identify the prevailing driving forces should still be in 
localized {\em differences} in surface conditions, including the 
variation in the surface reaction probabilities across the nanostructured surface 
induced by non-uniform exposure to virtually any 
constituent of the ``plasma cocktail" (electrons, ions, UV, radicals, etc.).

The plasma non-equilibrium plays a major role in the generation of a large variety of building units in the 
gas phase through the electron- and ion-assisted dissociation and remodeling of the original precursor species.
This approach is pursued for the development of organic nanomaterials  
for the next-generation organic nanoelectronic, photovoltaic and optoelectronic devices
\cite{Angel_Barranco}.  
However, the issue how exactly does the plasma non-equilibrium translates into the 
overwhelming variety of self-organized patterns on reactive plasma-exposed polymer surfaces, remains 
essentially open. Although some ideas can be drawn from the available knowledge on atomic-precision 
surface reconstructions using ion beams \cite{ion-beams-self-org, ion-beams-self-org1}, 
the unique features of soft matter 
discussed above need to be rigorously taken into account to obtain a consistent description. 

Moreover, care should be taken because the plasma treatment very often 
introduces defects to organic frameworks and additional effort may be required to 
heal these defects. Self-healing polymers have some advantage in this regard \cite{self-healing-poly}. 
Thus, there is a clear opportunity to understand the mechanisms of these structural defects and their reconstructions.
This knowledge may lead to the optimized plasma exposure, e.g., through the control of 
fluxes and reactivity of the main species, using remote plasmas, relative contribution of direct 
ion bombardment, etc. on one hand and using appropriate polymers with 
the stronger self-healing ability.

Surface functionalization of soft organic  matter also benefits from plasma effects. 
A combination of the effective production of the relevant functional groups (e.g., 
OH-, COOH-, NH$_x$, CF$_x$, etc.) and activation of the dangling bonds on the surface makes the 
plasma-assisted polymer surface functionalization particularly versatile.  
Since these surfaces are usually tailored to enable specific functionalities, the 
plasma surface processing should have deterministic features. For example, in  
bio-recognition of target antigens, it is important to tether relevant antibodies to the 
surface areas specifically prepared (e.g., patterned) for cell attachment.     
In the following subsection, we will consider the plasma interactions 
with biological objects of different sizes and structures.

\subsubsection{\label{bio} Biological objects}

\begin{figure}[t]
\begin{center}
\includegraphics[width=12cm,clip]{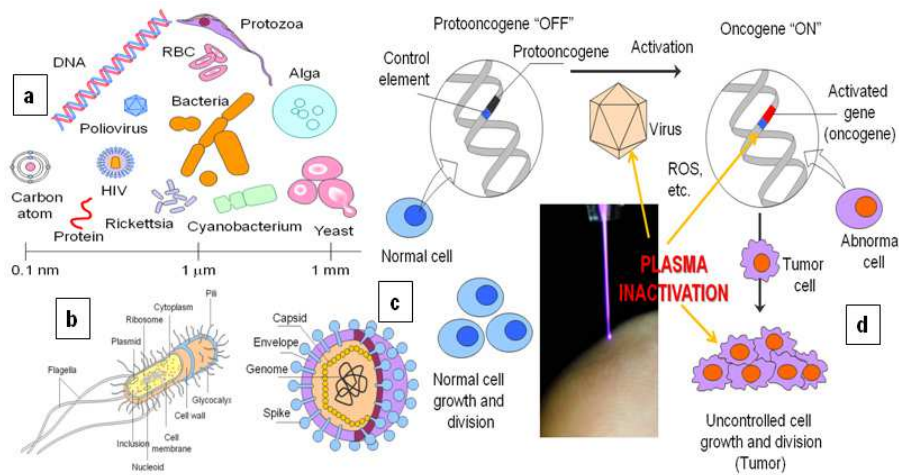} 
\caption{ \label{f28} Sizes of biological objects (a). Structure of 
a bacterial cell (b) and a virus (c). Mechanisms of interactions of 
plasmas and viruses in cancer activation and treatment (d). 
Inset shows a tiny ($\sim$10--100 $\mu$m) spot 
of an atmospheric plasma jet in direct contact with human skin.
Parts of graphs are re-used and modified with permission from Motfolio, 
Copyright \copyright Motfolio. 
} 
\end{center}   
\end{figure}

Interactions of gaseous low-temperature plasmas with biological objects has recently become
a highly-topical multidisciplinary area because of the rapid advances in the plasma biology, health care and 
medicine \cite{KongNJP2009, Laroussi-IEEE09, FridmanPPAP2010}. 
The key focus of these studies are the effects of low-temperature, mostly atmospheric-pressure plasmas on 
various biological objects of different sizes and structures sketched in Fig.~\ref{f28}(a). 
These effects are relevant to pathogen inactivation in food sterilization, oral hygiene, and treatment of 
infections, as well as controlling cellular responses in 
wound healing (e.g., blood coagulation), and more recently, aggressive cancer therapies.

The biological objects range from 
peptides and proteins with low-nm dimensions and chain-like  
structure to macroscopic hierarchically structured and responsive 
tissues and organs. The sizes of cells are in the micrometer range and vary between 
prokaryotic (Fig.~\ref{f28}(b)) and eukaryotic cells which also show very different 
internal and surface structures. These objects can be treated as microscopic 
substrates made of soft matter. 
These ``substrates" feature a large number of organelles (both on the surface and in the 
intracellular space) with sub-$\mu$m dimensions. Viruses are much smaller 
(typically $\sim$1--10s nm) and feature genetic material encased by a layer of capsid 
proteins and an outer shell (Fig.~\ref{f28}(c)).   

A nucleus of a typical low-$\mu$m size contains 
long double helical DNA structures as sketched in Fig.~\ref{f28}(d). Genes responsible for 
specific cellular responses are located in very small (typically nanometer) localized 
areas. For example, when an oncogene is activated (e.g., by oxidation 
involving reactive oxygen species (ROS) or by a virus-induced genetic mutation), 
intracellular signaling pathways (e.g., mediated by specific proteins) trigger faster and often
uncontrollable cell growth and division which leads to the development of cancerous tumors.   
When this happens, the cell first tries to repair any damage caused to its genetic material, and if 
this is not possible, activates apoptotic (programmed death) pathways mediated, e.g., by 
a ``guardian" protein {\it p53} \cite{NUS-Biomater2011}.
This programmed cell death follows a sequence of cellular and then nuclear disintegration steps.   
The products of apoptosis can be easily removed without 
any significant damage to the surrounding healthy cells. This approach is highly-promising 
to enable a highly-selective reducing of tumors on an otherwise healthy tissue \cite{KeidarBJC2011}.    
On the other hand, very different levels of cellular organizations make it possible to 
selectively inactivate bacterial pathogens without causing any damage to a healthy skin tissue.

What can the plasma do to help controlling so complex 
biological responses, both at the cellular and intracellular levels and how can 
microscopic plasma-specific effects  
be used? A surprisingly simple yet reasonable answer follows from 
the approach discussed in Sec.~\ref{BU}. 
Indeed, to induce the desired (deterministic) biological response (e.g., apoptosis), one should 
produce suitable biologically-relevant reactive species (e.g., reactive oxygen or nitrogen 
species, lipid peroxide, etc.) and then deliver them 
to the specified locations on the surface or in the interior of the cell, where 
relevant receptors that can trigger the desired signaling pathway are located. 
Given the microscopic sizes of the relevant surface areas, organelles, and proteins involved, 
nanoscale interactions of these objects with low-temperature plasmas may be particularly important.       

These interactions are fundamentally similar to the 
microscopic plasma effects in CNT nucleation and growth (Fig.~\ref{f10}). However, the specifics of the plasma 
interactions with biological objects are very different and the relevant studies of these 
effects are limited \cite{BabaevaJPD2010, HWLeeJPD2011}.      
Nevertheless, the plasma interactions have been studied for biological objects of 
different levels of organization, sizes, and dimensionality. 
While {\it in vitro} studies at the protein, virus, cell, and tissue levels are quite common, similar 
studies at the sub-cellular (e.g., organelle, receptor,
selected cell surface area, etc.) level are in their infancy.

Strong non-equilibrium of atmospheric plasmas very effectively 
produces reactive oxygen  and nitrogen 
species (ROS and RNS), such as O, O$^*$, O$_3^*$, OH, NO$^*$, etc.  
Similar to oxygen BUs in Fig.~\ref{f8}(b), these 
species are long-living and induce biological responses 
which are difficult to achieve using other species. For example, ROSs
can oxidize the DNA after permeating through the cell membrane and penetrating into the nucleus. 
Although measurements of the penetration depth of the plasma species into the cells are challenging, 
biological tests confirm the intracellular effects of the plasma-produced ROS and RNS \cite{XPLuPPAP2012}.

These species can penetrate as deep as $\sim$25 $\mu$m into a multilayer biofilm made of 
heat- and antibiotic-resistant {\it Enterococcus faecalis} bacteria and effectively inactivate them 
in a few tens of seconds \cite{plasma-flashlight}. 
Localized delivery of these species is also significantly affected 
by microscopic electric fields induced by the cell surface charging in a plasma \cite{BabaevaJPD2010}, 
which is conceptually very similar to the BU delivery and redistribution in the growth 
of surface-supported nanostructures in Fig.~\ref{f9}.

The energy dose delivered to the cells has to be controlled, as excessive overheating 
may lead to undesirable cell necrosis, loss of selectivity and even major damage of healthy cell tissues. 
For example, doses of the plasma exposure below $\sim$1 J/cm$^2$ are appropriate for sterilization and 
blood coagulation, while larger doses of $\sim$2--6 J/cm$^2$ are suitable to induce cellular apoptosis
or cell proliferation effects \cite{DobryninNJP2009}. Excessive doses 
of $\sim$10 J/cm$^2$ normally cause necrosis. These effects are quite similar to the 
control of energy delivery in the nucleation of metal oxide nanowires and processing of soft organic matter 
(Secs.~\ref{1D-oxide} and \ref{soft-matter}).

Interesting synergistic effects may arise when reactive radicals and plasma produced, coated, 
or functionalized nanoparticles are used 
simultaneously. For example, the rates of melanoma cancer cell death increased 5 times 
after antibody-conjugated Au NPs were added to open air plasmas \cite{IzaJPD2009}. 
The use of specific antibodies   
improves the selectivity of cancer cell treatment as well as the penetration of the 
plasma-produced reactive species into the intracellular space which is normally restricted by the cell membrane  
\cite{KongJPD2011}.

Below are a few examples of {\it in vitro} interactions of atmospheric plasmas 
with proteins, viruses, and cells. 
Amyloid fibrils, formed by the protein $\alpha$-synuclein,
which underlies Parkinson disease and the amyloid-$\beta$ peptide 
which is associated with Alzheimer disease have been successfully 
destroyed {\it in vitro} using atmospheric plasmas \cite{LaroussiAPL2010}.
A very effective inactivation of adenovirus cultures was also demonstrated and imaged 
using green fluorescence measurements upon infecting HEK 293A host cells with the plasma-exposed
viruses \cite{adenovirus}. 

Recently, an effective approach towards single-cell-level 
microplasma cancer therapy was demonstrated by confining the plasma plume
to the hollow central core of an optical fiber with the
inner diameter of approximately 15 $\mu$m \cite{Small-microplasma}. 
As discussed above, this area of the plasma localization is comparable 
with typical sizes of most of mammalian cells.

\begin{figure}[t]
\begin{center}
\includegraphics[width=11cm,clip]{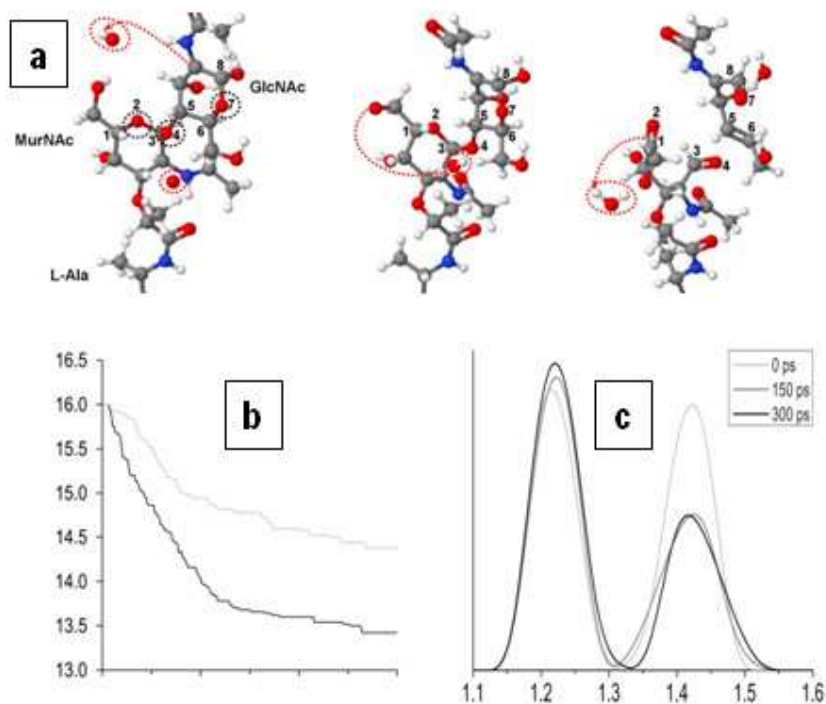} 
\caption{ \label{f-bio-new} Interaction of reactive plasma species with 
the peptidoglycan structure studied by atomistic simulations \cite{Erik-plasma-med}: 
(a) breaking mechanism of ether bonds in peptidoglycan structure by O-radical; 
(b) evolution of the number of ether bonds due to the interaction of O and O$_3$ species with PG;
(c) radial distribution function of ether bonds in PG.
Reproduced from \cite{Erik-plasma-med}. Copyright \copyright IOP Publishing Ltd and Deutsche Physikalische Gesellschaft. 
Published under a CC BY-NC-SA licence.} 
\end{center}   
\end{figure}

Therefore, understanding of atomic-level interactions
of the plasma-generated species with peptides and proteins is highly warranted. 
Recent numerical simulation study \cite{Erik-plasma-med} reports on the mechanisms of interaction of 
oxygen-containing radicals with peptidoglycan (PG). 
PG was used as a model system, as it forms the outer cell barrier of gram positive bacteria such as {\it Staphylococcus Aureus}. 

It was found that both O-radicals and O$_3$ molecules effectively break structurally important C-O ether bonds 
in the murein part of PG, leading to structural damage of the bacterial cell wall (Fig.~\ref{f-bio-new}). 
This underlying mechanism is based on multiple H-abstraction, leading to double bond 
formation and dissociation of ether bonds. Breaking of C-C and C-N bonds was observed as well. 

The subsequent study \cite{Erik-plasma-med2} revealed the mechanisms of interactions of H$_2$O$_2$ molecules with PG. 
In this case, the formation of HO$_2$ hydroperoxyl radicals is observed. These radicals are experimentally known 
to be strong bactericidal oxidants, leading to bacterial inactivation in aqueous environment.

This area presently remains largely unexplored and represents an opportunity for future studies. 
This can in part be attributed to the fact that numerical 
simulations of the interaction of the plasma with living cells 
are far more difficult compared to modeling a similar interaction with 
solid materials, due to the complexity of their structure and functions 
in an organism.

We note that in the case in Fig.~\ref{f28}(d), 
viruses may also need to be treated in the intracellular space to avoid their 
undesirable replication and release of the 
viral genetic material which may be a strong factor in enhancing the oncogene activation. 
Also, plasma exposure should not negatively affect the 
existing genomic, proteomic, and transcriptomic landscapes of the cells which in turn 
determine the cellular self-regulation pathways that may prevent the uncontrolled 
cell growth and division \cite{NUS-Biomater2011}.  
Solution of this problem may help tackling 
cancer, an issue with a century-long history.

\section{\label{Sec5} Nano-plasmas: interplay of the size and the fourth state of matter }

This section focuses on the second aim of the review (see Sec.~\ref{introC}), namely nano-plasmas near or from solid matter.  
We will also consider a typical example when nanoscale plasmas near a solid nanoparticle may be excited. 
The effects of the nanoscale spatial localization of the energy and ionized matter are of primary interest.
In Sec.~\ref{iPVD} we discuss the possibility to generate very dense and strongly-non-equilibrium 
plasmas in physical vapors of solids produced through the interaction of intense 
lasers with solid (e.g., metal) surfaces. 
In Sec.~\ref{nano-plasmas} we present an example when plasmons within a gold nanosphere help 
generating gaseous nano-plasmas confined within a liquid nano-bubble 
developing around the nanoparticle.

\subsection{\label{iPVD} Non-equilibrium nano-plasmas {\em of solids}}

Let us now consider strongly-non-equilibrium plasmas  
in physical vapors of solid materials. These vapors are produced by exposing solid target surfaces to
laser beams (laser ablation) or plasmas. The evaporated species (BUs) are ionized and 
then undergo energetic condensation onto a solid substrate. 
In pulsed laser deposition (PLD) used for the synthesis 
of a large number of nanofilms and nanostructures, 
the target is exposed to a pulsed laser beam. 
High-energy-density (HED) plasmas of dense plasma focus (DPF) or Z-pinch devices 
also rapidly evaporate material of solid targets to produce a variety of BUs. 

The degree of ionization of the BU vapor in both cases is very high, 
and is often close to unity. While HED plasmas require source gases, PLD may be 
conducted under high-vacuum conditions. In another example of the ionized physical vapor deposition 
(i-PVD), atoms of a solid material are sputtered from a target by 
the plasma ion bombardment \cite{AndersJPD07}. In this case, the vapor of the 
sputtered species is weakly ionized and less energetic and dense compared 
to the PLD and HED plasmas. 

Here we focus on non-equilibrium phenomena and plasma dimensions using PLD as an example.
This choice is motivated by the discovery of fullerenes C$_{60}$ 
in  hot carbon vapor plasmas produced by pulsed laser ablation of graphite and nucleation of 
carbon clusters in a thermalizing He flow \cite{KrotoNature1985}. 
Interestingly, this discovery, which has made an enormous impact on the present-day 
nanoscience and nanotechnology, was made in a ``plasma nanoscience-type" experiment which aimed at 
revealing the effects of plasmas and ion-molecule reactions on the formation of long-chain carbon molecules in 
circumstellar shells and interstellar space \cite{KrotoRMP1997}.

Laser ablation plumes contain 
stable neutrals (atoms, molecules, and clusters) 
as well as metastable species, ions, and electrons. An inherent pulsed nature of PLD  
opens new avenues for far-from-equilibrium surface processes which lead to  
many diverse metastable nanomaterials \cite{KomatsuJPD07}. When a solid target is
exposed to a laser beam, energetic ($\sim$1--100 eV) species are produced 
in a dense plasma plume. The plasma is  highly-ionized, in some cases reaching 
$\sim$100\% ionization degree.     
The flux of the charged species has a fountain-like shape and 
rapidly cools upon deposition onto a substrate. These two features are  similar 
to the HED plasma case \cite{MalhotraJPD09,RuzicJPD2011}. 
The temperature in the localized plasma 
plume can be as high as $\sim$20000 K. 
Noting that the plasma size is typically just a fraction of a millimeter, and using a 
typical ablation yield (for metals) of $\sim$10$^{15}$ atoms per pulse, one can estimate the 
ionized gas pressures of the order of 10$^9$ Pa \cite{PLD-RMP2000}.

Under such conditions, the Debye length (\ref{Debye-length}) 
of the plume plasma is a few orders of 
magnitude smaller than the axial plasma dimension, and can be 
in the range of {\em a few hundred nanometers}. Importantly, the ratio of the Debye length to the plume size 
increases only slightly during the plasma expansion \cite{Bulgakova}.
Surface processes (e.g., migration, nucleation, etc.)
can be enabled by using the energy of the impinging species and in many cases no external substrate heating 
is necessary. The degree of ionization of the ablated plasma as it expands and then impinges 
on the substrate is much lower ($\sim$ 1\% or even less) than in the ablated plume near the 
target \cite{PLD-RMP2000}, yet remains much higher compared to low-pressure  
discharge plasmas. 

PLD allows building units ablated from metal  to combine with the species generated 
in the plasmas of feed gases to produce, e.g., 
metal oxide or nitride compounds. 
Gases are also used to thermalize the plasma species through multiple collisions. For example, 
in the pioneering synthesis of C$_{60}$ buckminsterfullerenes \cite{KrotoNature1985}, the  
ablated carbon atoms produced in the plume plasma were 
entrained in a gas pulse confined to a narrow channel. 
Multiple collisions of C and He atoms led to the 
effective cooling, condensation, and nucleation of carbon species to 
produce carbon (most notably C$_{60}$) clusters 
emerging from the channel exit in a supersonic jet. 
This rapid cooling triggered a strongly non-equilibrium kinetic process which resulted in 
nucleation of C$_{60}$, which is less stable than graphite in 
thermodynamic sense.  

During the BU production phase through the interactions of the laser beam with the target surface, 
the degree of thermal non-equilibrium between the plasma species 
strongly depends on the duration of the laser pulse. 
Let us note  that the characteristic time of energy transfer from the electrons heated 
by the laser to the  ions, $\tau_{ei}$ is typically within the 1--10 ps range. 
Interestingly, depending on the relation between the laser pulse 
duration $\tau_p$ and $\tau_{ei}$, the plasma can be either non-equilibrium ($T_e \gg T_i$) 
or thermal ($T_e \sim T_i$). This condition is quite similar to the thermalization and non-equilibrium conditions 
for atmospheric-pressure plasmas (See Fig.~\ref{f31} and the associated discussion) \cite{Mariotti-Sankaran}. 
More importantly, these different conditions lead to the very different outcomes in terms 
of BU production, and eventually the nanoscale synthesis process.

For long pulses $\tau_p \gg \tau_{ei}$, the energy transfer from the electrons to the ions is 
effective and thermal equilibrium ($T_e \sim T_i$) is established. In this case, 
the material is heated through the heat conduction mechanism. This heating affects 
the surface layer of thickness 
\begin{equation}
\label{surflayer8}
l_h \sim \sqrt{\kappa_h \tau_p},
\end{equation} 
where $\kappa_h$ is the solid material-specific heat conduction coefficient. 

In this case, the bulk heating, melting, and evaporation of 
the solid target material take place in the {\em near-equilibrium} regime. 
Since the melts usually contain both atoms and atomic clusters, the produced cocktail of BUs in the 
evaporated plume contains atoms, clusters and even large droplets. A large abundance of 
droplets is in many cases undesirable for the synthesis of delicate nanostructures.
On the other hand, the ablated material preserves its stoichiometric elemental composition, which is  
retained by the deposited nanomaterials. 
  
In the opposite case of very short pulses ($\tau_p \ll \tau_{ei}$), the transfer of energy from 
the hot electrons to the lattice ions is ineffective and the electrons and ions find themselves 
out of thermal equilibrium, $T_e \gg T_i$.  This happens during the femtosecond (fs) 
laser ablation when the laser pulse duration 
is in $\sim$100s fs range. In this case, the energy delivery stops well before 
the material can be heated through heat conductance. The electrons within the laser 
penetration depth gain energy, which may exceed the threshold for the electron emission from the surface. 
As the electrons leave this layer, they pull the oppositely charged ions out of the target.
Moreover, the ions also experience strong repulsion due to the loss of the electron screening,
which also facilitates their exit from the material. This also leads to very high 
ionization degrees in femtosecond laser ablation plasmas.  
 
This strongly non-equilibrium ablation leads to several interesting features for the nanoscale synthesis.
Most importantly, purely ionic/atomic fluxes are achieved. The degree of ionization 
and energy (and also clustering) of these building units can be controlled by 
the electric fields, distance between the target and the substrate, 
as well as the pressure and composition of the background gas.  
This process is also energy efficient since the strong non-equilibrium conditions reduce the 
energy loss due to the material bulk heating. 
These very interesting non-equilibrium processes have made femtosecond laser 
ablation a common nanotool \cite{LouchevNT2006, KabashinNRL2010}, as was predicted earlier 
\cite{PLD-RMP2000}. 
   
Plasma-specific non-equilibrium and process kinetics-related effects are ubiquitous and critical in 
other cases involving ionized physical vapor of solids, e.g., HED plasmas, physical sputtering, 
ion beam-assisted evaporation, and a variety of hybrid techniques 
\cite{AndersJPD07, Helmersson, Rajdeep-review}.

\subsection{\label{nano-plasmas} Nano-plasmas meet plasmons} 

In the previous section, we did not specify the sizes of the solid targets from which atoms of ablated solid-state 
materials evaporate to form plasma plumes from the physical vapor phase. Importantly, the power density delivered
by the laser beam should be sufficient for the rapid transformation from the solid phase to the plasma phase. 
In other words, the laser intensity is expected to be high to produce the plasmas with possible sub-micrometer dimensions 
(nano-plasmas) near the solid target. 

Let us now pose the not so obvious but important questions: 

1) is it possible to generate nano-plasmas near solids using much lower 
energy than in Sec.~\ref{iPVD}? 

2) can reducing the size of solids to the nanoscales help   
achieving these goals? 

In other words, we will continue discussing the interplay of the plasma and size effects at the nanoscale.  

The answer to the first question is a yes if the energy delivered to the solid can be concentrated to a very small volume, for example, 
nanoparticles. The amount of the energy needed to generate plasmas near solids may be further reduced if some 
resonant phenomena within these nanoparticles are used to quickly ``expel" the energy out of the nanoparticle (e.g., to prevent 
melting and evaporation of the NP itself) and  further concentrate the energy to even smaller volumes.
As will be shown in the example below, this can be achieved by using plasmonic excitations 
in gold nanoparticles generated by low-intensity ultrafast femtosecond lasers. This also gives a positive answer
to the second question. 

The plasmonic effects originate due to the collective responses of the free-electron gas within the ordered lattice 
of metal ions to external electromagnetic fields. The eigenfrequency of collective 
electron oscillations (plasmons) with respect to the lattice ions 
 is determined by the free electron density and is described by the same equation as the plasma frequency $\omega_{pe}$ in 
(\ref{Debye-length}). This similarity in mathematical descriptions (within the framework of classical physics) of 
electron resonant phenomena has led to many parallels between the plasma-like effects in gaseous plasmas and solids with 
free electrons \cite{Pines1956, Hoyaux, Raimes, Jonscher, AdvPhys-PlasmaPlasmon2011,AmandaEPJD}. 
When plasmons are excited (e.g., by a laser with the frequency in the optical range matching the plasmon frequency)
within metal nanoparticles, the energy of resonant plasmon oscillations (localized surface plasmons, LSPs) 
is mostly concentrated within the nanoparticle.  

In other words, the plasmon oscillations are localized within the 
nanometer scale. However, it is very important to emphasize that these plasma-like oscillations 
originate without the nano-plasma generation through the common sequence of phase 
transitions. Indeed, the matter within the metal nanoparticle remains in the solid state 
despite significant electron delocalization. Thus, nano-plasma-like oscillations and the associated nanoscale 
electromagnetic energy localization may arise without nano-plasma generation within solids. 
This is why care should be taken not to over-simplify terminology to, e.g., to ``nano-plasma oscillations" 
without a clear explanation on what is actually meant.  
   
When the external electromagnetic field is off-resonance with the plasmon excitations, the field is 
scattered, focused, and strongly enhanced outside of the nanoparticle. 
This effect is used to generate nano-plasmas in a liquid near the nanoparticle in the example discussed below.
Further details of the effects of nanoscale energy localization which arise due to 
the plasma-like and size effects 
in nano-solids can be found elsewhere \cite{Stockman2011, AtwaterJAP2005, 
MaradudinPhysRep2005, ChemRev2009-LSP-Sensors, GreenNaturePhoton2012,
Akimov-OPEX-2009, Akimov-plasmonics2009, Silverinha, Hao-Schatz}.

\begin{figure}[t]
\begin{center}
\includegraphics[width=13.5cm,clip]{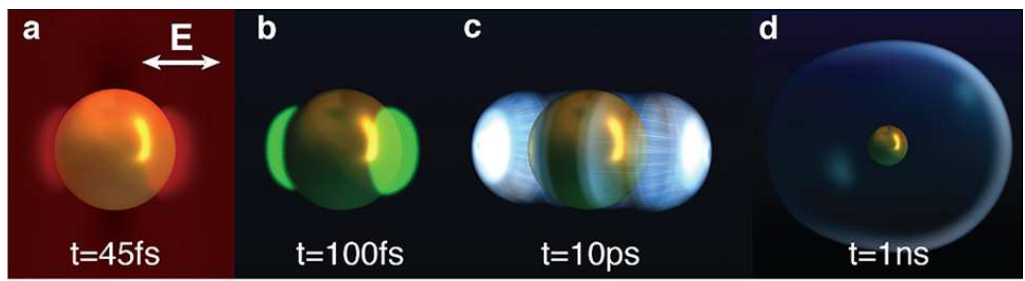} 
\caption{ \label{f-nanoplas} Sequence of physical processes leading to the
formation of a nano-plasma and then a vapor nano-bubble around a plasmonic 
nanoparticle exposed to an ultrafast infrared laser \cite{NanoplasmaNL2012}: 
(a) near-field is enhanced during the laser pulse; (b) Enhanced near-field ionizes the surrounding
water, creating a nano-plasma near the poles of the particle; (c) water heating due to the plasma
fast relaxation and  collisions with the plasma ions; a strong pressure
wave is then released in the water; (d) a nanoscale bubble is formed
around the particle. Time count starts from the onset of the laser
irradiation. Reprinted with permission from \cite{NanoplasmaNL2012}. 
Copyright \copyright (2012) American Chemical Society.} 
\end{center}   
\end{figure}

Let us now consider how nano-plasmas may be generated in a medium 
surrounding the plasmonic nanoparticle. In liquid media, these nano-plasmas in turn 
may generate nano-bubbles, the physical phenomenon that is usually referred to as 
nano-cavitation. These nano-bubbles are quite different from common vapor bubbles and
are actively pursued in medical imaging and cancer research. The nano-bubbles may be generated 
by heating the nanoparticles and the surrounding media in two different ways. 

The first way relies 
on the resonant heating of plasmonic nanoparticles and the excitation laser wavelength is 
chosen relatively close to the plasmon resonance where optical absorption dominates the scattering \cite{Lukianova}. 
The nano-cavitation mechanism is then determined by the heat transfer from the 
hot nanoparticle to the surrounding water. The nano-plasma oscillations are thus selectively excited 
to generate the heat to produce the nano-bubbles. 

The other way sketched in Fig.~\ref{f-nanoplas} relies on femtosecond (45 fs) laser excitation in the infrared 
wavelength (800 nm) where the scattering cross-section is much larger compared to the absorption \cite{NanoplasmaNL2012}. 
The 600-1100 nm wavelength excitations are particularly important to medical imaging and therapy because of 
good transparency of living tissues in this range.  

As can be seen in Fig.~\ref{f-nanoplas}(a), the near-field is 
strongly enhanced near the opposite poles of the gold nanoparticle, 
which is not heated because of the low, off-plasmon-resonance optical absorption. 
Consequently, the strong near-field ionizes the surrounding water as shown in Fig.~\ref{f-nanoplas}(b). 

It is quite possible that the gaseous 
molecules dissolved in the water are ionized first similarly 
to the plasma discharges in liquids mentioned in Sec.~\ref{nc-from-liquid}. 
However, because of the very small size of the nanoparticle ($\sim$100 nm), it is
not clear whether any small gaseous nano-bubbles surround the nanoparticle before 
the laser excitation. Therefore, the exact mechanisms of this ionization 
are currently unclear and represent an interesting opportunity for the future research.
  
Importantly, the plasma is confined to the near-field localization area, thus clearly becoming a
nano-plasma. This plasma may reach its peak density $\sim$4$\times$10$^{22}$ cm$^{-3}$ which rapidly reduces
by at least one order of magnitude within a few picoseconds \cite{NanoplasmaNL2012}. 
The energy absorbed in the nano-plasma is one order of magnitude higher compared to the 
energy absorbed by the nanoparticle. Localized energy transfer from the nano-plasma 
is then responsible for the generation of the 
pressure wave (Fig.~\ref{f-nanoplas}(c)) and eventually, cavitation nano-bubbles around the plasmonic 
nanoparticle (Fig.~\ref{f-nanoplas}(d)).  

Therefore, by tuning the resonant and non-resonant plasmon excitations 
within a nano-solid,
one can selectively 
enable very different nanoscale heating mechanisms 
of the surrounding nano-bubbles in water. The more recently demonstrated mechanism which involves 
the nano-plasma generation in the liquid medium around the poles of the plasmonic 
nanoparticle \cite{NanoplasmaNL2012}
is particularly interesting. Indeed, it shows that nano-plasmas 
may be generated at the nanometer-sized solid interface through 
plasma-like effects of nanoscale energy localization 
and opens many new opportunities for 
future research.

\section{\label{conclusions} Conclusion and outlook }

We hope that the examples presented in this review have convinced the reader that both dimensions 
of plasma nanoscience - the plasma-for-nano and nano-plasma not only 
represent very interesting research areas on their own but also in many cases interact at the 
interdisciplinary interface. This makes the 
field particularly fertile and sets it for the future 
expansion, both in breadth and depth. 

In the {\em plasma-for-nano} aspect, non-equilibrium and kinetic properties of the plasma offer many effective 
controls in nanoscale synthesis and processing thereby opening many interesting opportunities 
for non-equilibrium nanoscale synthesis and processing. 
In this way the  nano-architectures  can be tailored to feature 
unusual properties that originate from the 
exotic, non-equilibrium, kinetic-driven approach to nano-assembly.
  
The many plasma-specific effects also offer a plethora of 
opportunities to process nano-solids with 
different structural complexity. These solid materials 
range from simple 0D atomic clusters to complex 
and hierarchical hybrid structures.  
We have focused on the plasma-surface interactions with microscopic 
spatial localizations and followed the unique features of these interactions that 
arise because of the ionized state, reactivity, non-equilibrium, and other attributes of the 
plasmas. These attributes lead to the unique synthesis and processing 
of a broad range of materials systems and nanoscale objects 
with all possible dimensionalities.          

In the {\em nano-plasma} aspect, microscopic spatial localization of the plasma
enables many exotic physical effects. Micro-plasmas of gas discharges or even smaller plasmas 
of dense ionized physical vapors of solids feature
extremely non-equilibrium properties and ultrafast dynamics. These features 
make these plasmas very effective for the production of metastable nano-solids  
under far-from equilibrium conditions. Non-equilibrium nano-plasmas 
around nanometer-sized solid objects in liquids 
is a very interesting yet unexplored area, which is relevant to applications where nanoscale 
energy localization  is required. Size-dependent plasma-like (plasmonic) effects in nano-solids 
help generating the above nano-plasmas and are thus also of interest to plasma nanoscience research. 

An open question remains if it is possible to generate nano-plasmas within solids? 
Exotic states of matter such as warm dense matter 
generated in nanometer-thin metal films by ultrafast high-energy-density lasers, 
have attributes of both superheated nano-solids and strongly coupled 
nano-plasmas \cite{Ng-WDM, WDM2, WDM3, WDM4}. Future research may clarify 
if it is possible to control the relative importance of these two phases.  

The plasma nanoscience approaches should be tailored to lead to 
new materials with properties that are difficult to achieve otherwise.  
The functional properties should be targeted for applications in several 
important fields such as energy conversion and storage, biomedical, environmental, 
nanoelectronic and other devices.  

It is important that the new materials, devices, and processing 
technologies being developed using the plasma nanoscience approaches need to 
be energy-efficient as well as environment- and human-health friendly.   
For example, very high energy efficiency in nanoscale synthesis  
can be approached using control of energy and matter with nanometer and
nanosecond precision in strongly non-equilibrium plasmas \cite{PaiSciRep}.     
Plasma-based approaches are also capable to produce 
nanostructured materials directly from natural resources rather than 
commonly used expensive purified precursors \cite{Seo-Honey,MeyyaCarbon2012}.
The physics-based, deterministic approaches, similar to the ideas advocated 
in this review are quite likely to lead to positive outcomes.

The number of relevant reports is increasing very rapidly, and 
a substantially larger body of new knowledge on the plasma-specific 
effects and their utility in nanoscale applications may be expected in the near future. 
Numerical modeling of the plasma-specific effects involved 
is therefore highly-warranted to improve the level of our understanding of 
the many promising experimental results. 

The many examples in this 
review suggest that the interplay of the nanometer spatial localization and 
plasma-specific effects generates a plethora of interesting and rich physics,
which offer many exciting opportunities for future interdisciplinary 
research and the development of relevant applications. 
Finally, {\em nano-size matters - so does the fourth state of matter!}

\subsection{Note added in proof} 
Since the manuscript has been accepted for publication, 
we became aware of the following most recent relevant publications. 
Nanoscale plasma-surface interactions are very effective for the 
precise surface engineering of Si nanocrystals to enable specific quantum 
confinement effects \cite{[381]} and carbon nanotube growth without metal catalysts \cite{[382]}. 
By controlling the plasma process parameters, it is possible to tailor the volume 
fraction of Si nanocrystals embedded in amorphous Si matrix (nc-Si films), 
which dramatically improves the efficiency of recombination-free charge 
transfer (and also the open circuit voltage and photo-conversion efficiency) 
in nc-Si:H-based solar cells \cite{[383]}. Microwave plasma CVD was used to produce 
the previously unknown SiC tetrapod structures which show photoluminescence 
at wavelengths spanning the visible to near-infrared spectral range; 
other methods have previously failed to synthesise these nanostructures \cite{[384]}. 
Relevant effects have been discussed in Sec. 4.1 and 4.2.1. Self-organization of 
nanoparticles on plasma-exposed surfaces can now be monitored in real time through 
the very recent advent of in situ transmission electron microscopy-based platform 
for real-time characterization of nanoscale plasma-surface interactions \cite{[385]}. 
See Sec. 3.6 and 4.1.3 for relevant physics. Plasma-assisted electrochemistry 
improves control of the properties and size distributions of plasmonic 
Au nanoparticles through the unique plasma-specific mechanisms \cite{[386]}. 
Interactions of light with similar surface-supported nanoparticles lead to 
interesting plasma-like effects on the tunnel current in nanoislanded gold films \cite{[387]}. 
Relevant discussions can be found in Sec. 4.1.2 and 5.2. Surface functionalization and 
morphology control in networks of vertical graphene-like structures can be used to 
control their wettability and magnetotransport properties \cite{[388],[389]}. 
These 2D structures have been extensively discussed in Sec. 4.3.2. 
Plasma nanoscience approaches help achieving the goals of green and 
sustainable nanoscience \cite{[390]} briefly mentioned in Sec. 6. Plasmas enable 
reforming of diverse precursors ranging from fatty substances (e.g., butter) 
\cite{[391]} to petroleum asphalt \cite{[392]} to generate graphene nanosheets (see Sec. 4.3.2). 
These and similar graphene-like structures have shown superior performance in 
supercapacitor \cite{[391]} and fuel cell \cite{[393]} applications. Synergistic effects 
of atmospheric-pressure plasmas and nanometer-sized catalysts prove instrumental 
in the solution of environmental problems such as ozone destruction \cite{[394]}; 
plasma-catalysis is presently developing at a very rapid pace and represents 
a topical area of research. These randomly selected examples suggest that the 
number of relevant publications is increasing very rapidly and the reader is 
encouraged to use the basic information 
and references presented in this review to continuously monitor the progress 
in the fertile and rapidly developing plasma nanoscience research field.

\subsection{Acknowledgments} 
We sincerely thank many our colleagues for fruitful collaborations, discussions, and 
critical comments, with apologies of not being able to mention by name 
due to limited space. We also thank all authors of original figures for 
the permissions to reproduce.   
This work was partially supported by the Australian Research Council and CSIRO's
Science Leadership Program. 
K. O. is grateful to the University of Sydney (Australia) 
and Nanyang Technological University (Singapore) 
as Host Organizations for his ARC Future Fellowship as 
well as Huazhong University of Science and Technology (China), 
the University of Technology Sydney (Australia), and the University 
of Wollongong (Australia) for the visiting, adjunct, and honorary professor support.
We are particularly thankful to the anonymous referee
for the insightful discussion of states of matter and phase transitions
at nanoscales. We also thank every person who has ever contributed to the  
relevant areas and apologize for not being able to include all these results, 
although would certainly do that if that were physically possible.

\appendices

\section{\label{acronyms} Alphabetic list of acronyms used in the review}

0D - zero-dimensional \\
1D - one-dimensional \\
2D - two-dimensional \\
3D - three-dimensional \\
$a$-Si -- amorphous silicon \\ 
AD -- adsorption \\ 
ADH -- adsorption of hydrogen atoms \\ 
AFM –- atomic force microscopy \\
ALD -- atomic layer deposition \\
ALE -- atomic layer etching \\
BD -- bulk diffusion \\ 
BU -- building unit \\
C-dot -- carbon (quantum) dot \\ 
CNF –- carbon nanofiber \\
CNP –- catalyst nanoparticle \\
CNT –- carbon nanotube \\
CNTP –- carbon nanotip \\
CNW -- carbon nanowall \\ 
CVD -- chemical vapor deposition \\
DBD -- dielectric barrier discharge \\ 
DC -- direct current \\ 
DFT -- density functional theory \\
DNA -- deoxyribonucleic acid \\
DPF -- dense plasma focus \\  
DS -- desorption \\ 
DSH -- desorption of hydrogen atoms \\ 
EEDF -- electron energy distribution function \\
EFE -- electron field emission \\  
EV -- evaporation \\ 
fs -- femtosecond \\ 
FTIR -- Fourier transform infrared spectroscopy \\
GML -- graphene monolayer \\ 
GNR –- graphene nanoribbon \\
GNS –- graphene nanosheet \\
GO -- graphene oxide \\ 
GT -- Gibbs-Thomson (effect) \\
HED -- high energy density (plasma) \\ 
HEK -- human embryonic kidney (cell line) \\ 
HIN -- hydrogen-induced neutralization \\
HR -- hydrogen recombination \\ 
HRTEM –- high-resolution transmission electron microscopy \\
LEEM -- low-energy electron microscopy \\ 
LOM -- local order measure \\
LP -- lone pair (valence band) \\  
LSP -- localized surface plasmon \\ 
ID -- ion decomposition \\ 
IID -- ion-induced dissociation \\ 
i-PVD -- ionized physical vapor deposition \\ 
ITO -- indium tin oxide \\ 
MD -- molecular dynamics \\ 
MW -- microwave \\ 
MWCNT –- multi-walled carbon nanotube \\
$nc$-Si -- nanocrystalline silicon \\ 
NCD -- nanocrystalline diamond \\ 
NIR -- near-infrared \\ 
NND -- nearest neighbor distance \\ 
NP -- nanoparticle \\ 
NS -- nanostructure \\
NW –- nanowire \\
PECVD –- plasma-enhanced chemical vapor deposition \\
PG - peptidoglycan \\ 
PL - photoluminescence \\
PLD -- pulsed laser deposition \\ 
PLE - photoluminescence excitation \\
PMMA -- polymethylmetacrylate \\ 
PS -- polystyrene \\ 
PSS -- plasma-solid system \\
QD -- quantum dot \\
RF -- radiofrequency \\ 
RIE -- reactive ion etching \\ 
RNS -- reactive nitrogen species \\
ROS -- reactive oxygen species \\
SAED -- selected area electron diffraction \\ 
SD -- surface diffusion \\
SEM --  scanning electron microscopy \\
SERS -- surface enhanced Raman scattering \\  
STM -- scanning tunneling microscopy \\   
SWCNT –- single-walled carbon nanotube \\
TD -- thermal dissociation \\
TEM -- transmission electron microscopy \\  
THz -- terahertz \\ 
UHF -- ultra-high frequency \\
UNCD -- ultrananocrystalline diamond \\ 
USGs -- unsupported graphenes \\
UV -- ultraviolet \\
Vis -- visible \\
VLS -- vapor-liquid-solid \\ 
VSGs -- vertically standing graphenes \\ 
XPS -- x-ray photoelectron spectroscopy \\

\end{document}